\documentclass[12pt,preprintnumbers,tightenlines,superscriptaddress,nofootinbib,floatfix]{revtex4PATCH}

\usepackage[page,title]{appendix}
\usepackage[normalem]{ulem}
\usepackage{url}
\usepackage{graphicx}
\usepackage{amssymb}
\usepackage{bm}
\usepackage{mathrsfs}
\usepackage{amsmath} 
\usepackage{amsfonts}
\usepackage{color}
\usepackage{tikz}
\usepackage{slashed}
\usepackage{cancel}
\usepackage{hyperref}
\hypersetup{colorlinks,citecolor= black,linkcolor= black,urlcolor=black}
\usepackage{slashed}
\usetikzlibrary{arrows}
\usetikzlibrary{shapes}
\usetikzlibrary{trees}
\usetikzlibrary{matrix}
\usetikzlibrary{arrows} 			
\renewcommand{\draft}[1]{{\color{blue}#1}}
\usepackage{graphicx,epsfig,psfrag,bm,amssymb}
\usepackage{dcolumn}
\usepackage{bm}
\usepackage{color}
\usepackage{mathrsfs,amsfonts}

\usepackage{latexsym,cancel,amssymb,amsmath,mathrsfs,color}
\usepackage{graphicx}
\usepackage{epstopdf}
\usepackage{mciteplus}
\usepackage{latexsym}
\usepackage{amsthm}
\usepackage{amsmath}
\usepackage{amssymb}
\usepackage{hyperref}
\usepackage{bbm}
\usepackage{bm}
\usepackage{xfrac}
\usepackage{colordvi}
\usepackage{comment}
\usepackage{dcolumn}
\usepackage{times,latexsym,graphicx,wrapfig}
\usepackage{epsfig,lineno,bm}
\usepackage{slashed}
\newcommand{\beq}{\begin{eqnarray}}
\newcommand{\eeq}{\end{eqnarray}}
\newcommand{\be}{\begin{eqnarray}}
\newcommand{\ee}{\end{eqnarray}}

\newcommand{\Lag}{\mathcal{L}}

\newcommand{\benum}{\begin{enumerate}}
\newcommand{\eenum}{\end{enumerate}}
\newcommand{\bi}{\begin{itemize}}
\newcommand{\ei}{\end{itemize}}

\newcommand{\zp}{Z^\prime}
\newcommand{\Ap}{{A^\prime}}
\newcommand{\mAp}{m_{A^\prime}}

\newcommand{\zmax}{z_{\mathrm{max}}}
\newcommand{\zmin}{z_{\mathrm{min}}}

\newcommand{\Ebeam}{E_{\mathrm{beam}}}

\DeclareMathOperator\tr{tr}
\DeclareMathOperator\diag{diag}
\renewcommand{\zp}{Z^\prime}

\begin{document}


\title{Dark Matter, Millicharges, Axion and Scalar Particles, Gauge Bosons, and Other New Physics with LDMX }

\date{\today}

\preprint{FERMILAB-PUB-18-310-A, SLAC-PUB-17297}

\medskip 

\author{Asher Berlin}
\affiliation{SLAC National Accelerator Laboratory, Menlo Park, CA 94025, USA}
\author{Nikita Blinov}
\affiliation{SLAC National Accelerator Laboratory, Menlo Park, CA 94025, USA}
\author{Gordan Krnjaic}
\affiliation{Fermi National Accelerator Laboratory, Batavia, IL 60510, USA}
\author{Philip Schuster}
\affiliation{SLAC National Accelerator Laboratory, Menlo Park, CA 94025, USA}
\author{Natalia Toro}
\affiliation{SLAC National Accelerator Laboratory, Menlo Park, CA 94025, USA}

\begin{abstract}

The proposed LDMX experiment would provide roughly a meter-long region of instrumented tracking and calorimetry that acts as a beam stop for multi-GeV electrons in which each electron is tagged and its evolution measured. This would offer an unprecedented opportunity to access both collider-invisible and ultra-short lifetime decays of new particles produced in electron (or muon)-nuclear fixed-target collisions. In this paper, we show that the missing momentum channel and displaced decay signals in such an experiment could provide world-leading sensitivity to sub-GeV dark matter, millicharged particles, and visibly or invisibly decaying axions, scalars, dark photons, and a range of other new physics scenarios. 

\end{abstract}

\maketitle

\newpage
\tableofcontents
\newpage

\section{Introduction}\label{sec:Introduction}

Particle dark matter (DM) science is undergoing a revolution, driven simultaneously by recent advances in theory and experiment. 
New theory insight has motivated broadening the mass range for DM searches to include the entire MeV$-$TeV range and beyond, extending the traditional Weakly Interacting Massive Particle (WIMP) paradigm where the DM relic abundance is set by the freeze-out of annihilation reactions mediated by the weak interaction. 
The focus on the MeV$-$TeV range retains a healthy emphasis on the known mass scales of the Standard Model (SM).
It is also now widely recognized that it is important to search not only for DM itself, but also for particles that can mediate dark sector interactions with the SM, especially in the case of sub-GeV dark sectors. 
At the same time, ongoing advances on key experimental fronts promise to unlock much of the well-motivated and unexplored sub-GeV DM territory in the coming years~\cite{Battaglieri:2017aum}. Therefore, now is an especially exciting time to carefully scrutinize how to best leverage these opportunities to achieve as much science with strong discovery potential as possible. 

The Light Dark Matter eXperiment (LDMX) collaboration has recently proposed a new small-scale experiment to measure {\it missing momentum} in electron-nuclear fixed-target collisions at high luminosity with a beam in the $4 \text{ GeV}-16 \text{ GeV}$ range. The LDMX setup builds on the demonstration of an electron fixed-target missing energy search by NA64 at CERN~\cite{Banerjee:2017hhz}, and provides roughly a meter-long region of instrumented tracking and calorimetry that acts as a beam stop. The detector concept allows each individual electron to be tagged and its evolution measured as it passes through a thin target, tracking planes, and a high-granularity silicon-tungsten calorimeter. Not only does this enable a model-independent missing momentum and energy search, but it also offers an unprecedented opportunity to access remarkably short lifetime ($c\tau\sim 10 \ \mu m$) decays of new particles. Most existing studies of LDMX have focused on a specific (and important) class of sub-GeV DM models~\cite{izaguirre:2014bca,izaguirre:2015yja,alexander:2016aln,Battaglieri:2017aum}. Our analysis finds that LDMX is sensitive to a much broader range of both thermal and non-thermal DM, and simultaneously to new particles like axions with either photon or electron couplings, very weakly coupled millicharges, visibly and invisibly decaying dark photons and other gauge bosons, and Higgs-like scalars, among other new physics possibilities. Our findings can be divided into three broad categories of new physics studies in which we present sensitivity projections for LDMX (and a muon-beam variant \cite{Kahn:2018cqs}) and comparisons to many other existing and planned experiments: 
\begin{itemize}
  \item {\bf Dark Matter Particles:} In Sec.~\ref{sec:thermal-dm-section}, we provide new calculations of thermal freeze-out scenarios for many simple (and viable) sub-GeV scalar and fermion DM models coupled through a dark photon, as well as the natural generalizations of these models to those with other vector or scalar mediators. We also provide calculations of the relic abundance for viable freeze-in models with heavy dark photon mediators and low reheat temperatures, complementing existing calculations performed with ultra-light dark photons. For these models, as well as representative Strongly Interacting Massive Particle (SIMP) and asymmetric DM scenarios, we present sensitivity estimates for LDMX (using the missing momentum channel) and other applicable experiments. We find that important sensitivity targets can be reached for viable DM models spanning the entire keV-GeV mass range. We also illustrate the sensitivity of complementary experiments such as low-threshold direct detection, beam-dump, and $B$-factory experiments. In the case that DM freeze-out proceeds through non-resonant direct annihilations to SM particles (among the most predictive scenarios), the expected LDMX sensitivity is close to decisive when combined with Belle II.

  \item {\bf Millicharged Particles and Invisible Decays of New Particles:} In Sec.~\ref{sec:beyond_dm}, we provide calculations of millicharge production, rates for invisibly decaying gauge bosons, simplified models of sub-GeV scalars, and muonic forces motivated by the $(g-2)_{\mu}$ anomaly~\cite{Bennett:2006fi}. We show sensitivity estimates for LDMX (using the missing momentum channel) and other applicable experiments. We find that LDMX could provide leading sensitivity to millicharged particles below the $500 \ {\rm MeV}$ mass range. We also find that sensitivity to the invisible decays of dark photons and minimal gauged  $B-L$ (and similarly $L_i-L_j$, $B-3 L_i$) symmetries will be enhanced by many orders of magnitude compared to existing searches in the entire sub-GeV mass range, and cover unexplored parameter space that can address the $(g-2)_{\mu}$ anomaly.  
  
  \item {\bf Axion Particles, Dark Photons, and Visible Displaced Decays of New Particles:} In Sec.~\ref{sec:visible_signals}, we provide new calculations for the production and visible displaced decays of axion and dark photon particles. We consider both electron and photon coupled axions. In these cases, the signal is an electromagnetic shower in the back of the LDMX electromagnetic calorimeter (ECAL), where the residual potential backgrounds are dominated by hard neutron initiated hadronic showers. While the final background levels cannot be precisely determined without detailed experimental study, we find that for a well-motivated range of performance, LDMX will cover significant new territory for the minimal dark photon with mass less than $100\;{\rm MeV}$ and lifetime larger than $10 \ \mu m$. In addition, axion-like particles with similar mass and lifetime will be explored, including part of the parameter space for the QCD axion discussed in Ref.~\cite{Alves:2017avw}. Estimates of sensitivity to 
SIMP DM with displaced decay signatures are also given. For these scenarios, orders of magnitude in unexplored coupling and mass can be tested by LDMX.  Much of the parameter space for secluded DM models (discussed below) can also be explored by virtue of this sensitivity.
\end{itemize}
While we have tried to consider a broad survey of models discussed in the literature, covering many of the basic scenarios listed in the US Cosmic Visions report~\cite{Battaglieri:2017aum}, we have not been exhaustive. 
For example, in the minimal dark sectors investigated below, we have focused largely on the 
predictive parameter space with a mediator heavier than twice the DM mass. However, most of the signals we consider have a near threshold counterpart (with an off-shell mediator) that would extend beyond this regime, and this would be interesting to study in more detail in a future work. Additionally, sub-GeV supersymmetric hidden sectors~\cite{Morrissey:2009ur,morrissey:2014yma}, as well as variations of the strongly interacting models~\cite{Hochberg:2014kqa,Hochberg:2015vrg,Hochberg:2018rjs} can also be probed with the missing momentum and displaced visible 
decay approaches.

This paper is organized as follows. After briefly describing the LDMX setup in the next section, we discuss 
the theoretical motivation for DM and light mediators in Sec.~\ref{sec:theory_primer}.
In Sec.~\ref{sec:thermal-dm-section}, we analyze the thermal relic targets for 
several minimal dark sectors and highlight the sensitivity of LDMX and other experiments.
We show that the science case for LDMX extends beyond DM in Secs.~\ref{sec:beyond_dm} 
and~\ref{sec:visible_signals}, where we study the discovery potential 
of LDMX for a variety of beyond the SM (BSM) scenarios, including millicharges, gauge bosons, and 
scalars, in the missing momentum and visible channels, respectively.
We conclude in Sec.~\ref{sec:conclusions}.

\subsubsection{Experimental Comments}

As the above summary emphasizes, an important conclusion from this work is that LDMX-like experiments designed to measure missing momentum also provide sensitivity to long-lived particles with lifetimes typically shorter than what have been probed in beam dump experiments such as E141~\cite{Riordan:1987aw}, Orsay~\cite{Davier:1989wz}, E137~\cite{Bjorken:1988as}, NuCal~\cite{Blumlein:1990ay}, CHARM~\cite{Bergsma:1985qz}, and others. 
This is reflected in the range of reach projections shown in this paper, with new territory covered at larger couplings than beam dumps, yet well below the magnitude of couplings probed by collider experiments like BaBar~\cite{Lees:2014xha}, Belle~\cite{TheBelle:2015mwa}, or those at the LHC~\cite{Aaij:2017rft}. To see why this should be the case, it is worth reviewing a few experimental aspects of LDMX, as this will help the reader understand later sections of the paper.

\begin{figure}
  \centering
  \includegraphics[height=3.1cm]{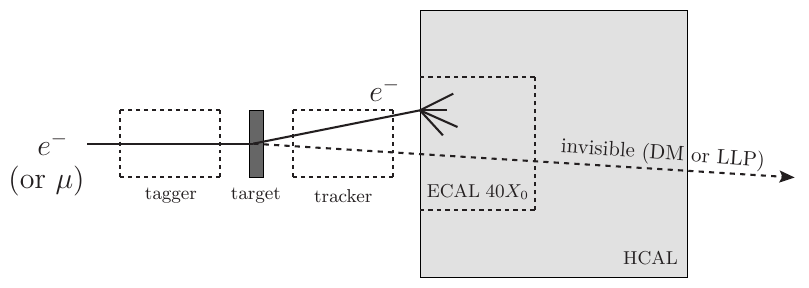}
  \includegraphics[height=3.1cm]{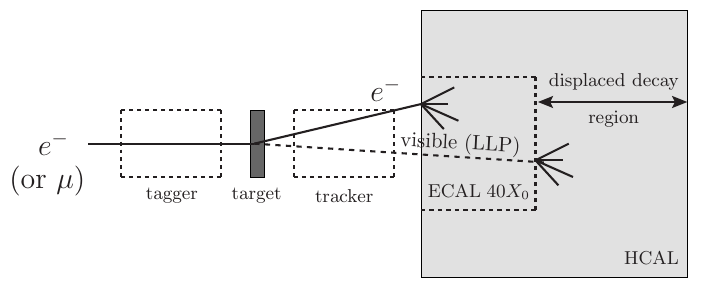}
  \caption{Schematic layout of an LDMX-like experiment. The {\it missing momentum} channel, in which most of the beam energy and momentum is lost in a reaction occurring in a thin upstream target, is illustrated on the left. The emitted particle either decays invisibly, e.g., to dark matter, 
  or it is long-lived and decays outside of the detector to SM final states. The {\it visible displaced decay} channel, in which a nearly full beam energy electromagnetic shower occurs far beyond the range of normal showers in the ECAL, is illustrated on the right. This signal is produced when a long-lived particle (LLP) 
  decays far inside the detector, initiating a displaced electromagnetic shower.
  \label{fig:ldmx_setup}}
\end{figure}

LDMX is designed primarily to measure missing momentum in electron-nuclear fixed-target collisions with a $4 \text{ GeV}-16 \text{ GeV}$ electron beam, though the use of a muon beam has also been suggested~\cite{Kahn:2018cqs}. To facilitate this measurement, the beam options under consideration are all high repetition rate (more than 40 MHz) and have a large beam spot (at least a few cm$^2$). In this way, an appreciable number of individual electrons can be separated and measured. The upstream part of the detector consists of a silicon tracker inside a dipole magnet, the purpose of which is to tag and measure the incoming momentum of each and every beam particle. The beam particles then impact a thin ($10 \%-30 \%$ of a radiation length) target. Tungsten is often the target considered. The target region defines the location where potential signal reactions are measured. A silicon tracker downstream of the target measures the recoil electron, and this is used to establish a measure of the momentum transfer in the collision. Downstream of this system are both an electromagnetic calorimeter (ECAL) and hadronic calorimeter (HCAL) designed to detect the presence of charged and neutral particles. Relative to the existing NA64 missing \emph{energy} experiment~\cite{Banerjee:2017hhz}, LDMX will leverage the larger available intensities of a primary electron beam to probe smaller couplings and the recoil beam electron momentum measurement to characterize signal and background events. 

New physics dominantly coupled to muons can be probed directly with a $\mu$ beam. Therefore we also consider the reach of the 
  recently-proposed $\mu$ beam version of LDMX~\cite{Kahn:2018cqs}. Relative to the experimental set-up described above, the muon missing momentum 
  experiment, LDMX $\mathrm{M}^3$, is limited to smaller beam intensities which can be compensated by a thicker target. Following Ref.~\cite{Kahn:2018cqs} 
  we take as fiducial LDMX $\mathrm{M}^3$ parameters a 50 radiation length tungsten target and a 15 GeV $\mu^-$ beam, 
  representative of the Fermilab muon beam capabilities. 
  Beam muon energy loss as it traverses the thick target must be determined in order to implement the missing momentum technique; this necessitates the use of an active target.
  The downstream tracking, ECAL and HCAL are similar to the ones in the $e^-$ beam experiment.

The signal of DM or other invisible particle production is a large energy loss by the electron (usually accompanied by sizable transverse momentum exchange), with no additional activity in the downstream calorimeters beyond that expected by the soft recoiling electron. This defines the {\it missing momentum} channel used in our studies, and a cartoon for a signal reaction of this type is shown in the left panel of Fig.~\ref{fig:ldmx_setup}. This channel's great strength is its inclusivity. LDMX's measurements in this channel will apply to a broad range of models over a range of mass extending from $\sim $ GeV to well below the keV-scale -- this is shown in Secs.~\ref{sec:thermal-dm-section} and~\ref{sec:beyond_dm}. In estimating the reach of the missing momentum approach at LDMX, we will 
assume that all backgrounds can be vetoed using the tracking, ECAL and HCAL systems described above. 
Most SM events that contain missing energy also result in a prompt visible energy deposition 
that can identified in the ECAL. Rare photonuclear reactions with forward going neutrons and no other energy 
depositions are more difficult, but can be vetoed with the HCAL.
The feasibility of this was first estimated in Ref.~\cite{izaguirre:2014bca}. Detailed detector studies of these and other 
backgrounds support these initial calculations~\cite{Akesson:2018vlm}. The current suite of simulations focuses on a 4 GeV beam; the rates of the 
potentially problematic processes decrease with larger beam energies. Even if the experiment cannot be 
made completely background free, moderate background counts do not significantly impact the sentivity of the 
experiment~\cite{Akesson:2018vlm}. 
The backgrounds from rare photonuclear reactions are greatly suppressed in a muon beam facility since the hard bremstrahlung rate for $\mu$ 
is reduced by $(m_e/m_\mu)^2$ relative to an $e^-$ beam~\cite{Gninenko:2014pea,Kahn:2018cqs}. 

While the missing momentum channel forms the basis of the LDMX design, the instrumentation required for this measurement also enables a second, complementary search for penetrating electromagnetic showers that occur far beyond the typical range of showers in the ECAL. Triggering on such events should be possible using energy deposition near the back of the ECAL or front of the HCAL. This defines what we refer to as the {\it visible displaced decay} channel in this paper, and a cartoon for a signal reaction is shown in the right panel of Fig.~\ref{fig:ldmx_setup}.  An analogous displaced-decay search has recently been performed by NA64~\cite{Banerjee:2018vgk}, but we emphasize that, unlike NA64, we consider here a visible decay search with the \emph{unmodified} LDMX detector.
Relative to the missing momentum channel, this channel is potentially more limited by reducible backgrounds that arise from very energetic neutral hadrons produced in hard electron or photon collisions with nuclei in the upstream part of the detector. These hadrons can initiate a displaced \emph{hadronic} shower when they interact or decay, which will sometimes fake the displaced electromagnetic shower expected from the decay of an exotic long-lived particle. A full experimental study of this signature has not been completed, but our estimates suggest that many signals with boosted decay lengths of $20 \text{ cm}-40 \text{ cm}$ would have sufficient yield to stand out over this background. We will estimate the sensitivity of LDMX to such reactions in Sec.~\ref{sec:visible_signals}. 

In practice, LDMX has presented a design intended for a first phase of running with roughly $4\times 10^{14}$ electrons on target (EOT) at 4 GeV, but has also considered scenarios for increasing this luminosity to about $\sim 10^{16}$ EOT using beams with energies up to 16 GeV. This corresponds to potential configurations in the US at DASEL~\cite{Raubenheimer:2018mwt}, Jefferson Lab, or in Europe at CERN~\cite{Akesson:2018yrp}. Moreover, the possibility of using a muon beam has also been suggested~\cite{Kahn:2018cqs}, and this would offer complementary reach for some models. In this paper, we largely show estimated reach for 8 GeV  and 16 GeV beam energies, and the full luminosity that LDMX is considering. 
In a few cases, we also show the reach for some models in a muon beam run with up to $10^{13}$ muons on target.

\section{Theory Primer}
\label{sec:theory_primer}

This section provides a brief high-level theory introduction to light DM and dark sectors. The main purpose of this section is to provide a logical organization to the landscape of dark matter/sector possibilities that have been previously investigated in the literature, which is summarized in Figs.~\ref{fig:DMflow} and \ref{fig:mediator_summary}. Our secondary aim is to explain the basic motivations behind our choice of models used later in the paper. 
However, the discussion in Secs.~\ref{sec:thermal-dm-section}-\ref{sec:visible_signals} is self-contained, so readers who are familiar with these topics may skip directly to Sec.~\ref{sec:thermal-dm-section}.

\subsection{Light Dark Matter}
\label{sec:theory_primer_A}

\begin{figure}[htbp]
 \includegraphics[width=17cm]{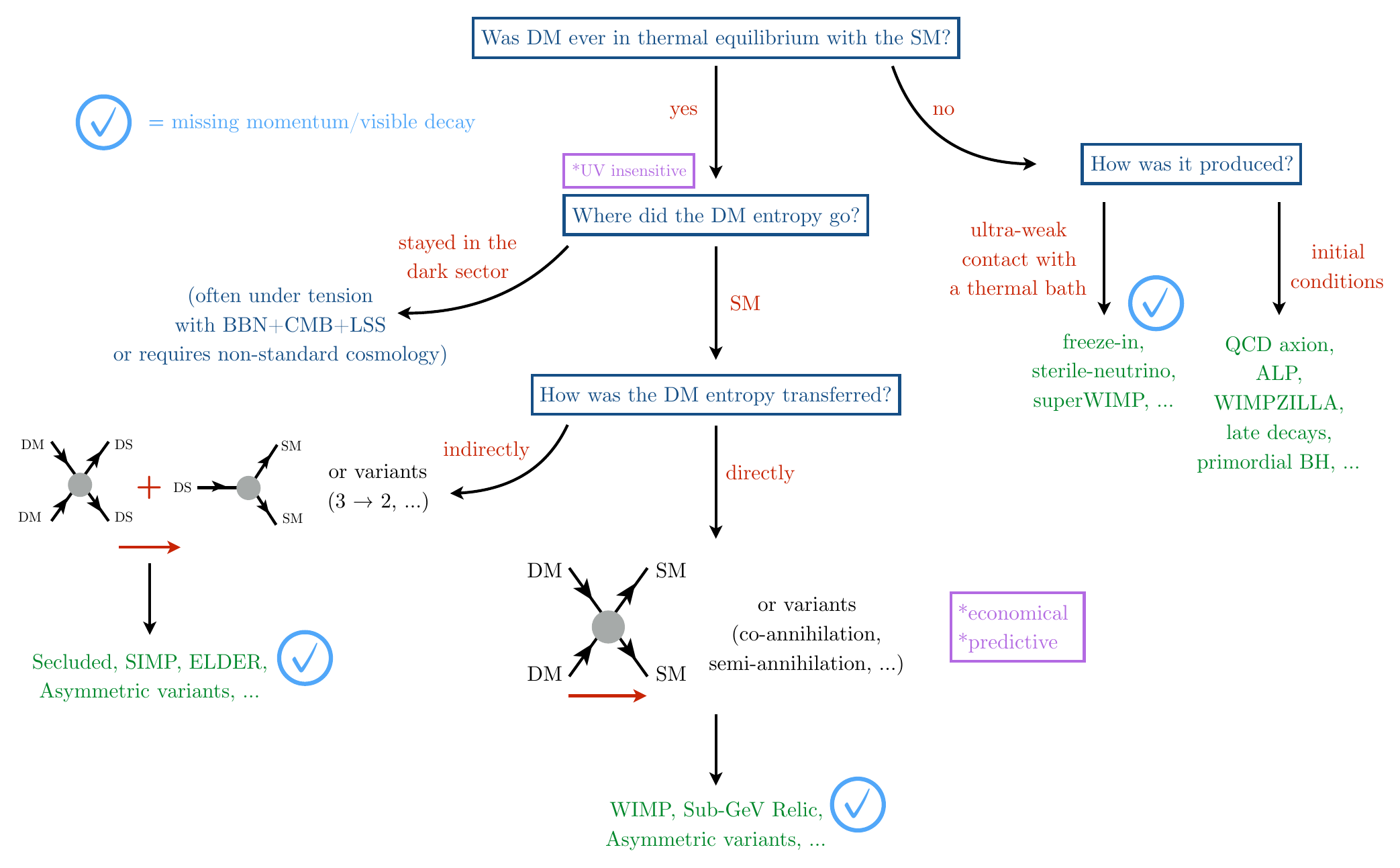}
\caption{The landscape of dark matter models, organized according to underlying principles and elementary questions. Early universe thermodynamics offers an especially simple way of understanding the important ways in which models are different, and how they relate to high-level questions about the origin of dark matter. 
If dark and visible matter are equilibrated in the early universe, dark matter has a large ($\sim T^3$) entropy, 
which must be reduced or transferred to visible particles to avoid overproducing dark matter.  
Blue checkmarks highlight branches for which we include representative models in this paper, as these often involve invisible or visible decays of light mediators. The abbreviations DM, DS, and SM are shorthand for dark matter, dark sector, and Standard Model particles, respectively. The red arrows indicate time flow for DM/DS processes in the early universe.
  \label{fig:DMflow}
}
\end{figure}

Many DM models have been proposed in recent years, underscoring the need to expand the scope of experimental searches. The sheer number of such models, and their apparent diversity, can give the daunting impression that anything goes, and that countless experiments will be needed to make any meaningful progress. 
Fortunately, very simple principles of (early universe) thermodynamics and lessons from the SM provide order to this landscape.  We highlight in Fig.~\ref{fig:DMflow} a small set of organizing principles and logical questions that allow one to characterize most models of DM into several overarching cosmological branches.

One of the most fundamental questions to ask of DM is {\it was it ever in thermal equilibrium with visible matter?} If the answer is negative, 
its abundance arises from cosmological initial conditions and/or from ultra-weak interactions with a thermal bath (see the right column of Fig. \ref{fig:DMflow}). 
In practice however, even tiny couplings of DM to the SM will bring the two into thermal contact. This occurs when interaction rates, $\Gamma$, exceed the expansion rate of the universe. Roughly speaking, $\Gamma\sim g_D^2 \, m_{\rm DM}$ at temperatures comparable to the DM mass, $m_{DM}$, where $g_D$ is some dimensionless coupling constant. For GeV-scale DM, this implies that equilibration is expected for couplings larger than
\be
g_D\gtrsim \sqrt{\frac{m_{\rm DM}}{M_\text{Pl}}}\sim 10^{-9} ~~~~\text{(equilibration)}
~,
\ee
where $M_\text{Pl} \sim 10^{18} \text{ GeV}$ is the Planck mass. Once equilibrated, DM number and entropy densities at early times are determined by the photon plasma temperature, $n_\text{DM} \propto s_{\rm DM}  \propto T^3$. 
Thus, unless the forces mediating dark-visible interactions are {\it extremely} feeble -- much weaker than the SM electroweak force -- DM equilibrates with the SM bath. In fact, this is often (but not always) a natural outcome of demanding that these scenarios are testable in the laboratory.
This fact has several far-reaching, model-independent implications:
\begin{itemize}
\item[] {\bf 1) Insensitivity to Initial Conditions:} Since the equilibrium DM distribution is set by the temperature, its subsequent evolution
is independent of earlier, unknown cosmological epochs (e.g. inflation, baryogenesis). 
\item[] {\bf 2) Necessary Entropy Transfer:}  Without a mechanism to significantly reduce its thermal abundance, the
DM number density would be comparable to the relic photon and neutrino number densities at late times. 
In this case, unless the DM is very light ($\lesssim 10$ eV and, thus, unacceptably hot),  its  energy density would greatly exceed the measured value 
at late times. 
Thus, it is essential for 
thermal DM to have an efficient entropy (or number density) depletion mechanism to avoid overproduction under standard cosmological
assumptions.\footnote{Under {\it standard} cosmological assumptions 
    the comoving DM number density is only diluted by heating of the plasma from SM particle annihilations and decays after DM freeze-out. 
    The amount of dilution is completely determined by the SM field content and particle masses.
    In a {\it nonstandard} cosmology, it is possible
  to reduce the DM abundance relative to SM particles further, e.g., through heating of the plasma via SM decays of new species or a phase transition.
In this scenario,
the DM entropy is not depleted, but the SM entropy is increased instead (see upper left arrow of the flowchart in Fig. \ref{fig:DMflow}). 
}
There are two main possibilities for this entropy depletion:
\begin{itemize}
\item \textbf{\emph{Transfer to Dark Sector (DS):}} If this large entropy is permanently transferred to other particles in the dark sector (e.g. dark radiation), there is generic tension with $\Delta N_{\rm eff}$ (the number of light relativistic species populated in the early universe) as inferred from measurements of the cosmic microwave background (CMB) and the successful predictions of Big Bang nucleosynthesis (BBN)~\cite{ade:2015xua}. 
\item \textbf{\emph{Transfer to Standard Model (SM):}} If the DM entropy is transferred to the SM, it can occur  {\it indirectly}, for instance, through DM DM $\cdots \to$  DS DS $\cdots$ processes followed by DS decays to SM particles, or  {\it directly} through, e.g., DM DM $\to$ SM SM annihilations, the latter of which yields
 predictive targets for the DM-SM interaction strength. 
Direct entropy transfers to the SM are outlined with the middle column in Fig.~\ref{fig:DMflow}
and include both the familiar WIMP paradigm and various sub-GeV dark sector models, which we 
explore further in Sec.~\ref{sec:thermal-dm-section}. 
\end{itemize}
\item[] {\bf 3) Bounded Mass Range:} Based only on the available cosmological and astrophysical data, the {\it a priori}  DM mass range is 
nearly unconstrained: $10^{-22} \, \, {\rm eV} \lesssim  m_{\rm DM} \lesssim 10 \, \, M_{\odot}$. However, if thermal equilibrium is achieved 
at early times under standard cosmological assumptions, the mass range becomes considerably narrower and more predictive, ${\rm MeV} \lesssim m_{\rm DM} \lesssim 100 \, {\rm TeV}$. If DM is lighter than an MeV, the entropy transfer (i.e. freeze-out) occurs during nucleosynthesis and often spoils
the successful predictions of BBN (see Ref.~\cite{Berlin:2017ftj} for an exception to this statement). 
If DM is heavier than $\sim 100$ TeV, the annihilation cross section governing the entropy
transfer generically violates perturbative unitarity~\cite{Griest:1989wd}, so nontrivial model building is required. 
\end{itemize}
\medskip
These features provide a unique and exciting degree of predictiveness.  Measurable quantities like the final abundance of DM are determined by particle physics alone and are insensitive to cosmological boundary conditions (aside from asymmetries in conserved quantum numbers).
Furthermore, by a lucky coincidence, the viable mass range of thermal DM roughly spans 
SM-like and terrestrially-accessible energy scales. 

Over the past several decades, the upper half of this range ($\sim$ GeV--100 TeV) has been the primary focus of the experimental community.
This focus was motivated largely by the so-called WIMP paradigm, in which weak-scale DM
naturally yields the necessary annihilation cross section for thermal freeze-out through the familiar SM electroweak interaction (the ``WIMP miracle"). Additional motivation for this paradigm came from a theoretical emphasis on supersymmetric extensions of the SM, whose 
most compelling versions naturally include WIMP DM candidates. 
However, in recent years, null results from such endeavors have cast doubt on its simplest incarnations, as exemplified by 
the latest limits from the LHC (e.g. in the missing energy plus jets channel designed to test supersymmetry ~\cite{Sirunyan:2017hci,Aaboud:2017phn}) and from direct detection experiments~\cite{Angloher:2015ewa,Akerib:2016vxi,Behnke:2016lsk,Amole:2017dex,Agnese:2017jvy,Cui:2017nnn,Aprile:2018dbl}. This has prompted the community to explore related scenarios that have comparably simple explanations for the origin and dynamics of DM~\cite{alexander:2016aln,Battaglieri:2017aum}. 

The community is now beginning to explore the lower half of this range ($\sim$ MeV-GeV) with an emphasis on a WIMP-like paradigm for light (sub-GeV) thermal DM. This remains a compelling and economical explanation for the missing mass of the universe. The proximity to the weak-scale makes this an obvious place to search for DM, though until recently it has been difficult to do so. Especially predictive scenarios arise if DM directly annihilated to SM species before freeze-out. In this case, new light forces that feebly couple to the lightest SM states (electrons, photons, and neutrinos) are directly motivated since they uniquely enable such processes in the early universe~\cite{Lee:1977ua,Boehm:2003hm}. Thus, light DM and light mediator particles go hand in hand, and models with both are often referred to as ``dark sectors." In this work, we pay particular attention to signals at LDMX arising from invisible or visible decays of such mediators. Although the main focus of Sec.~\ref{sec:thermal-dm-section} is to explore regions of parameter space directly motivated from considerations of thermal DM, non-thermal variants can also give rise to missing momentum/energy signatures that are detectable at experiments such as LDMX and Belle II. For example, in Sec.~\ref{sec:freezein}, we investigate these signals for models in which the DM is populated from freeze-in processes in cosmologies involving low reheat temperatures. 

\subsection{Dark Sector Mediators}
\label{sec:theory_primer_B}

Sub-GeV dark sectors and weakly-interacting particles are motivated by many models of physics beyond the SM, including models that address the hierarchy problem~\cite{Cheung:2009qd,Morrissey:2009ur,Katz:2009qq}, strong CP problem~\cite{Kim:1979if,Shifman:1979if,Dine:1981rt,Zhitnitsky:1980tq}, or the $(g-2)_\mu$ anomaly~\cite{Gninenko:2001hx, Pospelov:2008zw, Batell:2016ove}. Dark sectors are also natural in the context of string theory (see, e.g., Ref~\cite{Halverson:2018xge} for a review) and they are directly motivated by DM, especially in the sub-GeV mass range~\cite{Lee:1977ua,Boehm:2002yz,Boehm:2003hm,Pospelov:2007mp,Feng:2008ya}. 
\begin{figure}
  \centering
  \includegraphics[height=3.66cm]{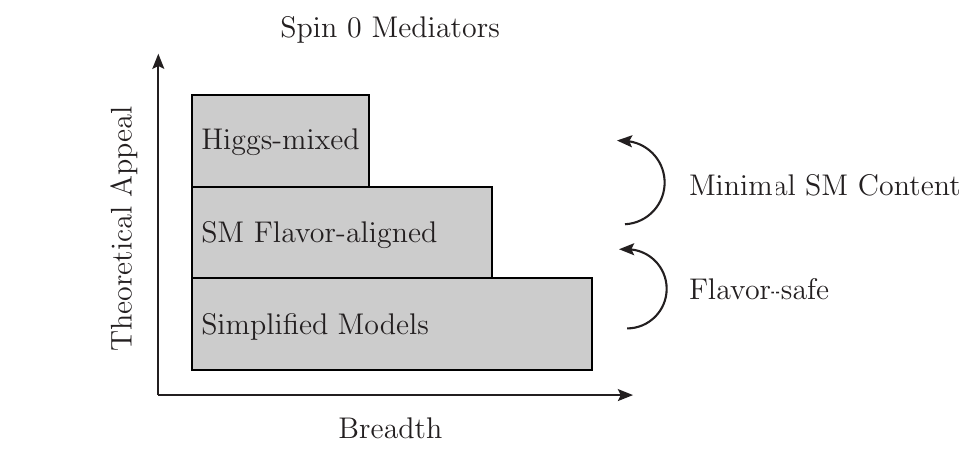}
  \includegraphics[height=3.66cm]{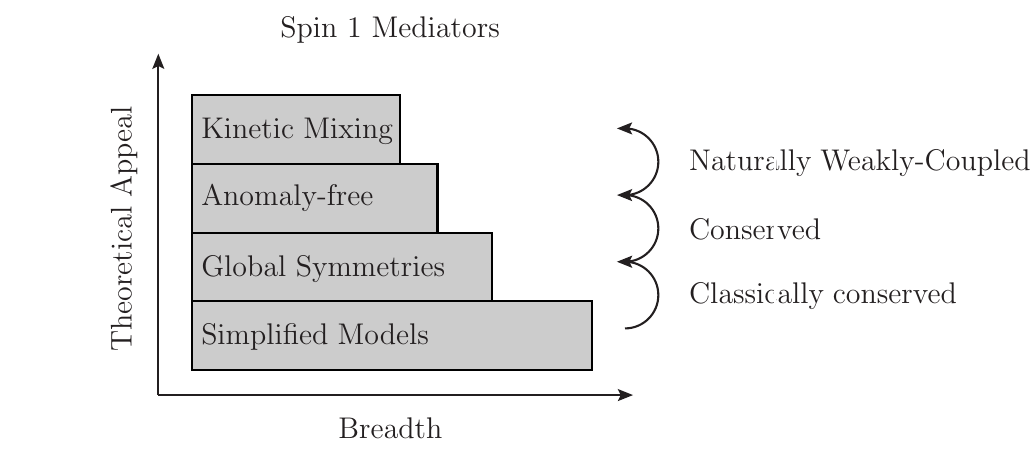}
\caption{
A summary of generic GeV-scale dark sector states that can interact with the SM via low-dimension operators. The width of each bar indicates the range of interaction types with the SM, while the vertical position conveys theoretical appeal (which includes input from existing experimental data). Models at the top are subsets of the models beneath them, with more theoretical appeal. In the left panel, we list possible scalar mediators. Spin-$0$ simplified models often lead to additional SM flavor violation, in conflict with observations. This tension is alleviated in scenarios where the new couplings are aligned with SM Yukawa matrices, as in models of Minimal Flavor Violation, axion-like particles, and flavor-specific mediators. Ultraviolet completions of these models usually involve new matter charged under the SM. The Higgs-mixed scalar is special in this regard because it automatically inherits mass-proportional interactions and is UV-complete in its minimal form. The right panel shows possible vector mediators. Spin-$1$ simplified models often lead to non-conserved and/or anomalous currents, which requires additional matter to resolve poor high-energy behavior of processes involving longitudinal gauge bosons. This is also the case if vector bosons instead couple to SM currents that are classically conserved, such as baryon ($B$) or lepton ($L$) number, but are still violated by non-perturbative processes in the SM. These issues are all avoided if the new vector boson couples to conserved, anomaly-free currents such as $B-L$ or $L_i-L_j$, though it is not especially natural to expect their interaction strength with the SM to be small. The most appealing interaction type for sub-GeV mass vector interactions is through kinetic mixing with the photon. This interaction is naturally small (as it can be generated by loops of heavy particles), gauge invariant, anomaly-free, and it is common in extensions of the SM.
\label{fig:mediator_summary}
}
\end{figure}

In practice, the main interest surrounding such particles in the context of DM is that they could comprise the DM itself, or act as mediators of interactions between DM and the SM. For dark sectors that are neutral under SM gauge forces, effective field theory offers a simple approach to classifying such interactions by operator dimension. Low dimension operators are expected to be most relevant at low energy (the universe today), and so we primarily focus on those. The allowed operators are also constrained by Lorentz symmetry and the spin of the dark sector particle(s). 
The lowest-dimensional possibilities for a given particle spin are:
\begin{itemize}
  \item{{\bf spin-0:}} The most general possibility for dimension-four interactions of a scalar particle, $\varphi$, with the SM is to assign arbitrary Yukawa-like couplings to SM fermions after electroweak symmetry breaking, $\varphi \bar f_{i} f_{j}$. A generic coupling assignment does not commute with the SM Yukawa matrices, and therefore leads to flavor-changing neutral currents (FCNC), whose existence is severely constrained. Issues with FCNCs are avoided if the interactions of the new scalar with SM fermions are at least approximately aligned with (commute with but not necessarily proportional to) the Yukawa matrices. This generates primarily flavor-diagonal interactions, $\varphi \bar f_i f_i$, and possibly small off-diagonal couplings that are consistent with FCNC searches. This broad class of models includes scalars with (N)MFV interactions~\cite{DAmbrosio:2002vsn,Agashe:2005hk}, two-Higgs-doublet models and their extensions~\cite{Batell:2016ove}, axion-like particles~\cite{Choi:2017gpf}, and flavor-specific mediators~\cite{Batell:2017kty}. SM gauge charges of the fermions also guarantee that the dimension-four low-energy scalar interaction, $\varphi \bar f_{i} f_{i}$, must, in fact, arise from higher-dimensional operators. Ultraviolet completions of these scenarios necessarily involve new particles that are charged under the SM gauge group. One exception to this is the Higgs portal coupling of a scalar mediator, $\varphi$, 
which mixes with the SM Higgs, $H$~\cite{Patt:2006fw}.  
Such mixing induces 
Higgs-like Yukawa interactions with SM fermions, 
but with a strength that is suppressed by 
the $\varphi-H$ mixing angle. This automatically gives interactions aligned with SM Yukawa couplings without the need for additional particles in the UV (hierarchy problems notwithstanding).

\item{{\bf spin-1/2:}} A neutral singlet fermion, $N$, can interact with SM particles through the lepton portal, $LHN$, which induces $N-\nu$ mixing after electroweak symmetry breaking. Since this mixing is proportional to small neutrino masses, the interaction strength is very suppressed. As a result, it is generically difficult for the minimal $N$-mediated processes to sustain thermal equilibrium in the early universe, and such models are often beyond the reach of direct terrestrial experiments, with the exception of high intensity proton colliders~\cite{Curtin:2018mvb} and beam dumps~\cite{Alekhin:2015byh}. 
Given these features and the focus on LDMX 
in this paper, we will not consider this possibility further. 

\item{{\bf spin-1:}} 
Simplified models of spin-$1$ mediators 
introduce arbitrary couplings of new vector particles 
to SM matter. Generic models 
with dimension-four interactions 
lead to FCNCs and to non-conserved and/or anomalous currents. 
These scenarios require additional matter to resolve poor high-energy behavior 
of reactions involving these new vector bosons. A theoretically appealing 
alternative is to couple vector bosons 
to SM currents that are classically conserved, such as baryon ($B$) or lepton ($L$) number. These classical symmetries are violated by non-perturbative processes in the SM. The corresponding anomaly gives rise to processes that are enhanced by factors of (energy/boson mass), leading to strong experimental constraints~\cite{Dror:2017ehi,Dror:2017nsg}; compared to interactions with 
tree-level-anomalous currents, the scale of the UV completions can be parametrically larger. 
An additional theoretical improvement is then coupling to conserved, anomaly-free currents such as $B-L$ and $L_i-L_j$, where $L_i$ is lepton number of generation $i$. 
Finally, the simplest and most appealing models of sub-GeV vector mediators are those where the vector mixes with SM hypercharge via kinetic mixing. This interaction is naturally small (as it can be generated by loops of heavy particles), gauge invariant and anomaly-free, and it is common in extensions of the SM. Such vectors (often referred to as ``dark photons") couple to the electromagnetic current but with a suppressed strength. 

\end{itemize}

These arguments are summarized in Fig.~\ref{fig:mediator_summary}, which 
schematically shows the hierarchy of minimal models with new spin-$0$ and spin-$1$
particles and highlights the breadth and theoretical appeal of the various possibilities.
We have chosen not to include mediators with spin larger than or equal to $3/2$, as these are severely constrained 
by Lorentz symmetry and basic principles of quantum mechanics (see, e.g., Ref.~\cite{Benincasa:2007xk}). 
Furthermore, in this work we focus on renormalizable interactions of dark sector states with SM particles (an exception being axion-like particles which interact through dimension-five operators). Higher-dimensional operators enable a variety of novel couplings, e.g., with non-Abelian dark sector fields~\cite{Juknevich:2009ji,Juknevich:2009gg,Forestell:2017wov}, but they are expected to be suppressed compared to the ones studied here.

\section{Predictive Light Dark Matter Models}\label{sec:thermal-dm-section}

Thermal DM below the weak-scale is a framework that retains the compelling cosmological explanations associated with the WIMP paradigm and motivates direct couplings to the SM that are often detectable with existing experimental technology. For these reasons, thermal DM models near or below the GeV-scale have received considerable attention in recent years~\cite{Battaglieri:2017aum}. In this section, we apply the lessons from Sec.~\ref{sec:theory_primer} to these models with special emphasis on {\it predictive} scenarios for which laboratory observables are directly related
to the processes that set the relic abundance at early times.  

Unlike the plethora of possibilities for weak-scale and heavier DM, the model building requirements for sub-GeV relics are qualitatively different and
naturally highlight a distinctive and broad class of models. Nearly all
\footnote{One notable exception
to this list involves a sub-GeV SIMP with a heavy $\sim$ 50 GeV mediator to transfer the DM entropy via scattering~\cite{Hochberg:2015vrg}. However,
aside from the heavy ($\gg$ GeV) mediator, even this scenario features SM neutrality, new forces, and CMB safety as listed here.
}
viable thermal DM scenarios below the GeV-scale  contain {\it all} of 
the following ingredients: 
\begin{itemize}
\item[] {\bf 1) SM Neutrality:} Unlike WIMPs, which can realize thermal freeze-out via SM gauge interactions, sub-GeV relics must not 
carry electroweak quantum numbers; new electroweak states are essentially ruled out for masses below $m_{Z}/2 \sim 45 \text{ GeV}$ by LHC, Tevatron, and LEP measurements ~\cite{Escudero:2016gzx,Kearney:2016rng,Egana-Ugrinovic:2018roi}. 
\item[] {\bf 2) Light New ``Mediators":}  Even if electroweak interactions were not excluded for new light particles, 
they would nonetheless be inefficient at depleting the large DM entropy as required by the arguments in Sec.~\ref{sec:theory_primer_A}. Indeed,
for a light WIMP, $\chi$, the $Z$-mediated cross section for $\chi \chi \to Z^*\to f \bar{f}$ annihilation is 
\be
\sigma v \sim G_F^2 \, m_\chi^2  \simeq 1.5 \times 10^{-29} {\rm \, cm^3  \, s^{-1}} \left(  \frac{m_\chi}{100\rm \, MeV}  \right)^2,
\ee
which falls short of the familiar thermal relic cross section 
$\sim \text{few} \times 10^{-26} \,  {\rm \, cm^3 \, s^{-1}}$~\cite{Steigman:2012nb}, thereby overproducing $\chi$ in the early universe. Thus, 
the presence of new force carriers more strongly coupled than the electroweak force is strongly motivated in models of
light thermal DM~~\cite{Lee:1977ua,Boehm:2003hm}. 
Furthermore, for DM annihilations directly to SM particles, 
successful transfer of the dark sector entropy to the SM implies that these ``mediators" 
couple to visible matter through the neutral, renormalizable ``portal" interactions presented in Sec.~\ref{sec:theory_primer_B}.

\item[] {\bf 3) CMB Safety:} An 
important question confronting light thermal DM is: {\it does it annihilate during the CMB era?} If so, there are strong  constraints on the power injected into the photon plasma during recombination. 
In particular, for $s$-wave annihilating DM, measurements of the CMB rule out $m_{\rm DM} \lesssim \mathcal{O}(10)$ 
GeV \cite{Slatyer:2009yq,ade:2015xua}.\footnote{
	Naively, the CMB is formed long after the DM has frozen out at $T \sim m_{\rm DM}/20$, so it would seem that DM annihilation
	 would already have stopped well before this point and that this bound does not apply. However, 
	 freeze-out merely implies that annihilations are out of equilibrium, not that they have stopped altogether. 
    A single GeV-scale DM annihilation deposits $\mathcal{O}(  {\rm GeV} )$ of energy into the plasma, 
   which is enough to ionize many hydrogen atoms.
    Thus, in the era of 
 	precision cosmology, even a feeble energy injection rate from rare DM annihilations at redshift $z \sim 1100$ can be meaningfully
  constrained for many scenarios.
 } Thus, most viable models of light thermal DM  have at least one of the following features:
 \begin{itemize}
 \item
 \textbf{\emph{Velocity suppressed annihilation at recombination:}} If the DM annihilation cross section is $p$-wave ($\sigma v \propto v^2$),  the 
 annihilation rate  will be smaller at low temperatures  as the velocity redshifts with Hubble expansion.
 \item 
 \textbf{\emph{Different DM population at recombination:}} If the DM population at freeze-out differs from the DM population during recombination, the annihilation
 rate can be parametrically different. Such a population shift can arise, for example, if the DM has a primordial particle-antiparticle asymmetry, 
 so that the antiparticles are depleted at early times and the particle population has no more annihilation partners during the CMB era. 
 Another possibility is DM freeze-out through coannihilation with another particle. If this particle is slightly heavier than the DM, then 
 its abundance is depleted by the time of recombination through scattering or decays. Thus, at late times DM has no partner to coannihilate with
 and energy injection into the plasma is shut off.
 \item 
\textbf{\emph{Annihilation to invisible particles:}} The Planck CMB bounds are based on visible energy injection at $T \sim$ eV, which reionizes the newly
recombined hydrogen and thereby modifies the ionized fraction of the early universe. DM annihilation to invisible particles (e.g. neutrinos or dark sector states) does not ionize 
hydrogen at an appreciable rate, thereby alleviating constraints from the CMB.
 \item 
\textbf{\emph{Kinematic barriers:}} Late time annihilations during recombination can be reduced
  if the annihilation final state is kinematically forbidden at low DM velocities~\cite{DAgnolo:2015ujb}. This can occur when the final state particles are slightly heavier than DM. Another possibility includes enhancing the rate in the early universe by annihilating through a resonance which is
  accessible at higher DM temperatures but not during recombination~\cite{Griest:1990kh}.
  Since the DM is already non-relativistic 
  during freeze-out, both of these scenarios require a mild tuning of the masses of DM and final/intermediate state particles 
  to ensure a sufficient annihilation rate during freeze-out.
 
 \end{itemize}
\end{itemize}

\noindent In this section, our primary goal is apply these insights to many viable {\it predictive} models of light thermal DM in which the mediator (``MED") decays {\it invisibly}  ($m_{\rm MED} > 2 \, m_{\rm DM}$)  and yields missing energy/momentum signatures at LDMX. For this mass hierarchy,  the relic density arises from  the freeze-out of ``direct ~annihilations," in which $s$-channel processes of the form
\be
~~~~~~~~~ \rm DM~ DM \to ~MED^* ~\to  SM ~SM   ~~~~~(direct ~annihilation),
 \ee
transfer the DM entropy 
 to the SM (following the middle column of Fig. 2). This yields  predictive targets in parameter space for terrestrial experiments.
 In Sec.~\ref{sec:dark-photon-benchmark-models}, we explore different spin and mass structures for the dark sector, under the canonical assumption of a kinetically mixed dark photon mediator. In Sec.~\ref{sec:other-mediators}, we generalize this discussion to include models of other spin-1 mediators and simplified models of viable spin-0 mediators. In Sec.~\ref{sec:simps}, we explore models beyond minimal thermal freeze-out that instead involve $3\rightarrow2$ and $2\rightarrow 1$ annihilations that can arise in a confining dark sector. As we will see, such scenarios can also give rise
 to invisible and semi-visible signatures at LDMX. 

For the opposite mass hierarchy regime  $(m_{\rm DM} > m_{\rm MED})$, the mediator decays {\it visibly}. In certain models, the DM entropy is transferred in a two-step process dubbed ``secluded annihilation" \cite{Pospelov:2007mp}. 
In this case, DM first annihilates to mediators
and then the mediators decay to SM particles, 
\be
\label{eq:secluded-schematic}
~~~~~~~~~ \rm DM~ DM \to  ~MED ~MED~~  ~~(secluded ~annihilation),
 \ee
 which is represented by the arrow labeled ``indirectly" in Fig. 2.
As long as the SM-MED coupling is sufficiently large to thermalize the dark and visible sectors, 
the DM abundance is independent of this coupling and there are no DM production targets for laboratory observables. However, in the absence of additional lighter field content in the dark sector, the mediator decays visibly to SM particles and can yield visible resonances, displaced vertices, or other (semi-)visible signatures commonly explored in accelerator based experiments~\cite{alexander:2016aln,Battaglieri:2017aum}. In Sec.~\ref{sec:visible_signals}, we show that such observables are also a key part of the LDMX scientific program. Furthermore, in  Sec.~\ref{ssec:secluded} we also point out that the traditional LDMX {\it invisible} signature can be sensitive to this secluded scenario through DM production via an off-shell mediator.  

Between these two domains ($m_{\rm DM} <  m_{\rm MED} < 2 \, m_{\rm DM}$) is a narrow but interesting mass range that combines many of the features of the two domains above.  Direct annihilations generally control freeze-out (the exception is a small region $m_{\rm DM} \approx  m_{\rm MED}$ where Boltzmann-suppressed secluded annihilation can dominate \cite{DAgnolo:2015ujb}) and informs predictive targets in coupling space for both mediator decays into SM particles and DM production via an off-shell mediator.  As illustrated in Fig.~2 of Ref.~\cite{izaguirre:2015yja}, the prospects for exploring the relic target in this domain, through a combination of both types of searches, are comparable to those for the more commonly considered mass hierarchy  $m_{\rm MED} > 2 \, m_{\rm DM}$.  A detailed analysis of this domain is complicated by the fact that fewer results and projections are available for DM production via off-shell mediators, and to our knowledge this case has not been analyzed since Ref.~\cite{izaguirre:2015yja} (from which the most notable omissions are the recent projections from  Belle II ~\cite{Battaglieri:2017aum} and LHCb \cite{Ilten:2015hya,Ilten:2016tkc} for visible dark photons). LDMX searches for both invisible and visible signals 
  are relevant in this regime. The former are briefly considered in Sec.~\ref{sec:dark-photon-benchmark-models} in the context of 
  dark photon models, while the latter are discussed in Sec.~\ref{sec:visible_signals}.

In the following sections, we briefly discuss several other DM scenarios outside the minimal direct/secluded annihilation paradigm (focusing for concreteness on the kinetically mixed vector mediator).  In Sec.~\ref{sec:asymmetric}, we summarize the target implied by CMB-safety for asymmetric DM in the kinematic range where direct annihilations dominate.  In Sec.~\ref{sec:simps}, we briefly discuss the sensitivity of LDMX and other experiments to strongly interacting DM models,  which have both different cosmologically motivated couplings and additional signal channels. 

In Sec.~\ref{sec:freezein}, we briefly discuss signals associated with nonthermal {\it freeze-in} production of DM~\cite{Hall:2009bx}. In this scenario, DM never reaches equilibrium with the visible sector, so the dark-visible interaction rate must be sufficiently suppressed during the early universe. Unlike the previous discussion, the DM \emph{nearly} thermalizes with the SM bath, thereby acquiring a sub-thermal abundance at late times through the process $f \bar{f} \to \chi \chi$, which is always slower than the Hubble expansion rate. Such freeze-in scenarios are inherently less predictive than thermal DM models, because one must assume that possible high-scale or early time contributions to their abundance (from inflation, reheating, etc.) are vanishingly small. Nonetheless, with these assumptions in place, it is possible to define production targets in terms of the  couplings that populate the DM from rare SM processes (see the right side of Fig. 2).
 For sub-GeV DM models, freeze-in production is often studied in the context of ultra-light ($\lesssim \text{keV}$) mediators (see Ref.~\cite{Battaglieri:2017aum}), though there is no {\it a priori} reason for this to be the primary focus; the mechanism itself does not require the mediator to be significantly lighter than the DM.  Indeed, over much of the viable freeze-in parameter space, these masses are comparable and the couplings required to populate the DM are too weak to be probed in terrestrial experiments. 
 However, two limiting regimes are exceptions to this general rule:
 \begin{itemize}
\item If the mediator is an ultra-light ($\lesssim \text{keV}$) vector\footnote{If, instead, the mediator is a singlet scalar with Higgs-portal mixing, it is
generically difficult to achieve the full DM abundance through freeze-in due to various 
astrophysical constraints \cite{Krnjaic:2017tio}.
} particle, DM-SM scattering rates at direct detection experiments are enhanced, and 
cosmologically motivated parameter space can be probed~\cite{Battaglieri:2017aum}. In this regime, the small coupling responsible for
cosmological production is partly compensated by the low-velocity enhancement in the scattering cross section $(\sigma_{\rm scatter} \propto v^{-4})$. However, this enhancement is not available in relativistic contexts, such as at accelerator experiments. 
\item
If instead, the mediator mass is larger than the DM mass and the cosmological reheat temperature\footnote{Low reheat temperatures are motivated in models involving gravitinos and/or moduli~\cite{Pradler:2006hh,Kohri:2005wn,Moroi:1999zb}.}, then freeze-in production of DM requires more appreciable couplings between DM and SM particles and accelerators can explore cosmologically interesting 
parameter space. In this regime, the non-negligible mediator mass makes direct detection probes more challenging by comparison. 
\end{itemize} 
In Sec.~\ref{sec:freezein}, we present LDMX projections for the latter scenario.




 \subsection{Predictive Dark Photon Models}\label{sec:dark-photon-benchmark-models}
 
 In this section, we study a family of models in which DM ($\chi$) directly annihilates to SM particles through an intermediate dark photon mediator ($\Ap$). $\Ap$ is the massive gauge boson of a broken $U(1)_D$ symmetry. In the gauge basis, it kinetically mixes with SM hypercharge. The relevant Lagrangian contains
\be
\label{eq:LagKinMix}
\mathscr{L}  \supset \frac{\epsilon}{2 \cos{\theta_W}} \, F^\prime_{\mu \nu} \, B^{\mu \nu} +  \frac{1}{2} \, m_{\Ap}^2 \, A^\prime_\mu {A^\prime}^\mu
~,
\ee
where $\theta_W$ is the weak mixing angle, $\epsilon$ is the kinetic mixing parameter, and $m_{\Ap}$ is the dark photon mass. In the low-energy theory, $\epsilon$ is a free parameter, but $\epsilon \ll 1$ is often expected to be generated by loops of particles charged under both hypercharge and $U(1)_D$~\cite{Holdom:1985ag,delAguila:1988jz}. In the limit that $m_{\Ap} \lesssim \text{GeV}$, the dark photon dominantly mixes with the SM photon; after diagonalizing the kinetic and mass terms, $\Ap$ inherits an $\epsilon$-suppressed interaction with the electromagnetic current, $J_\text{EM}$, and retains an unsuppressed coupling to the $U(1)_D$ current, $J_D$. In the mass eigenstate basis, these interactions take the form
 \be
 \label{eq:master-lagrangian}
- \mathscr{L} \supset 
A^\prime_\mu    \, (\epsilon \, e \, J_{\rm EM}^\mu +
 g_D \, J_{D }^\mu)
 ~,
 \ee
where 
\be
g_D \equiv \sqrt{4\pi \, \alpha_D}~,
\ee
is the $U(1)_D$ coupling constant. 
In Secs. \ref{sec:scalar-elastic}-\ref{sec:pseudo-dirac-large-delta}, we will introduce various DM sectors that couple directly to the dark photon. Although each of these models corresponds to a distinct form of $J_D$, the thermal abundance of DM retains the same dependence on the four model parameters $\{ \epsilon, \alpha_D, m_\chi, \mAp \}$ and can be treated in full generality.

 In these models, DM remains in chemical equilibrium with the SM through $\chi \chi \leftrightarrow A^{\prime *} \leftrightarrow f \bar{f}$ before freezing out while non-relativistic. In the limit that $\mAp \gg m_\chi$, the annihilation rates for the models in
 Secs. \ref{sec:scalar-elastic}-\ref{sec:pseudo-dirac} have the same parametric dependence,
 \be
 \label{eq:define-y}
 \sigma v (\chi \chi \to f \bar{f} ) \propto \frac{\epsilon^2 \alpha_D m_\chi^2}{\mAp^4} \equiv \frac{y}{m_\chi^2}~~,~~~ 
 y \equiv \epsilon^2 \alpha_D  \left(  \frac{ m_\chi}{\mAp}     \right)^4,
 \ee
 where we have defined the dimensionless interaction strength, $y$.  

\begin{figure}[t!]
\includegraphics[width=3.in,angle=0]{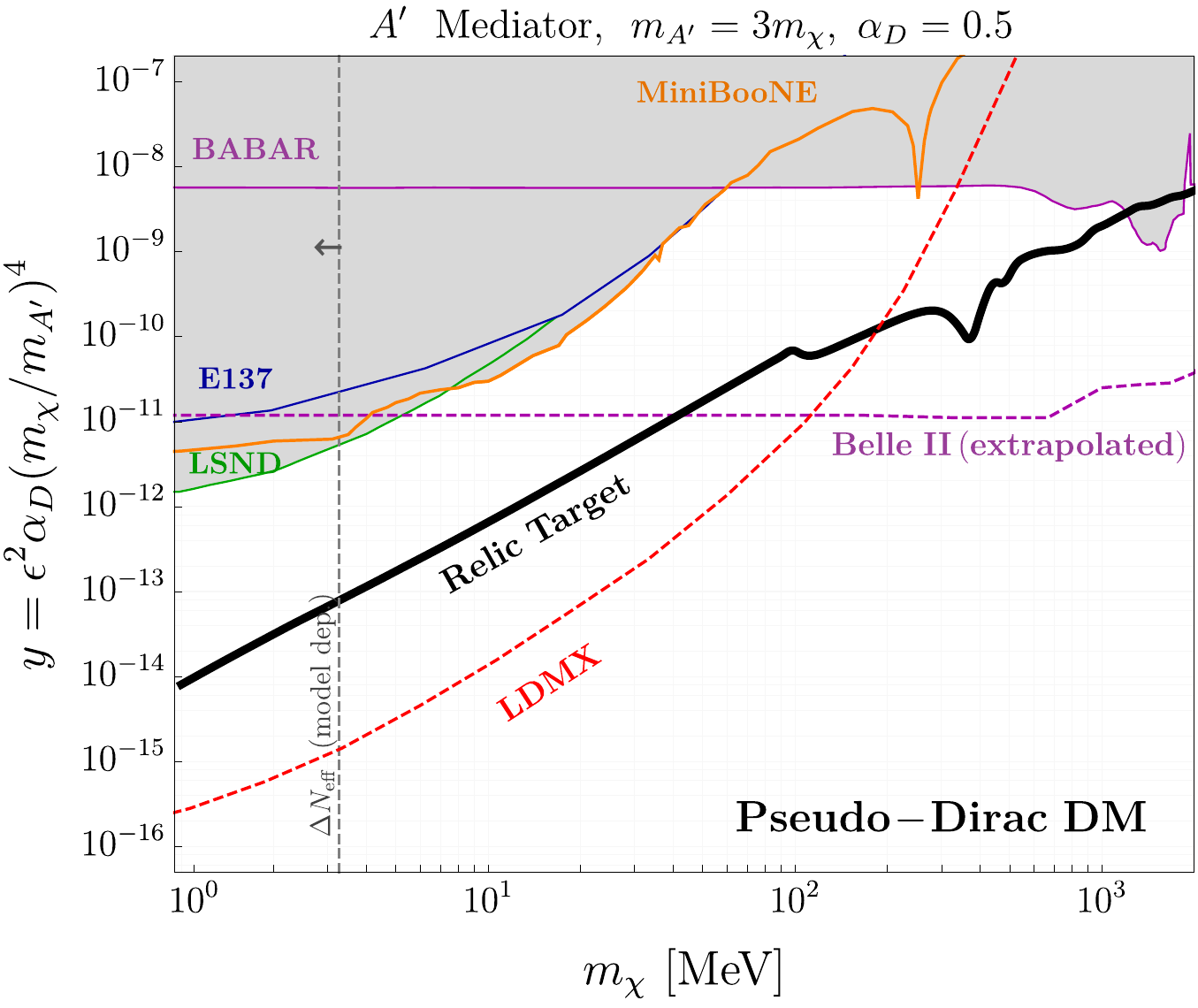}
\includegraphics[width=3.in,angle=0]{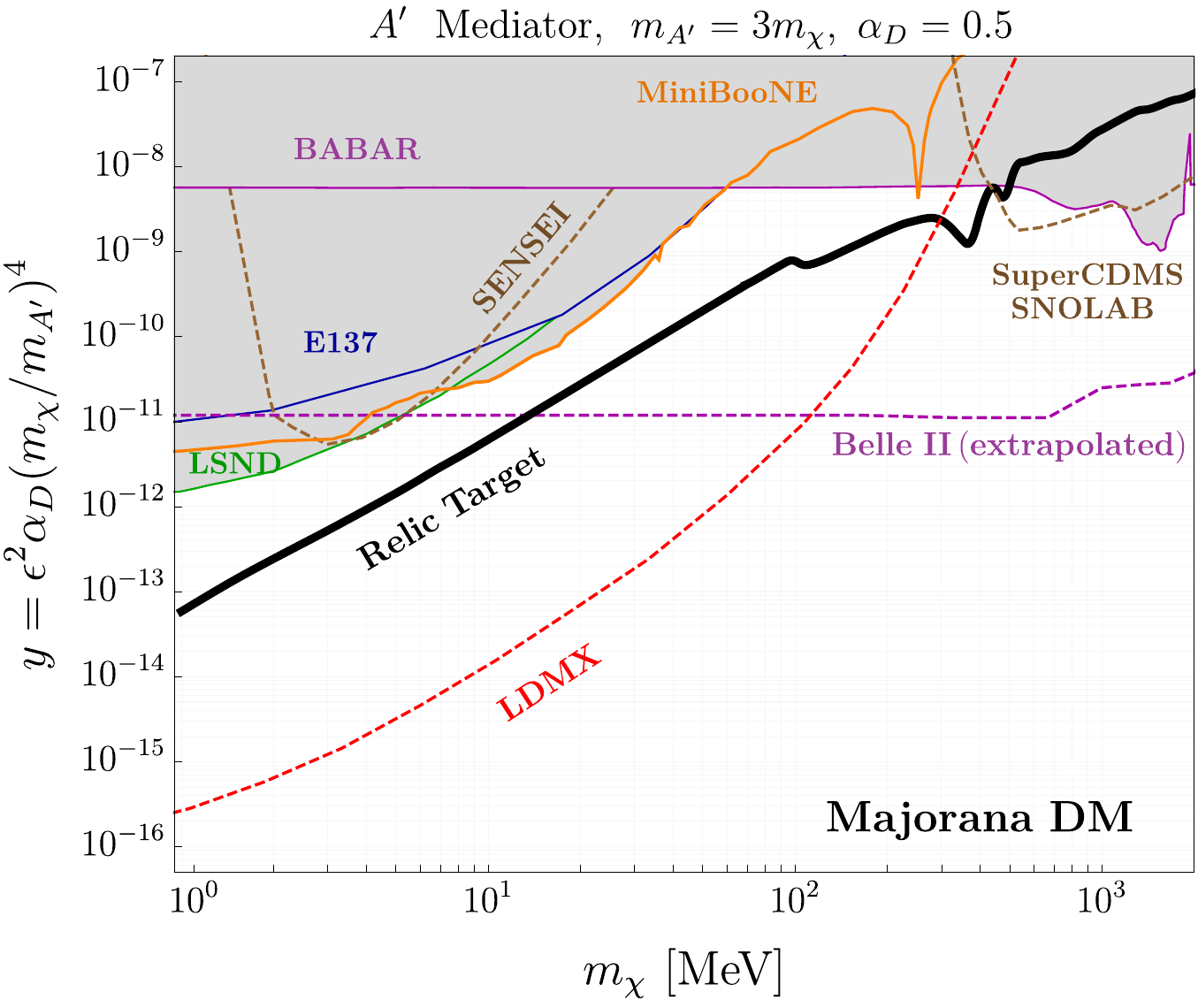}
\\
\includegraphics[width=3.in,angle=0]{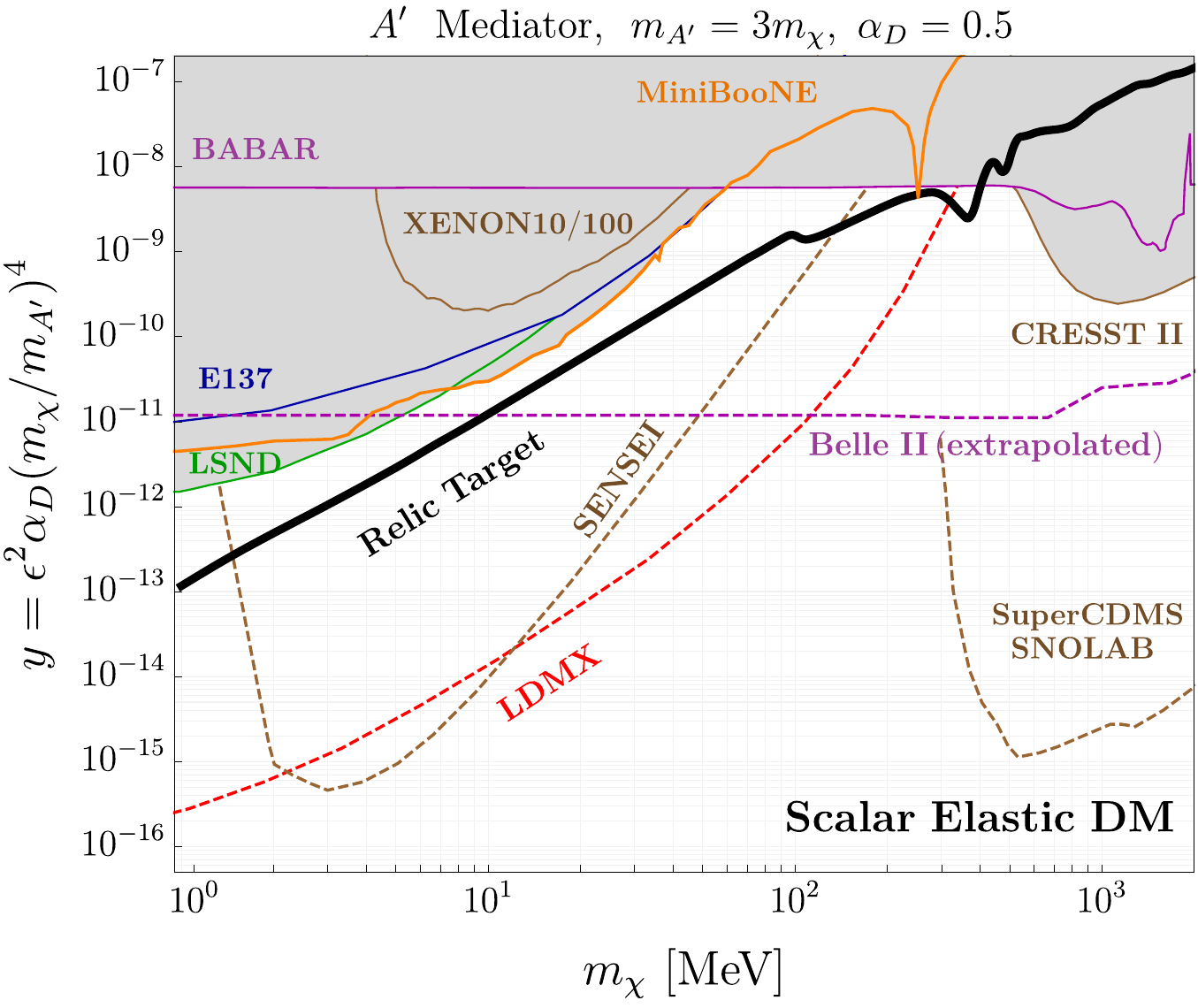}
\includegraphics[width=3.in,angle=0]{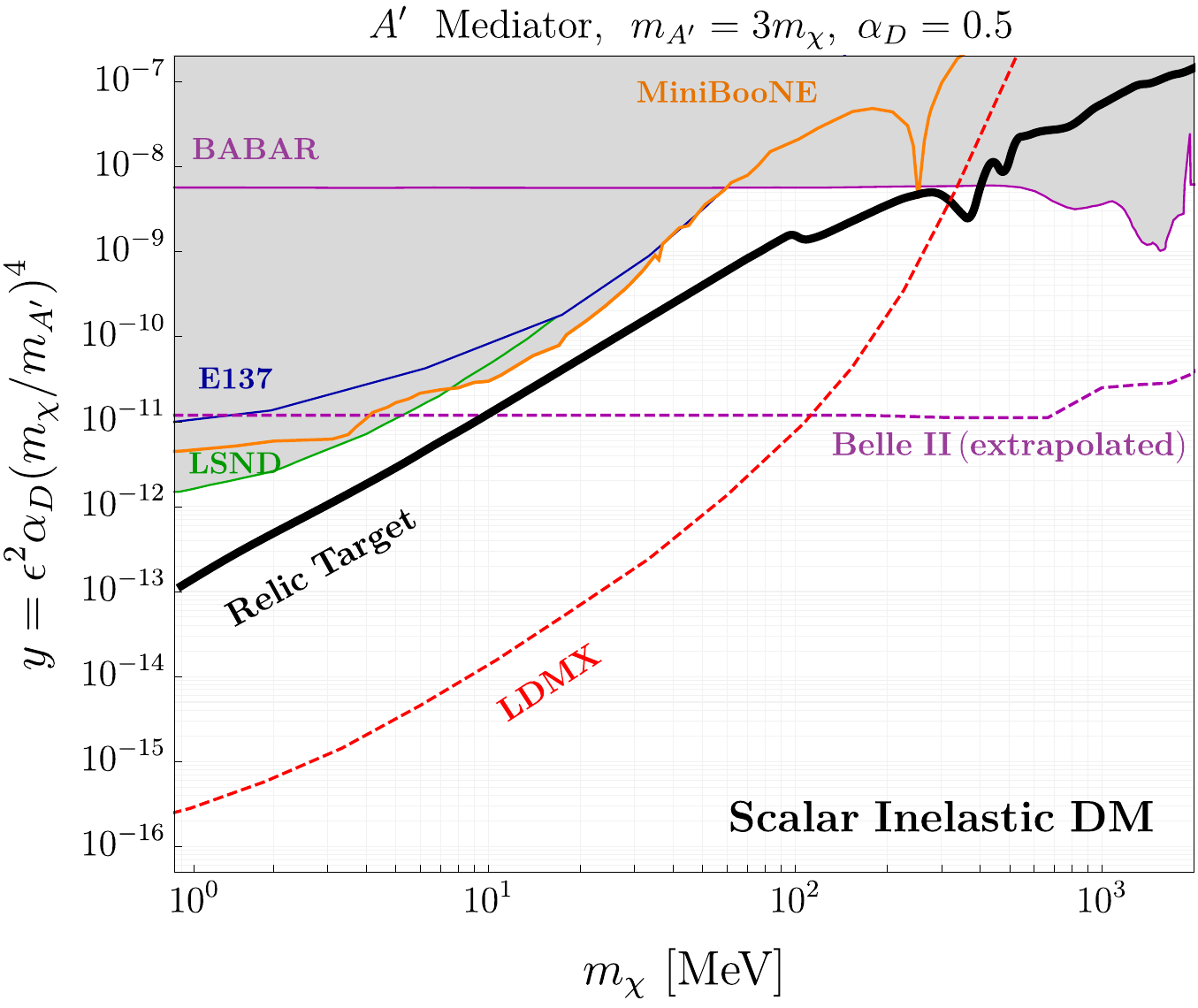}

\caption{
Thermal targets for representative dark matter candidates coupled 
to kinetically mixed dark photons. 
The black curve in each panel represents the parameter
 space for which the abundance of $\chi$ is in agreement with the observed dark matter energy density.
 In each model, $\chi$ freezes out through direct annihilations to SM fermions, i.e., $\chi\chi \to A^{\prime *}  \to f \bar{f}$. The shaded
 region above the purple curve is excluded by the  BaBar $\gamma+$ missing energy search~\cite{Izaguirre:2013uxa,Essig:2013vha}.
 The LSND proton~\cite{deNiverville:2011it} and E137 electron beam
 dump searches (green and blue curves, respectively) constrain DM production and scattering in a downstream detector~\cite{Batell:2014mga}. 
 The dashed purple 
 curve is the projected sensitivity of a $\gamma$+missing energy search at Belle II presented in 
Ref.~\cite{Battaglieri:2017aum} and computed by rescaling the 20 fb$^{-1}$ background study up to 50 ab$^{-1}$~\cite{HeartyCV}. 
 The dashed green curve labeled SENSEI is a direct detection projection assuming a  silicon target with 100\ g\ cdot yr of exposure 
 with $2e^-$ sensitivity~\cite{Battaglieri:2017aum}.
 The red dashed curve is the LDMX projection for a $10\%$ radiation length tungsten target and an 8 GeV beam presented in Ref.~\cite{Battaglieri:2017aum} , which was 
 scaled up to $10^{16}$ EOT relative to a background study  
 with $4 \times 10^{14}$ EOT. The vertical dashed curve in the upper left panel is the limit on $\Delta N_{\rm eff}$  from Table I of Ref.~\cite{nollett:2013pwa}, 
 which constrains $m_\chi < 3.27$ MeV for (pseudo)Dirac particles; similar bounds apply to the other scenarios for 
 $m_\chi < 1$ MeV. Note that electromagnetically coupled particles {\it decrease} $N_{\rm eff}$ at the time of recombination, so this effect can be compensated with additional dark radiation. }
 \label{fig:thermal-targets}
\end{figure}

In Fig.~\ref{fig:thermal-targets}, we present the DM parameter space in the $y - m_\chi$ plane for the various DM models to be discussed in Secs. \ref{sec:scalar-elastic}-\ref{sec:pseudo-dirac}. Along the black contour, $\chi$ freezes out with an abundance in agreement with the observed DM energy density.\footnote{
 Note that the thermal targets shown here for dark photon mediators in Fig.~\ref{fig:thermal-targets} are shifted slightly upwards relative to results shown in Sec.~VI of Ref.~\cite{Battaglieri:2017aum}.
 This difference is due to a coding bug that affected previous results and has been corrected in this work.
} We have performed this calculation by numerically solving the relevant Boltzmann equations governing the DM number density~\cite{Edsjo:1997bg,Gondolo:1990dk}. We have included hadronic contributions to the general thermally-averaged DM annihilation cross section to SM final states, $\langle \sigma v (\chi \chi \to A^{\prime *} \to \text{SM}) \rangle$, utilizing the data-driven methods of Ref.~\cite{Patrignani:2016xqp,Ilten:2018crw} (see also the discussion in, e.g., Ref.~\cite{Izaguirre:2015zva}).

 Excluded regions are shown in solid gray in Fig.~\ref{fig:thermal-targets}. These include constraints from searches for DM production and scattering in a detector placed downstream of the beam dumps LSND~\cite{Auerbach:2001wg,deNiverville:2011it}, E137~\cite{Bjorken:1988as,Batell:2014mga}, and MiniBooNE~\cite{Aguilar-Arevalo:2017mqx}, as well as a monophoton search for invisibly decaying dark photons at BaBar~\cite{Lees:2017lec} and the direct detection experiments XENON10/100 and CRESST II~\cite{Angle:2011th,Essig:2012yx,Angloher:2015ewa,Essig:2017kqs}. We also highlight the possible reach of Belle II (obtained by scaling the 20 fb$^{-1}$ projection by factor of 2500 to 50 ab$^{-1}$, and assuming statistics-limited sensitivity)~\cite{Battaglieri:2017aum, HeartyCV}, future versions of the direct detection experiments SENSEI (assuming a  silicon target with 100\ g $\cdot$ yr of exposure 
 and $2e^-$ sensitivity) and SuperCDMS~\cite{Battaglieri:2017aum}, as well as a missing momentum search at LDMX. The projected LDMX sensitivity in Fig.~\ref{fig:thermal-targets} corresponds to a $10\%$ radiation length tungsten target scaled up to an $8$ GeV beam and $10^{16}$ EOT relative to a background study with a $4$ GeV beam and $4 \times 10^{14}$ EOT~\cite{Battaglieri:2017aum}. This is a reasonable extrapolation because the photonuclear background 
   rate \emph{and} the background veto inefficiency dramatically decrease with a larger beam energy.

\subsubsection{Scalar Elastic Dark Matter}\label{sec:scalar-elastic}

If $\chi$ is a complex scalar with unit charge under $U(1)_D$, then the DM current that couples to $\Ap$ in Eq.~(\ref{eq:master-lagrangian}) is given by
\be
J^\mu_D =  i (\chi^* \partial^\mu \chi - \chi \partial^\mu \chi^*)
~.
\ee
The non-relativistic cross section for DM annihilations to a pair of light ($m_\ell \ll m_\chi$) SM leptons is given by
\be
\label{eq:scalar-annihilation}
\sigma v (\chi \chi^* \to A^{\prime *} \to \ell^+ \ell^- ) \simeq   \frac{8 \pi}{3} ~ \frac{  \alpha_\text{em} \, \epsilon^2 \, \alpha_D \, m_\chi^2 \, v^2}{   (4 m_\chi^2- m^2_{A^\prime})^2 
} \simeq 
\frac{8 \pi \, \alpha_\text{em} \, v^2}{3} ~ \frac{ y}{ m^2_\chi}
~,
 \ee
 where $v$ is the relative DM velocity,  and in the last step we have taken the $m_{A^\prime} \gg m_\chi$ limit.
 Since this is a $p$-wave process, the annihilation rate is strongly suppressed by the small DM velocity at late times, alleviating constraints from energy injection during recombination. We incorporate DM annihilations to hadronic final states through the approximate relation, 
\be
\label{eq:ApDecaytoHadrons}
\sigma v (\chi \chi^\star \to A^{\prime *} \to \text{hadrons}) \simeq R(s) \times \sigma v (\chi \chi^\star \to A^{\prime *} \to \mu^+ \mu^-)
~,
\ee
where $R \equiv \sigma (e^+ e^- \to \text{hadrons}) / \sigma(e^+ e^- \to \mu^+ \mu^-)$ is the data-driven parameter from Ref.~\cite{Patrignani:2016xqp}. 
 
At direct detection experiments, the non-relativistic $\chi - f$ elastic scattering cross section is approximately
\be
\label{eq:sigmav-scalar-elastic}
\sigma (\chi f \to \chi f) \simeq \frac{16 \pi \, \alpha_\text{em}  \, \epsilon^2 \,  \alpha_D \,  \mu_{\chi f}^2}{ (q^2 + m^2_{A^\prime})^2 }
~,
\ee
where $q$ is the three-momentum transfer and $\mu_{\chi f}$ is the DM-target reduced mass. Since there is no  suppression in the non-relativistic scattering limit, this scenario is 
the most favorable for direct detection experiments. Constraints from XENON10/100 and projections for LDMX, SENSEI and SuperCDMS are shown in the bottom-left panel of Fig.~\ref{fig:thermal-targets}~\cite{Battaglieri:2017aum}.

\subsubsection{Scalar Inelastic Dark Matter}\label{sec:scalar-inelastic}
A variation on the scalar elastic model described above can arise if $\chi$ acquires additional mass terms that explicitly break $U(1)_D$ (the scalar analogue of Majorana masses) such as
\be
\label{eq:lag-scalar-inelastic}
-\mathscr{L} \supset 
 m_\chi^2 \, |\chi|^2  + \mu_\chi^2 \, \chi^2 + \text{h.c.}
 ~,
\ee
where the $\mu_\chi$ mass term may arise after the spontaneous symmetry breaking of $U(1)_D$ (e.g., by coupling to a dark Higgs with $U(1)_D$ charge +2). 
Diagonalizing this system yields the mass eigenstates $\chi_{1,2}$, which couple off-diagonally (inelastically)  to the dark photon in Eq.~(\ref{eq:master-lagrangian}) through the current
\be
J_D^\mu =   \chi_1 \partial^\mu \chi_2 - \chi_2 \partial^\mu \chi_1
~.
\ee
Since DM couples purely off-diagonally to the dark photon in these scenarios, scattering processes (such as $\chi_1 f \to \chi_2 f$) at direct detection experiments are kinematically suppressed if the $\chi_{1,2}$ fractional mass-splitting is larger than $\mathcal{O} (10^{-6})$~\cite{TuckerSmith:2001hy}. However, for fractional mass-splittings that are smaller than $\mathcal{O}(10^{-1})$ the cosmology and accelerator phenomenology discussed throughout this work is left unchanged. This is evident in the bottom-right panel of Fig.~\ref{fig:thermal-targets}, which aside from the lack of sensitivity of direct detection experiments, is identical the case of elastic scalar DM. Note that, like the scalar elastic case
described above, this model also features $p$-wave annihilation and is safe from CMB bounds.

\subsubsection{Majorana Elastic Dark Matter}\label{sec:majorana-elastic}
For a Majorana fermion coupled to a dark photon, we write the $U(1)_D$ current of Eq.~(\ref{eq:master-lagrangian}) as
\be
J_D^\mu = \frac{1}{2} \, \overline \chi \gamma^\mu \gamma^5 \chi
~,
\ee
where the conventional factor of $1/2$ is meant to counteract the additional factor of $2$ in the Feynman rule for identical particles. 

As in Sec.~\ref{sec:scalar-elastic}, the non-relativistic cross section for DM direct annihilations to leptons is approximately 
\be
\sigma v (\chi \chi \to A^{\prime *} \to \ell^+ \ell^- ) \simeq   \frac{8 \pi}{3} ~ \frac{ \alpha_\text{em} \, \epsilon^2 \, \alpha_D \, m_\chi^2 \, v^2}{   (4 m_\chi^2- m^2_{A^\prime})^2 }
 \simeq 
\frac{8 \pi \, \alpha_\text{em} \, v^2}{3} ~ \frac{ y}{ m^2_\chi}
~,
 \ee
which is identical to the form in Eq.~(\ref{eq:scalar-annihilation}). As a result, this scenario is similarly CMB-safe. At direct detection experiments, the non-relativistic $\chi - f$ elastic scattering cross section is
\be
\sigma (\chi f \to \chi f) \simeq \frac{8 \pi \, \alpha_\text{em} \, \epsilon^2 \, \alpha_D \, \mu_{\chi f}^2}{m_{\Ap}^4} ~~ \frac{3 m_\chi^2 + 2 m_\chi m_f + m_f^2}{(m_\chi + m_f)^2} ~~ v^2
\ee
where $v$ is now the relative $\chi$-$f$ velocity. The velocity suppression in this rate significantly weakens prospects for
direct detection experiments, as seen in the top-right panel of Fig.~\ref{fig:thermal-targets}.

\subsubsection{Pseudo-Dirac Inelastic Dark Matter (Small Splitting: $\Delta \ll m_\chi$) }\label{sec:pseudo-dirac}

We now consider a Dirac pair of two-component Weyl fermions ($\eta, \xi$) that have opposite unit charge under $U(1)_D$, and possess both a $U(1)_D$ conserving (breaking) Dirac (Majorana) mass term, $m_D$ ($m_M$). In analogy to Sec.~\ref{sec:scalar-inelastic}, the relevant mass terms are
\be
- \Lag \supset m_D \, \eta \, \xi + \frac{1}{2} \, m_M \left( \eta^2 +  \xi^2 \right) + {\text{h.c.}}
\ee
In the limit that $m_D \gg m_M \neq 0$, the mass eigenstates ($\chi_{1,2}$) correspond to a pseudo-Dirac pair given by
\be
\chi_1 &\simeq \frac{i}{\sqrt{2}} ~ (\eta - \xi)  ~~,~~ \chi_2 &\simeq \frac{1}{\sqrt{2}} ~ (\eta + \xi)
~,
\ee
with nearly degenerate masses, $m_1 \lesssim m_2$, where
\be
m_{1,2} \simeq m_D \mp m_M
~.
\ee
For later convenience, we define the dimensionful mass-splitting,
\be
\Delta \equiv m_2 - m_1 \simeq 2 \, m_M
~.
\ee
 In this mass basis, $\chi_{1,2}$ couple off-diagonally to the dark photon. In four-component notation where $\chi_{1,2}$ are Majorana fermions, the current of Eq.~(\ref{eq:master-lagrangian}) becomes
\be
J_D^\mu = i\, \bar\chi_1\gamma^\mu\chi_2
~.
\ee
As in Sec.~\ref{sec:scalar-inelastic}, scattering off of SM fermions ($\chi_1 f \to \chi_2 f$) is kinematically suppressed for mass-splittings larger than $\Delta / m_1 \gtrsim \mathcal{O}(10^{-6})$~\cite{TuckerSmith:2001hy}.

For $\Delta \ll m_1$, we define $m_\chi \equiv m_1 \simeq m_2$. In this limit, the  coannihilation cross section to 
light SM leptons via an intermediate $\Ap$ is approximately
\be
\label{eq:sigmavIDM}
\sigma v(\chi_1\chi_2 \to A^{\prime *} \to \ell^+ \ell^-) \simeq \frac{16 \pi \, \alpha_\text{em} \, \epsilon^2  \,  \alpha_D \, m_\chi^2}{(4 m_\chi^2- m^2_{A^\prime})^2 
} \simeq \frac{16 \pi \, \alpha_\text{em} \, y}{m_\chi^2}
~,
\ee
which is valid for $m_\chi \gg m_\ell$. In the second equality, we have taken the $m_{A^\prime} \gg m_\chi$ limit and used the definition of $y$ in Eq~(\ref{eq:define-y}). As in all the previous DM models discussed above, this form is 
approximately valid away from resonances 
and particle thresholds. Although this is an $s$-wave process, the effective thermally-averaged form of the cross section that enters the Boltzmann equation 
is exponentially reduced at temperatures below the $\chi_{1,2}$ mass splitting, i.e., $T \lesssim \Delta$, due to the Boltzmann suppression in the number density of the slightly heavier $\chi_2$. The reduced density of $\chi_2$ at the time of recombination suppresses the annihilation rate and therefore the energy injection in the CMB.  
If $\Delta>2\,m_e$, then $\chi_2$ is depleted via decay to $\chi_1 e^+e^-$ , leading to a completely negligible $\chi_2$ abundance at recombination; for $\Delta<2\,m_e$, the depletion of $\chi_2$ through scattering leaves some residual freeze-out abundance of $\chi_2$, but over much of the parameter space this abundance is sufficiently small to completely alleviate the CMB constraints. In the bottom-right panel of Fig.~\ref{fig:thermal-targets}, we compute the thermal target for this scenario 
assuming that the mass-splitting, $\Delta$, is negligible, i.e., $\Delta \lesssim \mathcal{O}(0.1) m_\chi$.  

\subsubsection{Pseudo-Dirac Inelastic Dark Matter (Large Splitting: $\Delta > 2m_e$) }\label{sec:pseudo-dirac-large-delta}
If the pseudo-Dirac model described in Sec.~\ref{sec:pseudo-dirac} features a larger 
mass-splitting, the cosmology is slightly altered and 
the heavier state, $\chi_2$, is potentially unstable on accelerator timescales, motivating
novel discovery opportunities in the search for visible $\chi_2$ decay products at fixed-target and collider experiments~\cite{Izaguirre:2014dua,Izaguirre:2015zva,Izaguirre:2017bqb,Berlin:2018pwi}.  Furthermore, for $m_{A^\prime}  > m_1  + m_2$ and $\Delta \lesssim \mathcal{O}(1)m_1$, DM freeze-out is dominantly controlled by coannihilations into SM fermions ($\chi_1\chi_2 \to A^{\prime *} \to f \bar{f}$), as discussed in the previous subsection. In this case, $\chi_2$ decays through an off-shell dark photon into $\chi_1$ and a pair of SM leptons. In the limit that $m_{\Ap} \gg m_{1,2} \gg m_\ell$ and $\Delta \ll m_1$, the corresponding decay rate takes the approximate form 
\be
\label{eq:decaychi2}
\Gamma(\chi_2 \to \chi_1 \, \ell^+ \ell^- ) \simeq \frac{4\, \alpha_\text{em} \, \epsilon^2\, \alpha_D \, \Delta^5}{15\pi \, m_{\Ap}^4}
\propto y \, (\Delta/m_1)^5 \, m_1
~.
\ee
For fixed values of $\Delta / m_1$, the proper lifetime of $\chi_2$ is dictated by the same couplings that control the DM relic abundance, i.e., $y$ and $m_1$.

A key conceptual difference between the small and large mass-splitting regimes is that the cosmology is highly sensitive to $\Delta$ if the splitting is comparable to (or larger than) the $\chi_1$ freeze-out temperature, $T_{f} \simeq m_1/20$. 
For  $\Delta \gtrsim  T_f$, the population of the heavier $\chi_2$ is Boltzmann suppressed at freeze-out compared to $\chi_1$, i.e., $n_{\chi_2} / n_{\chi_1} \propto e^{-\Delta/T_f}$.
As a result, $\chi_1$ has fewer potential coannihilation partners during freeze-out relative to the case of 
smaller splittings. To compensate for this depletion, the requisite $\chi_1\chi_2 \to A^{\prime *} \to f \bar{f}$ coannihilation cross section  
must increase {\it exponentially} to obtain an abundance of $\chi_1$ that is in agreement with the observed DM energy density.

\begin{figure}[t!]
\includegraphics[width=10cm]{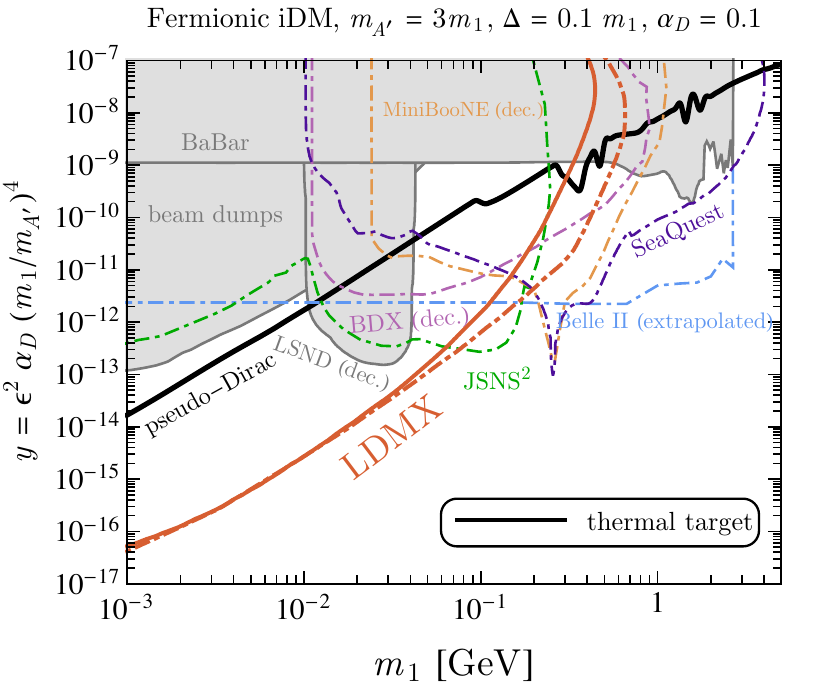}
\caption{ 
Parameter space for pseudo-Dirac DM. The mass eigenstates, $\chi_{1,2}$, couple off-diagonally to the dark photon, $A^\prime$, and freeze-out through coannihilations to SM particles.
The heavier state in the pseudo-Dirac pair is unstable 
and decays via $\chi_2 \to \chi_1 f \bar{f}$. These displaced visible decays can be searched for at
accelerator experiments. Here we present various projections for LDMX (for a 8/16 GeV electron beam assuming $10^{16}$ EOT and a 10\% tungsten/aluminum target in solid/dot-dashed red, respectively) and SeaQuest~\cite{Berlin:2018pwi}, JSNS$^2$~\cite{Jordan:2018gcd}, BDX, and MiniBooNE~\cite{Izaguirre:2017bqb,battaglieri:2016ggd}. Also shown are constraints from LSND, BaBar~\cite{Izaguirre:2015zva,Izaguirre:2017bqb,Berlin:2018pwi}, Belle II~\cite{Izaguirre:2015zva}, and LEP~\cite{Hook:2010tw}. We do not show constraints derived from the electron beam dump E137 since they suffer from uncertainties pertaining to the energy threshold of the analysis~\cite{Berlin:2018pwi}.
}
\label{fig:iDMplot}
\end{figure}

Nonetheless, one can still define thermal targets for each choice of $\Delta$  as $y \equiv \epsilon^2 \alpha_D (m_1/m_{A^\prime})^4$ 
and in Fig.~\ref{fig:iDMplot} we show a representative example for $m_{\Ap} / m_1 = 3$, $\Delta = 0.1 \, m_1$, and $\alpha_D = 0.1$. In this figure, many of the beam dump and $B$-factory 
constraints are identical to those in the bottom-right panel of Fig.~\ref{fig:thermal-targets}; however, there are now 
additional constraints and future projections for experiments able to detect displaced visible $\chi_2 \to \chi_1 \ell^+ \ell^-$ decays, which offer the greatest sensitivity at high mass and splitting.

\subsubsection{Resonant and Forbidden Regimes: $1 < m_\Ap/m_\chi < 3$}
\begin{figure}[t!]
\includegraphics[width=3.in,angle=0]{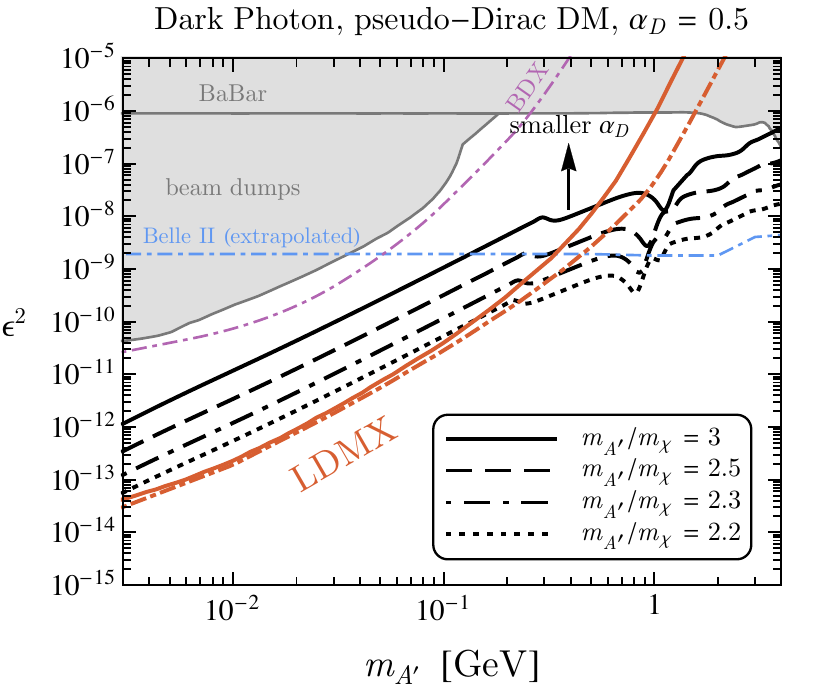}
\includegraphics[width=3.in,angle=0]{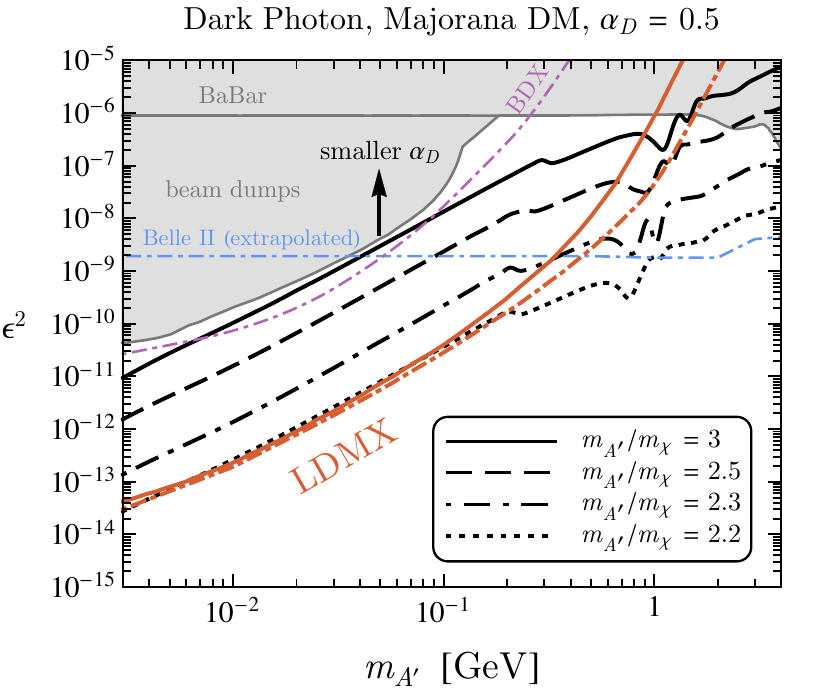}
\includegraphics[width=3.in,angle=0]{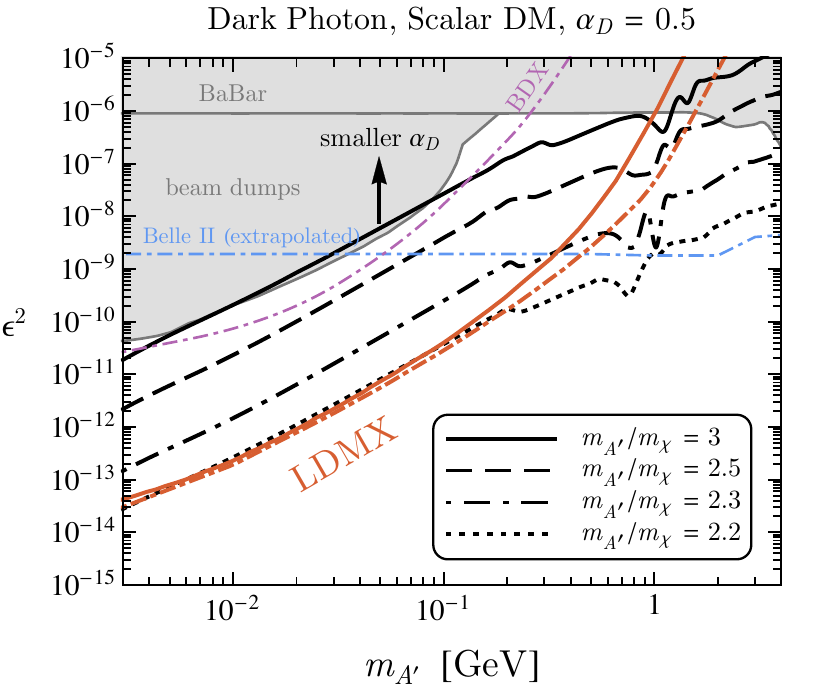}
\caption{Thermal targets for a subset of the dark photon mediated models in Fig.~\ref{fig:thermal-targets}, but 
presented in the $\epsilon^2 - m_{A^\prime}$ plane with fixed $\alpha_D = 0.5$. The different thermal 
targets (black contours) correspond to various choices of $m_{A^\prime}/m_\chi $ just above the resonance ($m_{A^\prime} \approx 2 \, m_\chi$) where $\chi$ freezes out through annihilations to SM fermions, $\chi \chi \to A^{\prime *} \to f \bar{f}$. The thermal targets presented here are consistent with the results of 
Ref.~\cite{Feng:2017drg}. The shaded gray regions are excluded from previous experiments, such as a BaBar monophoton analysis~\cite{Lees:2017lec}, and beam dump searches at LSND~\cite{deNiverville:2011it},  E137~\cite{Bjorken:1988as,Batell:2014mga}, and MiniBooNE~\cite{Aguilar-Arevalo:2017mqx}. In dot-dashed blue is the projected sensitivity of a monophoton search at Belle II presented in 
Ref.~\cite{Battaglieri:2017aum} and computed by rescaling the 20 fb$^{-1}$ background study up to 50 ab$^{-1}$~\cite{HeartyCV}. Also shown in dot-dashed purple is the projected reach of the beam dump experiment BDX~\cite{battaglieri:2016ggd,Izaguirre:2013uxa}.  The projected sensitivity of LDMX is shown in solid (dot-dashed) red, assuming $10^{16}$ EOT from a 8 (16) GeV electron beam and a $10\%$ radiation length tungsten (aluminum) target.
\label{fig:resonant_ap_thermal_targets}
  }
\end{figure}

\begin{figure}
\includegraphics[width=0.48\textwidth]{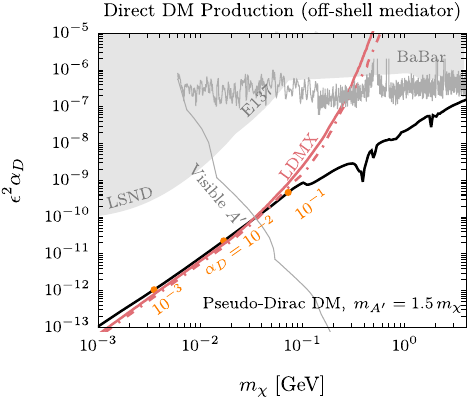}
\includegraphics[width=0.48\textwidth]{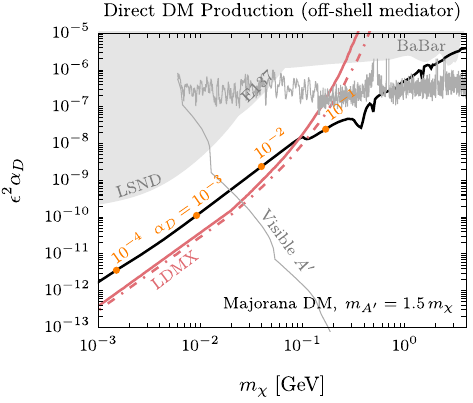}
\caption{Thermal targets for a subset of the dark photon-mediated models in Fig.~\ref{fig:thermal-targets}, shown in 
the $m_\chi -\epsilon^2 \alpha_D$ plane for $\mAp/m_\chi = 1.5$. The relic abundance of $\chi$ makes up 
all of the DM along the solid black line. The orange points label the values of $m_\chi$ for a given $\alpha_D$ 
below which the ``forbidden'' annihilation channel $\chi\chi\rightarrow \Ap\Ap$ determines the relic abundance. When 
this happens, the abundance is determined by $\alpha_D$ alone and the model ceases to predict a target for the $\chi$-SM 
coupling. The solid (dot-dashed) red curve shows the sensitivity of LDMX with $10^{16}$ EOT and an 8 (16) GeV 
electron beam. Because $\mAp < 2m_\chi$, pairs of $\chi$ particles are produced directly through 
electron bremsstrahlung through an off-shell $\Ap$ at LDMX. Regions shaded in gray are excluded by 
the BaBar monophoton search~\cite{Lees:2017lec}, and by beam-dump experiments LSND~\cite{deNiverville:2011it} and E137~\cite{Bjorken:1988as,Batell:2014mga}.
Possible constraints from searches for visibly-decaying $\Ap$ are shown by a thin gray line for $\alpha_D = 0.5$~\cite{Andreas:2012mt}.
These are model dependent since they assume a 100\% $\Ap$ branching fraction into the SM.
\label{fig:offshell_ap_thermal_targets}
}
\end{figure}

   In the previous sections we have focused on the mediator/DM mass ratio of $\mAp / m_\chi =3$, which 
   allows on-shell decays of $\Ap$ to DM and avoids resonant enhancement of the annihilation rate in the early universe.
   In Figures~\ref{fig:resonant_ap_thermal_targets} and~\ref{fig:offshell_ap_thermal_targets} we show the effects of relaxing this 
   assumption by varying $m_\chi/\mAp$ 
   to illustrate qualitatively different regions of parameter space. Fig.~\ref{fig:resonant_ap_thermal_targets} explores 
   the sensitivity of LDMX to scenarios with $\mAp \approx 2m_\chi$ in which 
   DM annihilation is resonantly enhanced, such that sufficient annihilation rates can be achieved for smaller 
   $\chi$-SM couplings~\cite{Feng:2017drg}. Near this resonance, the annihilation cross-section becomes insensitive to $\alpha_D$ 
   so we show the thermal target in the $\epsilon^2 - \mAp$ plane. 
   Away from the resonance, we have fixed $\alpha_D=0.5$. This is a conservative assumption since 
   smaller values of $\alpha_D$ move the targets to larger couplings as indicated by the black arrows. 
   The LDMX sensitivity is shown for $10^{16}$ EOT with 8 and 16 GeV electron beams as 
   the red solid and dot-dashed lines, respectively. Existing constraints 
   are indicated by gray shaded regions.

   So far we have focused on the range $\mAp \gtrsim 2m_\chi$. In Fig.~\ref{fig:offshell_ap_thermal_targets} we consider the complementary mass range with $m_\chi \lesssim \mAp \lesssim 2m_\chi$, fixing $\mAp/m_\chi = 1.5$.
   This choice forbids on-shell $\Ap$ decays to DM. In accelerator experiments, dark matter then must be produced directly through an off-shell $\Ap$.
   In this mass range direct DM annihilation to the SM, $\chi \chi \rightarrow \bar f f$ competes with the 
   ``forbidden'' annihilation channel $\chi \chi \rightarrow \Ap \Ap$~\cite{DAgnolo:2015ujb,Cline:2017tka}, whose rate depends on $\alpha_D$ and sensitively on the $\chi$--$\Ap$ mass ratio, but not on $\epsilon$. If direct annihilation dominates, the requirement of 
   obtaining the correct relic abundance determines a thermal target in $m_\chi -\epsilon^2 \alpha_D$ parameter space.\footnote{When $\mAp < 2m_\chi$, 
   $y$ is no longer a preferred variable since the $s$-channel propagator in $\chi \chi \rightarrow \bar f f$ is dominated by $s\approx 4m_\chi^2$ instead of $\mAp^2$.}
   If, on the other hand, the forbidden-channel annihilation is strong enough to produce the observed DM abundance, the target for the $\chi$-SM coupling $\epsilon$ turns into an upper bound. 
   Fig.~\ref{fig:offshell_ap_thermal_targets} shows the predictive thermal targets as the thick black lines.
   For a given $\alpha_D$ there is a minimum value of $m_\chi$ for which this target exists; for lower 
   DM masses the forbidden channel dominates and there is no thermal target. These lower bounds are indicated by orange dots for several 
   values of $\alpha_D$.
   As before, the LDMX sensitivity is shown for $10^{16}$ EOT with 8 and 16 GeV electron beams as 
   the red solid and dot-dashed lines, respectively. We see that LDMX is sensitive to direct DM 
   production.
   The gray regions are excluded by BaBar~\cite{Lees:2017lec}, E137~\cite{Bjorken:1988as,Batell:2014mga} and LSND~\cite{deNiverville:2011it}.
   The beam dump constraints for DM production through an off-shell $\Ap$ were evaluated using the method of Ref.~\cite{Izaguirre:2017bqb}.
   For $\mAp < 2m_\chi$, an on-shell $\Ap$ dominantly decays to SM particles, enabling searches for visible 
   signals. The resulting bounds are model-dependent since they are significantly weakened if the $\Ap$ can decay into any dark-sector final states. In the absence of such a channel, the constraints on a visibly-decaying $\Ap$ are shown in Fig.~\ref{fig:offshell_ap_thermal_targets} 
   for $\alpha_D = 0.5$ as the thin gray line~\cite{Andreas:2012mt}. Smaller $\alpha_D$ would result in moving the bounds from visible searches down in the 
   $m_\chi - \epsilon^2 \alpha_D$ plane.  Bump-hunt constraints are, therefore, weakest for large $\alpha_D$. However, displaced-decay searches have a \emph{maximum} effective $\epsilon$ for given mediator mass and because of this,  lowering $\alpha_D$ down towards $\sim 10^{-3}$ opens up more thermal parameter space by pushing the gray curve down.

\subsection{Predictive Dark Matter with Other Mediators}
\label{sec:other-mediators}
In this section, we generalize the above discussion to include spin-1 mediators (Sec.~\ref{ssec:spin1}) and spin-0 mediators (Sec.~\ref{ssec:spin0}) with 
more general couplings to the SM.  In the vast majority of these models, the electron coupling dominantly controls DM freeze-out. 
Hence, direct searches for these mediators through electron couplings is a well-motivated and powerful technique.   The leptophilic scalar and baryonic coupled vector are extreme examples of this; even though the electron coupling is highly suppressed in these models, it is the coupling that controls freeze-out for light DM. Furthermore, LDMX can probe interaction strengths motivated by thermal freeze-out.  The exception to this rule is if the dominant annihilation channel for light dark matter is into neutrinos, as is the case, for example, for a vector coupled to $L_\mu - L_\tau$.  These scenarios motivate a muon-beam variant to LDMX \cite{Kahn:2018cqs}, and are discussed in 
Sec.~\ref{ssec:muphilic}.  For simplicity, we group together in this discussion the different possibilities for DM spin (as discussed in Sec.~\ref{sec:dark-photon-benchmark-models}) that are compatible with CMB bounds for each choice of mediator. 

\subsubsection{Predictive Dark Matter with Other Spin-1 Mediators}
\label{ssec:spin1}

We now consider variations to the dark photon mediated models of Sec.~\ref{sec:dark-photon-benchmark-models} by coupling the DM sector to SM currents other than electromagnetism. This is similar in spirit to the recent works of Refs.~\cite{Ilten:2018crw,Bauer:2018onh}, which recasted the existing bounds and projected sensitivities of future and proposed experiments to visibly decaying $\zp$ gauge bosons for the scenarios of Eq.~(\ref{eq:U(1)list}). As a representative set, we will focus on new light forces corresponding to the following gauged global symmetries of the SM,
\be
\label{eq:U(1)list}
U(1)_{B-L}~~,~~U(1)_{B-3L_i} ~~,~~ U(1)_{L_i - L_j} ~~,~~ U(1)_B
~,
\ee
where $B$ corresponds to baryon number and $L_i$ denotes lepton number of generation $i = e , \mu , \tau$. Gauged symmetries of lepton family number differences, $U(1)_{L_i - L_j}$, are anomaly-free and require no additional particles beyond the associated gauge boson to preserve consistency of the quantum theory at high energies (though they are broken at a minute level by neutrino mass mixing). This is to be contrasted with the others of  Eq.~(\ref{eq:U(1)list}). The addition of neutrinos uncharged under the SM gauge group render $U(1)_{B-L}$ and $U(1)_{B - 3 L_i}$ anomaly-free, while the inclusion of new heavy SM-chiral fermions can cancel off anomalous triangle diagrams for $U(1)_B$ at high energies. In this section, we will focus on this representative set of gauge theories, with a particular emphasis on the DM thermal target parameter space. To contrast these theories with those involving the well-studied kinetically mixed dark photon ($\Ap$), we refer to the corresponding spin-1 force carriers of Eq.~(\ref{eq:U(1)list}) as $\zp$.

Similar to the previously considered models, we imagine that the $\zp$ couples predominantly to the DM currents of Sec.~\ref{sec:dark-photon-benchmark-models} with strength $g_D \sim \mathcal{O}(1)$ and feebly to the SM. In analogy to the dark photon-electron coupling, $\epsilon \, e$, we define the $\zp$-SM coupling strength as
\be
\label{eq:epsdef}
\epsilon_{BL} \equiv g_{\zp} / e ~,~ \ldots
~,
\ee
and similarly for the others of Eq.~(\ref{eq:U(1)list}), where $g_{\zp}$ is the usual gauge coupling constant of a typical $B-L$, $B-3 L_i$, $L_i - L_j$, or $B$ gauge boson. In models involving kinetically mixed dark photons, the hierarchy $\alpha_D \gg \alpha_\text{em} \, \epsilon^2$ is a natural outcome of the \emph{indirectness} of the SM sector's  coupling to $U(1)_D$ via kinetic mixing (in contrast to the direct $U(1)_D$ charge of the DM). However, for the models of Eq.~(\ref{eq:U(1)list}), this hierarchy is less straightforward, especially if both the DM and SM sectors are directly charged under the $\zp$. In minimal realizations of the gauge symmetries of Eq.~(\ref{eq:U(1)list}), this hierarchy would instead require a very large DM charge $Q_\text{DM} \gg Q_\text{SM}$. 
 The hierarchy $\alpha_D \gg \alpha_\text{em} \, \epsilon_{BL}^2$ (and similarly for the other symmetries) can arise more plausibly if the DM sector is directly charged under a distinct $U(1)_D$ that kinetically mixes with a $U(1)$ of Eq.~(\ref{eq:U(1)list}). For instance, if the $U(1)_D$ gauge boson ($X$) kinetically mixes with a $\zp$ of Eq.~(\ref{eq:U(1)list}) through
\be
\mathscr{L} \supset - \, \frac{\epsilon_X}{2} \, X^{\mu \nu} \, Z^\prime_{\mu \nu} + Z_\mu^\prime \, J_{\zp}^\mu
~,
\ee
then, in the mass eigenstate basis, $X$ inherits a suppressed coupling to the $\zp$ current ($J_{\zp}$)
\be
\mathscr{L} \supset \epsilon_X \left( m_X / m_{\zp} \right)^2 ~ X_\mu \, J_{\zp}^\mu \equiv g_{\zp}^\text{eff} ~ X_\mu \, J_{\zp}^\mu
~,
\ee
where we have taken the limit that $\epsilon_X \ll 1$ and $m_X \ll m_{\zp}$. In this case, $\epsilon_{BL}$ and the others of Eq.~(\ref{eq:epsdef}) are defined by replacing $g_{\zp}$ with $g_{\zp}^\text{eff}$. Since we are mainly interested in the low-energy phenomenology of these models, we will assume a large DM-$\zp$ coupling and a small SM-$\zp$ coupling (leading to the weakest bounds from DM searches in relation to the thermal targets), while remaining agnostic about the origin of this coupling structure.  In Sec.~\ref{ssec:bl_neutrinos} we will consider the alternate case where SM decays of a $B-L$ gauge boson dominate.

In this section, we will emphasize mediators that couple to electrons at tree-level (aside from $U(1)_B$).  We defer a discussion of thermal targets in a gauged $U(1)_{L_\mu -L_\tau}$ scenario to Sec.~\ref{sec:muon-thermal-dm}, which presents the reach of an LDMX-like experiment with high-intensity {\it muon } beams. For the various models of Eq.~(\ref{eq:U(1)list}) that couple to electrons, the DM thermal targets typically differ by $\mathcal{O}(1)$ factors due to different available final states for the annihilation processes. However, in models of $U(1)_B$ gauge bosons, thermal freeze-out is dramatically altered for $m_\chi \lesssim \mathcal{O}(100) \text{ MeV}$ since DM annihilations to pion final-states are kinematically suppressed. In this case, annihilations to electrons only occur through the radiatively generated kinetic mixing between $U(1)_B$ and $U(1)_Y$, which is induced by loops of hadronic particles. As is convention in the literature~\cite{Tulin:2014tya,Ilten:2018crw,Dror:2017ehi,Dror:2017nsg}, we take this effective kinetic mixing to be $\epsilon \sim e \, g_{\zp} / 16 \pi^2$. 

\begin{figure}[t!]
\includegraphics[width=7.8cm]{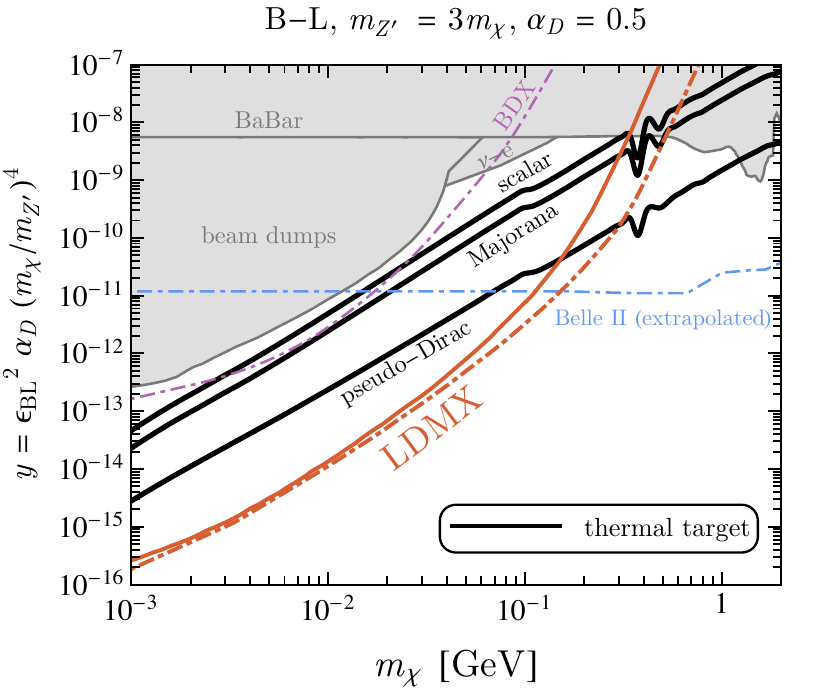}
\includegraphics[width=7.8cm]{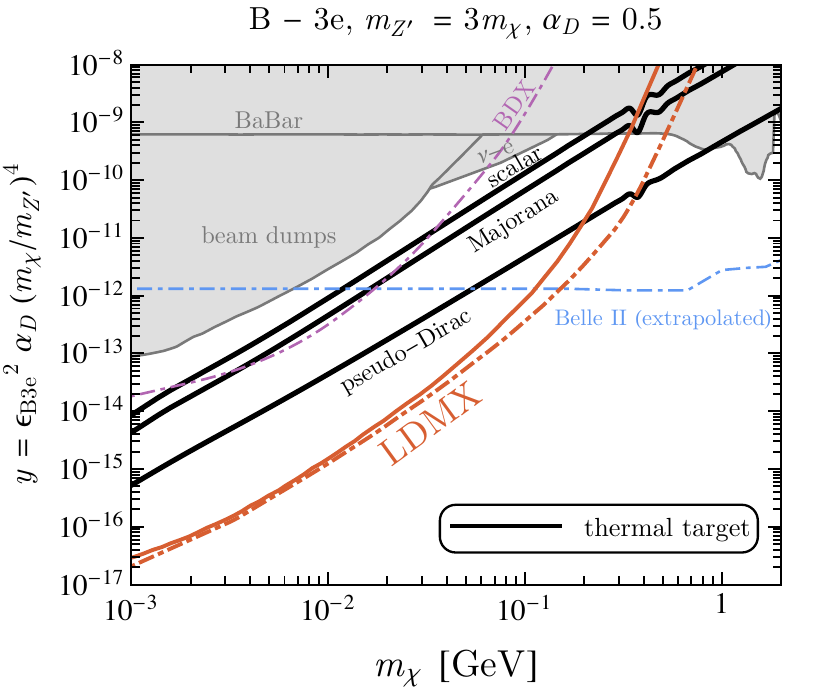}
\includegraphics[width=7.8cm]{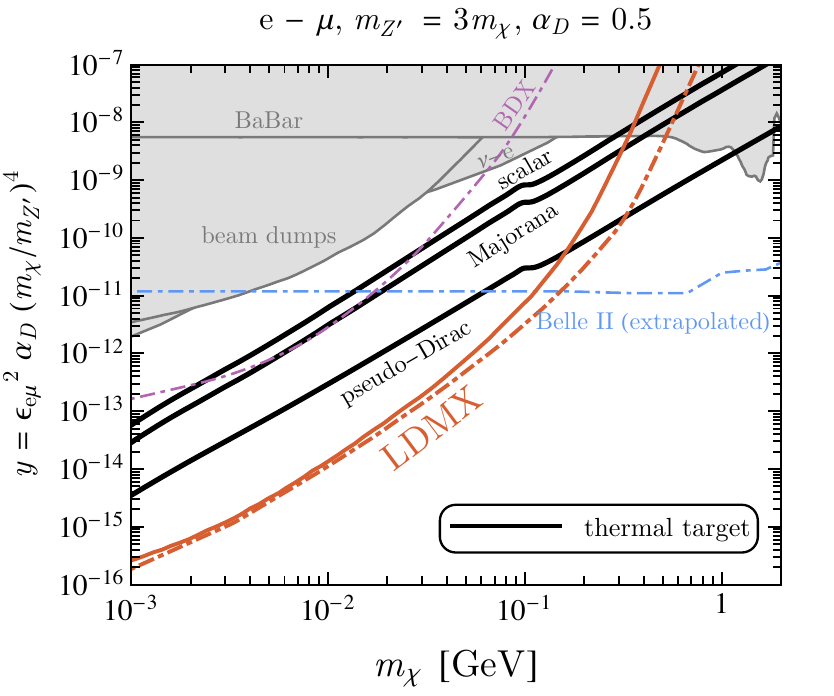}
\includegraphics[width=7.8cm]{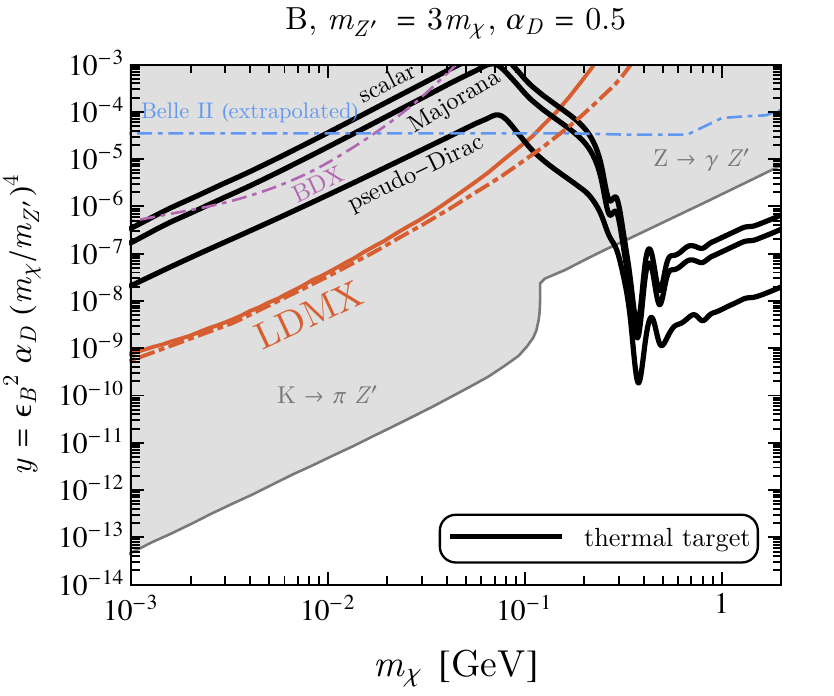}
\caption{As in Fig.~\ref{fig:thermal-targets}, thermal targets for the representative dark matter candidates of Sec.~\ref{sec:dark-photon-benchmark-models} but instead coupled 
to $U(1)_{B-L}$ (top-left), $U(1)_{B-3e}$ (top-right), $U(1)_{e - \mu}$ (bottom-left), and $U(1)_B$ (bottom-right) $\zp$ gauge bosons, fixing $m_{\zp} = 3 m_\chi$ and $\alpha_D = 0.5$. The black line corresponds to parameter space where the relic abundance of $\chi$ agrees with the observed dark matter energy density. The shaded gray regions are excluded from previous experiments, such as a BaBar monophoton analysis~\cite{Lees:2017lec}, and beam dump searches at LSND~\cite{deNiverville:2011it},  E137~\cite{Bjorken:1988as,Batell:2014mga}, and MiniBooNE~\cite{Aguilar-Arevalo:2017mqx}. Also shown in dot-dashed blue is the projected sensitivity of a monophoton search at Belle II presented in 
Ref.~\cite{Battaglieri:2017aum} and computed by rescaling the 20 fb$^{-1}$ background study up to 50 ab$^{-1}$~\cite{HeartyCV}. Future direct detection experiments will have sensitivity to the cosmologically motivated regions of parameter space shown for scalar DM (see Fig.~\ref{fig:thermal-targets}). We also show constraints derived from the observed $\bar{\nu} - e$ scattering spectrum at TEXONO~\cite{Deniz:2009mu, Bilmis:2015lja}, and for the baryonic current, $U(1)_B$, bounds from considerations of enhanced anomalous decays into $\zp$ final states~\cite{Dror:2017ehi,Dror:2017nsg}. 
 The projected sensitivity of LDMX is shown in solid (dot-dashed) red, assuming $10^{16}$ EOT from a 8 (16) GeV electron beam and a $10\%$ radiation length tungsten (aluminum) target. }
 \label{fig:vectortargets}
\end{figure}

We calculate the DM relic abundance by numerically solving the relevant Boltzmann equation, which incorporates the cross section for DM annihilations to SM fermions through an intermediate $\zp$, i.e., $\chi \chi \to Z^{\prime *} \to f \bar{f}$ . The calculation of this cross section for leptonic final states is nearly identical to that of the previously considered dark photon models. Slight variations arise, for instance, from the fact that new $U(1)$'s involving lepton number couple directly to SM neutrinos. Incorporating annihilations to hadronic final states is less straightforward. In doing so, we adopt the publicly available results from the data-driven approach of Ref.~\cite{Ilten:2018crw}.

Our results are presented in Fig.~\ref{fig:vectortargets}, which shows constraints in the $y - m_\chi$ plane on the invisibly decaying $\zp$ gauge bosons of Eq.~(\ref{eq:U(1)list}) for various DM models. The freeze-out parameter, $y$, is defined similarly to Eq.~(\ref{eq:define-y}), after making the replacement $\epsilon \to \epsilon_{BL}$ for $U(1)_{B-L}$ and likewise for the other models. As in Fig.~\ref{fig:thermal-targets}, we fix $m_{\zp} / m_\chi = 3$ and $\alpha_D = 0.5$ and highlight existing constraints from BaBar and the beam dumps LSND, E137, and MiniBooNE, which are derived from DM production and scattering in a secondary detector. Also shown are projections from Belle II and LDMX, where the latter assumes a 8 or 16  GeV electron beam, $10^{16}$ EOT of luminosity, and a 10\% radiation length tungsten (solid red) or aluminum (dot-dashed red) target. Unlike the dark photon portal of Sec.~\ref{sec:dark-photon-benchmark-models}, $U(1)_{B-L}$, $U(1)_{B-3e}$, and $U(1)_{e - \mu}$ forces mediate interactions between neutrinos and electrons, which are constrained from observations of $\bar{\nu}-e$ scattering at TEXONO and other neutrino experiments~\cite{Deniz:2009mu, Bilmis:2015lja}.

The constraints, projections, and cosmologically motivated parameter space for these three leptonically-coupled models are qualitatively very similar to those for the dark photon scenario of Sec.~\ref{sec:dark-photon-benchmark-models}. Minor differences arise, for instance, when the $\zp$ does not directly couple to SM hadrons, which strongly suppresses constraints from the proton beam dump experiments LSND and MiniBooNE so that the strongest constraint at low masses come from the electron beam-dump E137 and $\bar\nu-e$ scattering. Models incorporating $U(1)_B$ forces are the most strikingly different for two reasons. First, the $\zp$ of $U(1)_B$ only radiatively couples to electrons, as discussed above. As a result, thermal freeze-out requires much larger couplings for $m_\chi \lesssim \mathcal{O} (100) \text{ MeV}$ since annihilations in the early universe are strongly suppressed for DM masses below the pion threshold. Second, the presence of non-vanishing anomalies in gauged $U(1)_B$ theories leads to the enhanced growth of the exotic decays, $K \to \pi \zp$ and $Z \to \gamma \zp$, at high-energies, which are strongly constrained from current measurements~\cite{Dror:2017ehi,Dror:2017nsg}. It is important to note that these limits are model-dependent, but regardless, strongly disfavor $U(1)_B$-coupled thermal DM lighter than $\sim \text{few} \times 100 \text{ MeV}$.

 \subsubsection{Predictive Dark Matter with Spin-0 Mediators}
 \label{sec:scalarmeddm}
 \label{ssec:spin0}
 
  In this section, we focus on another variation of the models previously considered in Sec.~\ref{sec:dark-photon-benchmark-models}. In particular, we will investigate the cosmologically motivated parameter space for DM that annihilates to SM leptons through the exchange of a spin-0 mediator, which we denote as $\varphi$. Compared to the canonical dark photon, the most analogous version of a spin-0 mediator that is on similar theoretical footing is a new SM neutral scalar that directly couples to the SM Higgs through the trilinear or quartic interactions $\varphi |H|^2$ and $\varphi^2 |H|^2$. Below the scale of electroweak symmetry breaking, $\varphi$ mass-mixes with $H$, inheriting couplings analogous to the SM Higgs-fermion couplings, i.e., $\sim \sin{\theta} \, (m_f /v)$, where $v \simeq 246 \text{ GeV}$ is the SM Higgs vev and $\sin{\theta}$ describes the strength of $\varphi - H$ mixing. Hence, for a given mixing angle, $\varphi$ couples to SM fermions proportional to mass. In these models, most regions of cosmologically motivated parameter space are ruled out by measurements of the invisible width of the SM Higgs and invisible exotic decays of heavy-flavor mesons, such as $B \to K \varphi$ (see Ref.~\cite{Krnjaic:2015mbs} and references therein). 
 
These constraints are alleviated if $\varphi$ does not couple to hadrons or the SM Higgs. For these reasons, we will focus on light spin-0 mediators that couple dominantly to leptons, proportional to mass with either parity-even or parity-odd interactions, i.e.,
\begin{align}
\label{eq:simplescalarmed}
&\mathcal{L} \supset g_\varphi ~ \varphi ~ \sum\limits_{\ell} (m_\ell / m_e) ~ \bar{\ell} \ell ~~~~ \text{(parity-even)}
\nonumber \\
&\mathcal{L} \supset g_\varphi ~ \varphi ~ \sum\limits_{\ell}  (m_\ell / m_e) ~ \bar{\ell} i \gamma^5 \ell ~~~~ \text{(parity-odd)}
~,
\end{align}
where the sum is over all or some subset of the SM leptons. In Eq.~(\ref{eq:simplescalarmed}), we have normalized the interaction terms by the $\varphi-e$ coupling, $g_\varphi$.
This low-energy Lagrangian can arise in a gauge invariant manner through the dimension-five operator
\be
\frac{\varphi}{\Lambda} ~ \bar{E}_L  e_R \, H 
~,
\ee
where $\Lambda$ is a scale associated with new physics. Various UV-completions of this scenario have been considered, e.g., in two-Higgs-doublet models involving additional singlet scalars or vector-like quarks~\cite{Chen:2015vqy,Batell:2016ove}. In these scenarios, strong model-dependent constraints often arise from irreducible scalar couplings to SM hadrons. Throughout this section, we adopt the simplified model of Eq.~(\ref{eq:simplescalarmed}) in order to describe the relevant phenomenology with a particular focus on electron-couplings. As in Eqs.~(\ref{eq:epsdef}), we define the $\varphi - e$ coupling strength as
\be
\epsilon_\varphi \equiv g_\varphi / e
~.
\ee
%

\begin{figure}[t!]
\includegraphics[width=8cm]{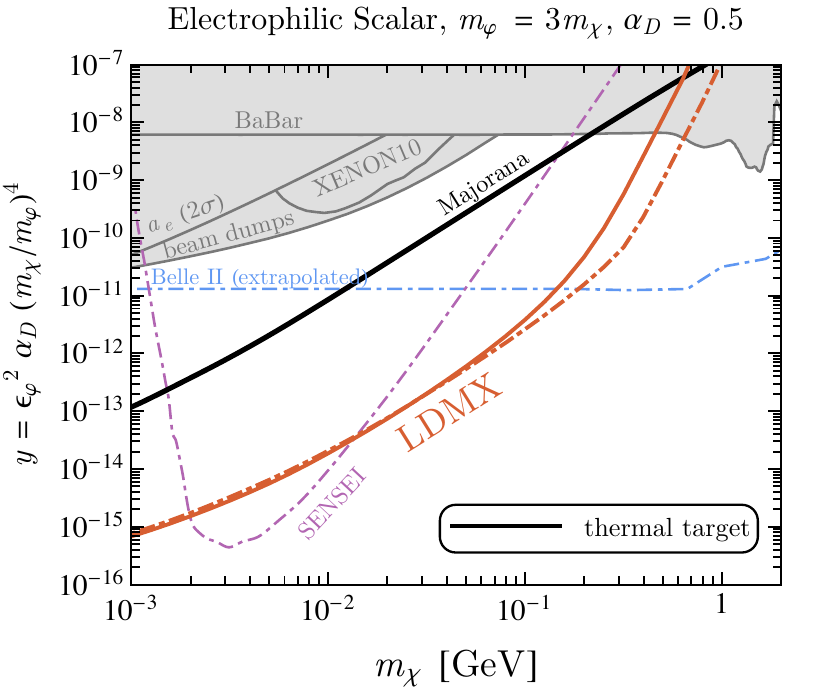}
\includegraphics[width=8cm]{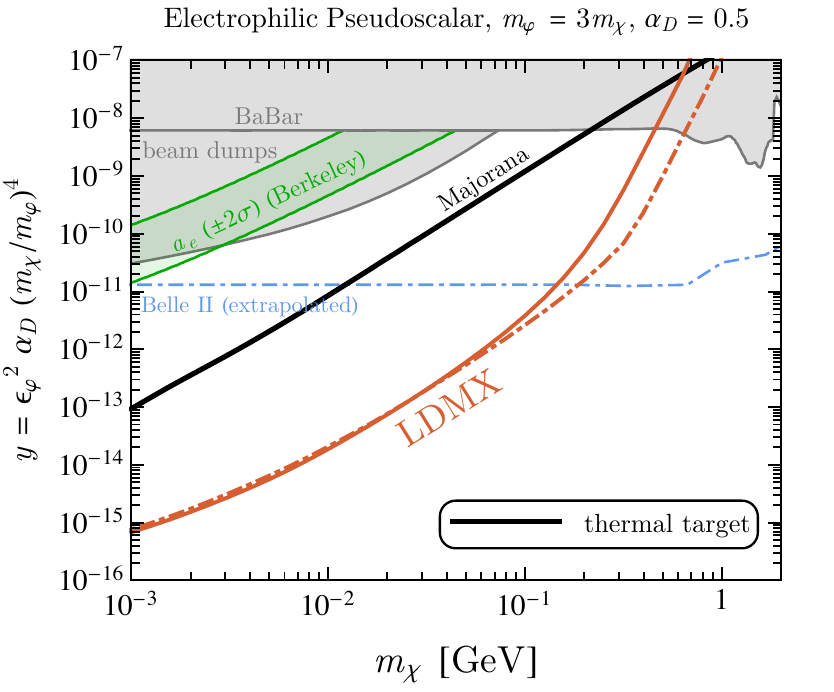}
\includegraphics[width=8cm]{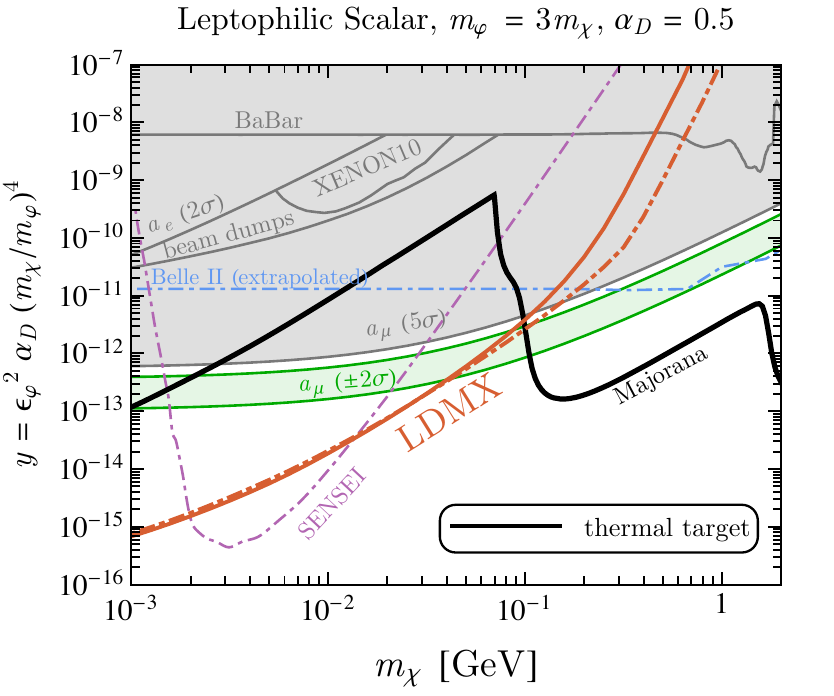}
\includegraphics[width=8cm]{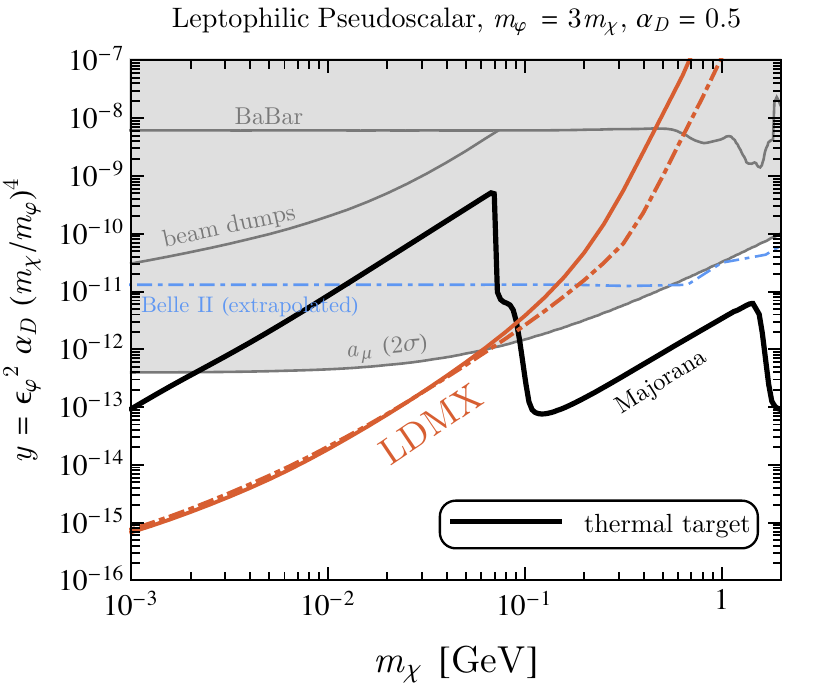}
\caption{Thermal targets for Majorana dark matter that couples to an electrophilic (top row) or leptophilic (bottom row) spin-0 mediator, $\varphi$. In each model, we assume that the $\chi-\varphi$ interaction is parity-even, fixing $m_{\varphi} = 3 m_\chi$ and $\alpha_D = 0.5$. In the left (right) column, $\varphi$ couples to SM leptons through parity-even (parity-odd) interactions. The black line corresponds to parameter space where the relic abundance of $\chi$ agrees with the observed dark matter energy density. The shaded gray regions are excluded from previous experiments, such as the BaBar monophoton analysis~\cite{Lees:2017lec}, beam dump search at E137~\cite{Bjorken:1988as,Batell:2014mga}, and the XENON10 direct detection experiment~\cite{Angle:2011th,Essig:2012yx,Angloher:2015ewa,Essig:2017kqs}. Also shown in dot-dashed blue is the projected sensitivity of a monophoton search at Belle II presented in 
Ref.~\cite{Battaglieri:2017aum} and computed by rescaling the 20 fb$^{-1}$ background study up to 50 ab$^{-1}$~\cite{HeartyCV}. Future direct detection experiments, such as SENSEI, will have sensitivity to the cosmologically motivated regions of parameter space shown for parity-even $\varphi-e$ couplings~\cite{Battaglieri:2017aum}. We also show constraints derived from the observed magnetic moment of the electron and muon as well as regions favored to explain recently reported anomalies~\cite{Bennett:2006fi,Hagiwara:2011af,Davier:2010nc,Jegerlehner:2009ry,Parker191}. The projected sensitivity of LDMX is shown in solid (dot-dashed) red, assuming $10^{16}$ EOT from a 8 (16) GeV electron beam and a $10\%$ radiation length tungsten (aluminum) target. 
 }
 \label{fig:scalarmedtarget}
\end{figure}

We take the DM to be comprised of a Majorana fermion, $\chi$, that couples to $\varphi$ with the parity-even interaction
\be
\mathcal{L} \supset \frac{1}{2} ~ g_\chi ~ \varphi ~ \bar{\chi} \chi
~,
\ee
and define $\alpha_D \equiv g_\chi^2 / 4 \pi$ and the freeze-out parameter, $y$, analogous to Eq.~(\ref{eq:define-y}). DM freeze-out proceeds through annihilations to SM leptons via $\chi \chi \to \varphi^* \to \ell^+ \ell^-$. The non-relativistic annihilation cross section for either the parity-even or parity-odd interactions of Eq.~(\ref{eq:simplescalarmed}) is given by
\be
\sigma v (\chi \chi \to \varphi^* \to \ell^+ \ell^-) \simeq  \frac{g_\varphi^2 \, \alpha_D \, (m_\ell / m_e)^2 \, m_\chi^2 \, v^2}{2 \, (4 m_\chi^2 - m_\varphi^2)^2} \simeq 2 \pi \, \alpha_\text{em} \, (m_\ell / m_e)^2 \, v^2  ~ \frac{y}{m_\chi^2}
~,
\ee
where in the first and second equality we have taken $m_\chi \gg m_\ell$ and $m_\varphi \gg m_\chi$, respectively. As in many of the models previously discussed, this process is suppressed by the relative DM velocity, $v$, which reduces the annihilation rate in the non-relativistic limit below constraints derived from the CMB. For parity-even couplings, the DM-electron elastic scattering cross section, relevant for direct detection experiments such as XENON10 and SENSEI, is approximately
\be
\sigma (\chi e \to \chi e) \simeq \frac{4 g_\varphi^2 \, \alpha_D \, \mu_{\chi e}^2}{m_\varphi^4}
~,
\ee
whereas the corresponding rate for parity-odd couplings is velocity-suppressed in the non-relativistic limit. 

In Fig.~\ref{fig:scalarmedtarget}, we explore the parameter space for DM that couples to a spin-0 mediator with parity-even (scalar) or parity-odd (pseudoscalar) interactions. We also illustrate scenarios in which $\varphi$ couples to all SM leptons proportional to mass (leptophilic) or exclusively to electrons (electrophilic). As in the previous sections, we highlight existing constraints from the electron beam dump E137, and projections from Belle II and LDMX. In calculating the production of $\varphi$ at electron fixed-target experiments we have used the approximate analytic expressions of Ref.~\cite{Liu:2016mqv}.

In the top row of Fig.~\ref{fig:scalarmedtarget}, we present the DM models that couple to electrophilic scalar and pseudoscalar mediators. These models are qualitatively similar to those involving dark photon mediators in Sec.~\ref{sec:dark-photon-benchmark-models}, aside from a significant suppression in the constraining power of the electron beam dump E137. We have calculated the rate of DM production and scattering at E137 analytically, following the analysis in Ref.~\cite{Essig:2017kqs}. The reduced sensitivity of E137 stems from a suppression in the DM-electron elastic scattering rate in the downstream detector, which can be understood as the decoupling of a scalar potential in the ultra-relativistic limit. To see this more explicitly, one can compare the spin-averaged amplitude squared for $\chi e \to \chi e$ via scalar or vector mediators. In the relativistic limit such that the energy of the incoming DM particle ($E_\chi$) is much greater than the electron recoil energy ($E_e$), we find
\be
\label{eq:E137suppression}
\frac{|\mathcal{M}|^2 (\text{scalar exchange})}{|\mathcal{M}|^2 (\text{vector exchange})} ~ (\chi e \to \chi e) \sim \frac{E_e}{m_e} ~ \frac{m_\chi^2}{E_\chi^2} ~~ \text{max}(1 , \frac{E_e \, m_e}{m_\chi^2})
~.
\ee
For E137, $E_\chi$ and $E_e$ are typically comparable to the beam energy, $E_\text{beam} = 20 \text{ GeV}$, and threshold recoil energy, $E_\text{th} \sim \mathcal{O}(1) \text{ GeV}$, respectively. Substituting these typical kinematic scales into Eq.~(\ref{eq:E137suppression}) implies that compared to vector-mediated scattering, scalar-mediated scattering at E137 is suppressed by a few orders of magnitude for $m_\chi \lesssim 100 \text{ MeV}$.

The bottom row of Fig.~\ref{fig:scalarmedtarget} corresponds to DM that is coupled to a leptophilic scalar or pseudoscalar. In this case, strong bounds from measurements of the magnetic moment of the muon rule out most of the cosmologically viable parameter space for DM masses below the muon threshold. For electrophilic pseudoscalars and leptophilic scalars, we have also highlighted relevant parameter space that is favored by the recent anomalies in the measurements of the electron~\cite{Parker191} and muon~\cite{Bennett:2006fi,Hagiwara:2011af,Davier:2010nc,Jegerlehner:2009ry} magnetic moments.

\subsubsection{$(g-2)_{\mu}$ Motivated Muon-Philic Dark Matter}\label{sec:muon-thermal-dm}
\label{ssec:muphilic}

In this section, we apply the thermal DM analysis from Sec.~\ref{sec:thermal-dm-section} to a $U(1)_{L_\mu - L_\tau}$ gauge boson mediator ($Z^\prime$) motivated by the $(g-2)_{\mu}$ anomaly, and consider scenarios where DM freeze-out is controlled by such interactions. 
These models motivate muon-beam versions of LDMX~\cite{Kahn:2018cqs} and NA64~\cite{Gninenko:2014pea,Chen:2018vkr}.
A representative Lagrangian for this scenario is
\be \label{eq:lag1}
{\cal L} \supset
  - \frac{1}{4} { F^{\prime} }^{\alpha \beta}  F^{\prime}_{\alpha \beta}    + \frac{m_{{\zp}}^2}{2} {Z^{\prime} }^\beta  Z^{\prime}_\beta  - Z^\prime_\beta   \left(     J^{\beta}_{\mu-\tau} 
 + J^{\beta}_{\chi} \right)  , ~~
\ee
where $  F^{\prime}_{\alpha \beta} $ is the $\zp$ field strength,  
$m_{{\zp}}$ is its mass, 
and the SM current is 
\be
\label{eq:Jmuon}
J^{\beta}_{\mu -\tau} = g_{\mu - \tau} \left(    \bar \mu\gamma^\beta \mu +  \bar \nu_\mu \gamma^\beta P_L \nu_{\mu}  -
							  \bar \tau\gamma^\beta \tau -  \bar \nu_\tau \gamma^\beta P_L \nu_{\tau}    \right). 
\ee
As in Sec.~\ref{sec:thermal-dm-section}, we consider
several possible DM currents
\be
\label{eq:dm-currents-mutau}
J_\chi^\mu= g_\chi \times  
\begin{cases}
                        i \chi^* \partial_\mu  \chi + \text{h.c.} ~ & {\rm Complex ~ Scalar} \\
              \frac{1}{2}  \overline \chi \gamma^\mu \gamma^5 \chi ~  &{\rm Majorana~ Fermion} \\
                            i \, \overline \chi_1 \gamma^\mu \chi_2  ~  & {\rm Pseudo\!\!-\!\!Dirac~ Fermion} \\
                     \overline \chi \gamma^\mu \chi ~ & {\rm Dirac~ Fermion} \\
                    \end{cases}
\ee
where, in addition to the complex scalar, Majorana fermion, and pseudo-Dirac considered above in Sec.~\ref{sec:dark-photon-benchmark-models}, we have also included a pure Dirac fermion because, for an $L_\mu-L_\tau$ mediator, it is possible 
for the $s$-wave $\chi \bar \chi \to \nu \nu$ annihilation process to be safe from CMB limits on DM annihilations
during recombination; these bounds assume that the annihilation yields {\it visible} final states (see the discussion in Sec.~\ref{sec:thermal-dm-section}). 

\begin{figure}[t]
\includegraphics[width=3.2in,angle=0]{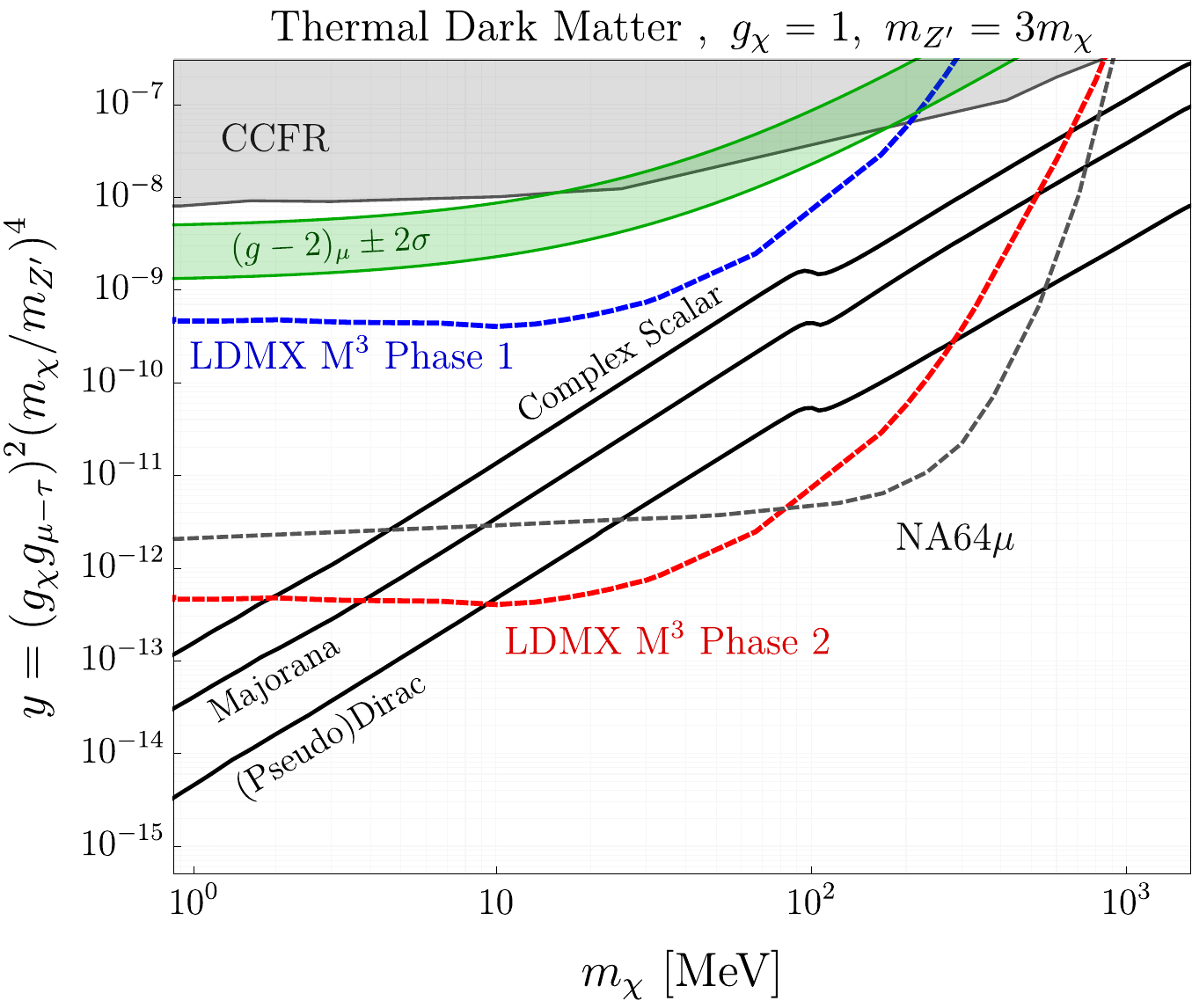}
\includegraphics[width=3.2in,angle=0]{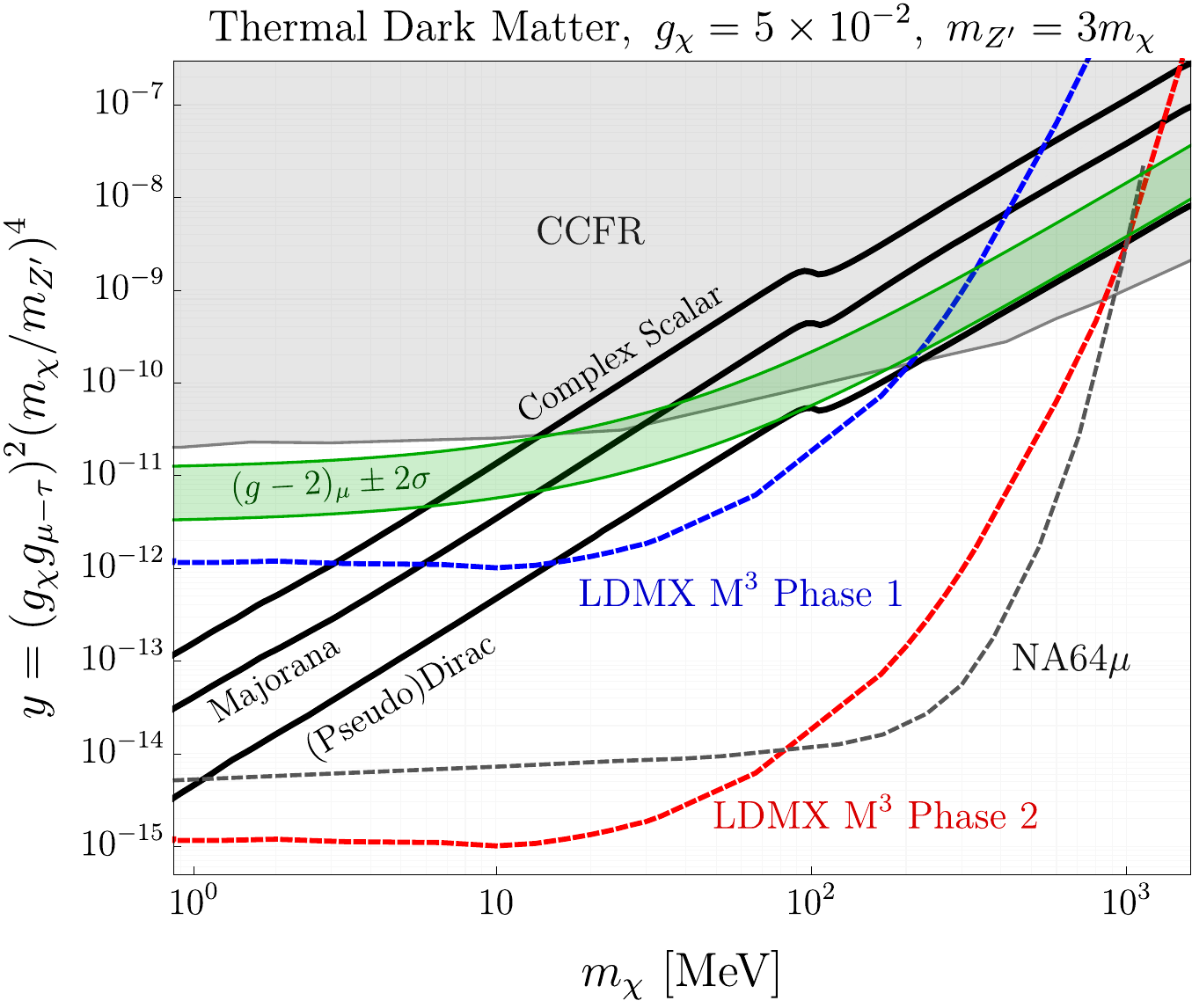}
\caption{
Predictive thermal dark matter charged under $U(1)_{L_\mu - L_\tau}$ that undergoes
direct annihilation via $\chi \chi \to Z^{\prime *} \to f \bar{f}$, where $f = \mu, \nu_\mu, \tau, \nu_\tau$.
The vertical axis is the product of couplings that governs relic abundance for a given choice of DM mass and spin (see text). Also plotted are CCFR constraints from measurements of neutrino trident production~\cite{Mishra:1991bv,Altmannshofer:2014pba}, as well as the projected sensitivity at 
NA64~\cite{Gninenko:2014pea}. There are also bounds from $\Delta N_{\rm eff}$ (not shown) that arise from $\chi\bar\chi \to \nu \nu$ annihilation during BBN -- these bounds vary from 1-10 MeV depending on the choice of dark matter candidate spin~\cite{Boehm:2013jpa,Nollett:2014lwa}. 
For the pure Dirac model, the annihilation
cross section $\chi\bar\chi \to \mu^+\mu^-$ or $\tau^+\tau^-$ is $s$-wave, so this process is ruled out by CMB energy injection bounds for $m_\chi > m_\mu$~\cite{ade:2015xua}. Both plots presented here are taken from Ref.~\cite{Kahn:2018cqs}.
}

\label{fig:muon-thermal-targets}
\end{figure}
\begin{figure}[t]
\includegraphics[width=3.2in,angle=0]{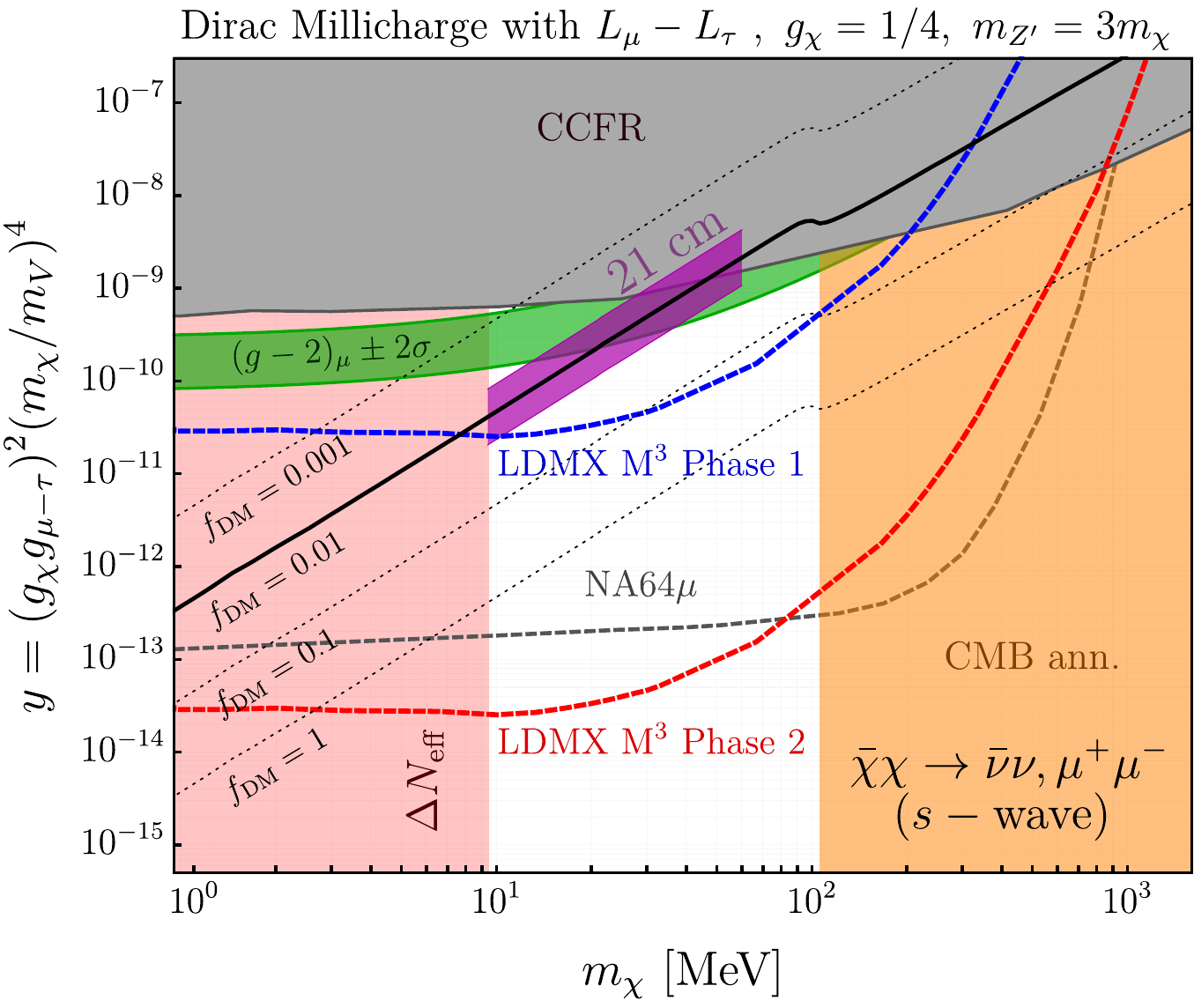}
\includegraphics[width=3.2in,angle=0]{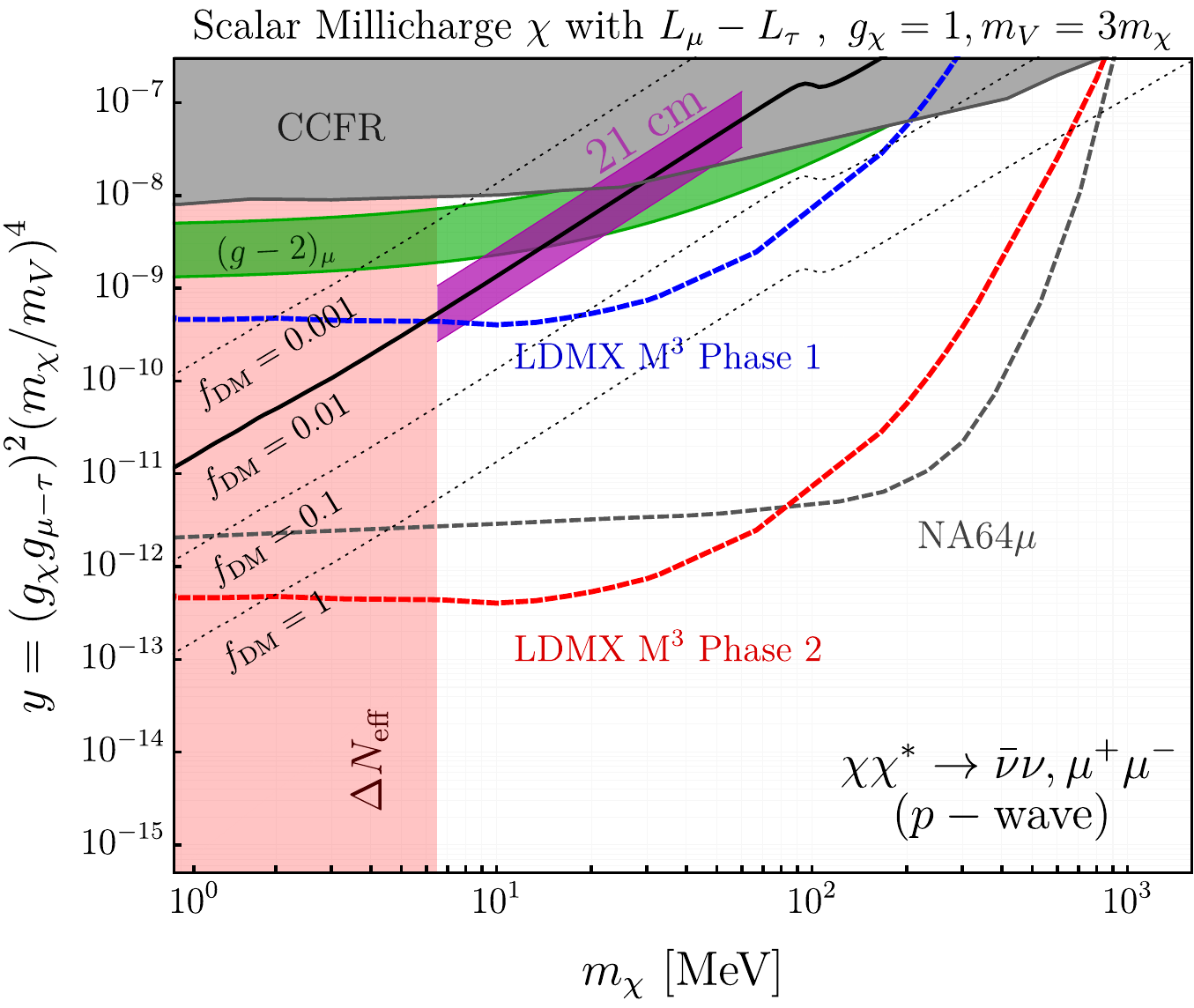}
\caption{
Constraints on fermion (left) and scalar (right) dark matter which carries {\it both } a QED millicharge and additional charge under 
a gauged $L_\mu - L_\tau$ force. This combination of interactions is motivated by the dark matter interpretation
of the EDGES 21-cm excess, which requires the millicharge to explain the anomaly and the additional 
force in order to generate the requisite DM fraction $f_{\rm DM} \simeq 10^{-2}$~\cite{Berlin:2018sjs}.
The purple band represents parameter space which can explain the amplitude of the observed 21-cm absorption feature. 
 Future measurements at NA64~\cite{Gninenko:2014pea} and LDMX-M$^3$~\cite{Kahn:2018cqs} are expected to be sensitive to this scenario. In the green band, this model also resolves the  $(g-2)_{\mu}$ anomaly ~\cite{Bennett:2006fi,Hagiwara:2011af,Davier:2010nc,Jegerlehner:2009ry}. The 
 shaded gray region is constrained by the CCFR experiment~\cite{Altmannshofer:2014pba,Mishra:1991bv}.
Both plots presented here are taken from~\cite{Berlin:2018sjs}.
 }
 \label{fig:edges}
\end{figure}

For models we consider here, the relic density is set via $\chi \chi \to f \bar{f}$ annihilation where $f = \mu, \nu_\mu, \tau, \nu_\tau$ and the thermally averaged cross section can be written in the velocity expansion limit $\langle \sigma v\rangle = \sigma_0 x^n$ 
\be \label{eq:sigma0}
~~~~~~\sigma_0 =   \frac{g_\chi^2 g_f^2    (m_f^2 + 2m_\chi^2)}{k \pi \left[  (m_{{\zp}}^2 -   4 m_\chi^2 )^2    + m_{{\zp}}^2 \Gamma_{\zp}^2    \right] } ,
\ee 
where $g_{\chi, f}$ is the $\zp$ coupling to $\chi$ or $f$,  and  $k = 2,\, 12$, and $6$ for  
a (pseudo-)Dirac particle, a complex scalar, or a Majorana fermion, respectively.  
In the $m_{{\zp}} \gg m_\chi \gg m_f$ limit (away from the $m_{\zp} = 2m_\chi$ resonance), 
for each of the models in Eq.~(\ref{eq:dm-currents-mutau}) we have
\be
\sigma_0 \propto  \frac{g_\chi^2 g_f^2 m_\chi^2}{m_{{\zp}}^4} = \frac{y}{m_\chi^2}~, ~ y \equiv (g_\chi g_f)^2 \left( \frac{ m_\chi }{ m_{{\zp}} } \right)^4.
\ee
Demanding $\Omega_\chi = \Omega_{\rm DM}$ defines thermal relic density targets for each model in Eq.~(\ref{eq:dm-currents-mutau}) (see Ref.~\cite{Kahn:2018cqs} for details). In Fig.~\ref{fig:muon-thermal-targets}, we present these thermal targets in the $y-m_\chi$ parameter
space plotted alongside constraints from the CCFR experiment~\cite{Altmannshofer:2014pba,Mishra:1991bv} and a
green band within which $Z^\prime$ can resolve the $(g-2)_\mu$ anomaly~\cite{Bennett:2006fi,Hagiwara:2011af,Davier:2010nc,Jegerlehner:2009ry}. Also shown are projections for an LDMX-style {\it muon-beam} missing-momentum experiment (labeled LDMX M$^3$) from Ref.~\cite{Kahn:2018cqs}  and projections from NA64 in a muon beam~\cite{Gninenko:2014pea}. 

In Fig.~\ref{fig:edges}, we also show the LDMX M$^3$ projections for a variation of this scenario in which a $\sim 10^{-2}$ fraction of the DM carries both $L_\mu - L_\tau$ charge {\it and} a QED millicharge. This scenario is motivated by the 3.8$\sigma$ anomaly in 
the observed 21 cm absorption feature reported by the EDGES collaboration~\cite{Bowman:2018yin}. As reported in 
\cite{Barkana:2018lgd,Munoz:2018pzp}, if MeV-scale DM with a millicharge scatters off baryons at redshift $z \sim 20$, it can efficiently cool the hydrogen
population and thereby enhance the 21 cm absorption line. However, achieving the required $\sim 10^{-2} \, \Omega_{\rm DM}$ abundance fraction of 
these particles requires forces beyond the minimal millicharge interactions that cool the hydrogen. In Ref.~\cite{Berlin:2018sjs} it was found 
that the only viable scenarios in which this additional force yields a predictive thermal target involve preferential couplings to the
 second and third SM fermion generations. In Fig.~\ref{fig:edges}, the black contours represent the $f_{\rm DM} =$ const. parameter space
 for which the $Z^\prime$-mediated interactions set a fixed DM fraction. 
 Intriguingly, there is also parameter space for which the EDGES favored region overlaps with the green band where 
 the $Z^\prime$  also resolves the persistent $(g-2)_\mu$ anomaly.

\subsection{Secluded Dark Matter}   
\label{ssec:secluded}

In models of DM with ``secluded annihilation" ($m_{\rm DM} > m_{\rm MED}$), the DM  transfers its entropy via DM DM $\to$ MED MED annihilation as in Eq.~(\ref{eq:secluded-schematic}).
As long as the mediator-SM coupling is sufficient to thermalize the dark and visible sectors, this annihilation rate depends only on
the DM-mediator coupling and is independent of its SM coupling, which governs laboratory observables.\footnote{If the SM-mediator 
coupling is {\it insufficient} to thermalize these sectors, the mediator is generically long-lived and can 
come to dominate the energy density of the universe, in which case the DM abundance does depend on the mediator-SM coupling 
which governs the mediator's entropy dump into the SM -- see Refs.~\cite{Berlin:2016vnh,Berlin:2016gtr}. } Thus, unlike the models presented in preceding sections,  this scenario offers no predictive thermal targets, but the mediator's SM decays motivate a robust program of dark-force searches; the  LDMX projections for {\it visible} searches is presented 
in Sec.~\ref{sec:visible_signals}.

Like all other sub-GeV  thermal DM scenarios, the secluded regime is subject to the CMB-safety requirement, which  dictates the 
viable mediator options for achieving the DM-SM entropy transfer. 
If the mediator is a canonical dark photon (like the models in Sec. \ref{sec:dark-photon-benchmark-models} but with the opposite $\chi/A^\prime$ mass hierarchy), then for all
choices of DM spin, the $\chi \chi \to A^\prime A^\prime $ annihilation rate is $s$-wave, which is ruled out by CMB constraints as discussed
in the preamble of this section (a dark photon can still mediate secluded annihilation in asymmetric DM models).  
Secluded annihilation of Dirac fermion DM into Higgs-mixed scalar mediators is allowed by CMB constraints, but the constraints on these models are so severe that the prospects for tests at either accelerator or direct detection experiments are poor.

There are better detection prospects for the cases where the mediator couples non-minimally to SM matter.  While spin-1 mediators that can decay to electrons (as in Sec.~\ref{ssec:spin1}) are still at odds with CMB bounds, secluded annihilation into vector mediators that decay (almost) exclusively into neutrinos are allowed. 
These include sub-muon-mass mediators coupled to, for example, $L_\mu - L_\tau$ (see Sec.~\ref{ssec:muphilic}).  (Secluded annihilation into sub-electron-mass mediators coupled to lepton number is safe with respect to CMB energy-injection constraints, but these are ruled out by their large effects on the effective number of neutrino species $N_{\rm eff}$).
In addition, Dirac fermion DM coupled to a leptophilic spin-0 mediator is compatible with both direct searches and CMB bounds in an accessible region.

While the secluded scenario is not predictive, it does offer several potential signatures that \emph{may} be observable, if we are fortunate to live in the right parameter regions.  These include: 
\begin{itemize}
\item Mediator decays into SM final states (which, as noted above, must be invisible decays into neutrino final states if the mediator is a vector, but could also be into visible final states for scalar mediators).
\item DM production via an off-shell mediator (leading to an invisible final state in LDMX). However, the small DM-mediator coupling required for this mechanism to produce the correct relic abundance ($\alpha_D \lesssim 10^{-6}$  for DM lighter than 100 MeV) and modest couplings of most viable mediators to electrons and protons make this channel relatively challenging to detect compared to on-shell mediator production.
\item Direct detection of DM-electron or DM-nucleon scattering, especially in the case of an extremely light scalar mediator where kinematic enhancement of non-relativistic scattering can partially compensate for the small couplings noted above. 
\end{itemize}

\subsection{Asymmetric Dark Matter}
\label{sec:asymmetric}

\begin{figure}[t]
 \includegraphics[width=8cm]{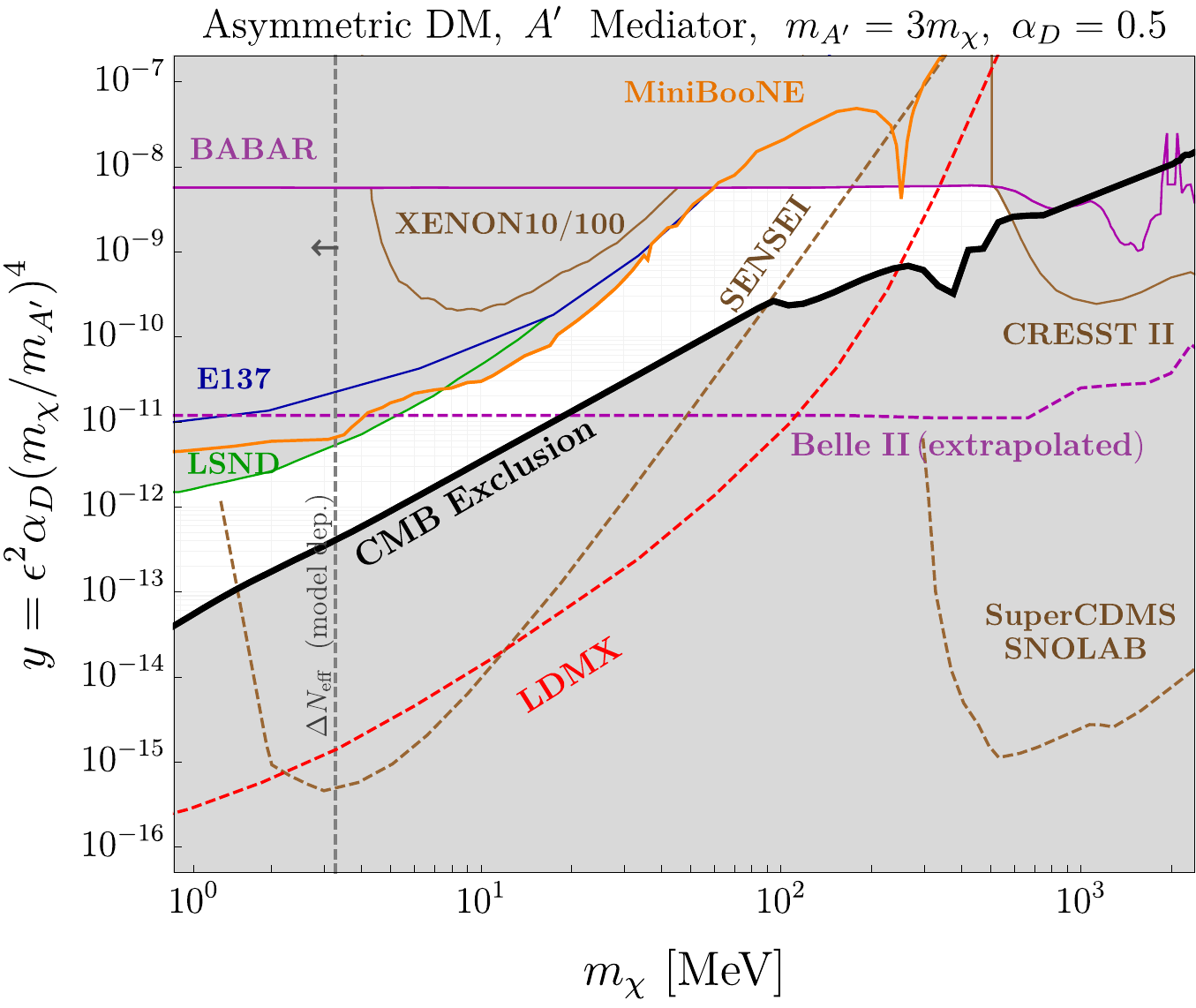}
\caption{ Viable parameter space for a particle-antiparticle asymmetric population
of Dirac fermion DM particles with a kinetically mixed dark photon mediator $A^\prime$. Although this figure is similar to the panel of plots shown in Fig.~\ref{fig:thermal-targets}, the key difference is that the black curve here represents a {\it lower} bound 
on the annihilation rate; points below this curve are excluded by CMB constraints from DM annihilation during recombination~\cite{ade:2015xua}. See also~\cite{Graesser:2011wi,Lin:2011gj} for a discussion.}
\label{fig:asymm-dm}
\end{figure}

If the DM carries an approximately or exactly
 conserved global quantum number, analogous to baryon or lepton number in the SM,
its population can acquire a particle-antiparticle asymmetry at late times. Unlike in symmetric DM models,  
the cosmological abundance is not governed by particle-antiparticle annihilation during freeze-out, but 
by other particles and interactions that satisfy the Sakharov conditions in the early universe~\cite{Sakharov:1967dj}. 
A representative model consists of a Dirac fermion $\chi$ coupled to a dark photon $A^\prime$ of mass
$m_{A^\prime}$ from a hidden $U(1)_D$ gauge
group as described in 
Sec.~\ref{sec:thermal-dm-section}
\be
\label{eq:asymm-lag}
{\cal L }  =  i \overline \chi D_\mu  \gamma^\mu \chi + m_\chi \overline \chi \chi~~,~~ D_\mu = \partial_\mu + i g_D A^\prime_\mu
\ee
where $g_D \equiv \sqrt{4 \pi \alpha_D}$ is the gauge coupling. 

The key difference with respect to the models in Sec.~\ref{sec:thermal-dm-section} is that the $\chi$ population is asymmetric $\Omega_{\chi} \ne \Omega_{\bar \chi}$ in the present day halo. 
Although this asymmetry is not set by the interactions in Eq.~(\ref{eq:asymm-lag}), this scenario still 
lends itself to a predictive target because the DM has a thermal abundance which would overclose the universe in 
the absence of an annihilation interaction. Indeed, since the asymmetric component is fixed by a conserved
quantum number, the $\chi \bar \chi$ annihilation rate must be {\it larger} than in the symmetric scenario to annihilate away even more
of the antiparticles.

The $\bar \chi \chi \to  f\bar{f}$  annihilation cross section is $s$-wave, 
\be
\sigma v(\chi_1\chi_2 \to f \bar{f}) = \frac{16 \pi \epsilon^2 \alpha \alpha_D m_\chi^2}{(4 m_\chi^2- m^2_{A^\prime})^2 + m^2_{A^\prime} \Gamma^2_{A^\prime}} \simeq \frac{16 \pi \alpha y}{m_\chi^2},
\ee
where the $y$ variable here follows the convention in Sec.~\ref{sec:thermal-dm-section}.
Thus, if there is a non-negligible $\bar \chi$ antiparticle density present during recombination, this 
scenario is constrained by Planck measurements of CMB temperature anisotropies~\cite{ade:2015xua}.
This bound is presented as the black curve labeled ``CMB exclusion" in Fig.~\ref{fig:asymm-dm}. Note that, because
the antiparticle density at late times is exponentially suppressed, $n_{\bar \chi} \propto e^{-\langle \sigma v \rangle}$, CMB limits are weaker for larger cross sections, so the CMB imposes a {\it lower} bound on $y$~\cite{Graesser:2011wi,Lin:2011gj}.  Note also that, unlike in symmetric models with thermal targets, every point on this plane can 
accommodate the total DM abundance (albeit with a different particle asymmetry), whereas for the symmetric parameter 
space, only points on the thermal targets in Sec.~\ref{sec:thermal-dm-section} correspond to this possibility.

\begin{figure}[htbp]
 \includegraphics[width=10cm]{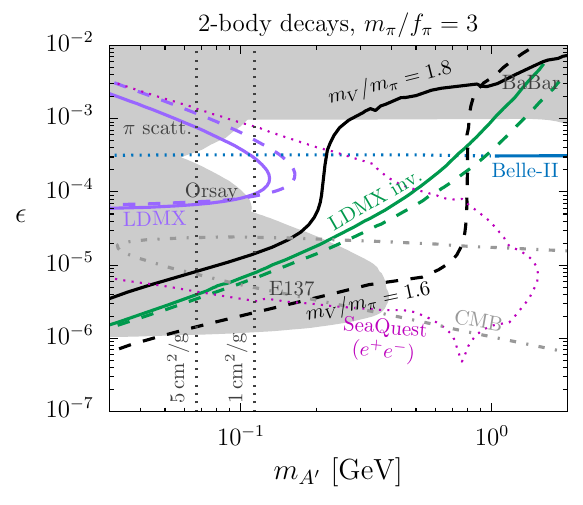}
\caption{
  Projected reach of an LDMX-style experiment to missing momentum (green solid and dashed lines) and 
  visible late decay (purple solid and dashed lines) in a model with a strongly interacting 
  dark sector.
  The invisible and visible channels are described in detail in Sections~\ref{sec:simps} and \ref{sec:visible_simps}, respectively.
  The solid (dashed) lines correspond to $8$ ($16$) GeV electron beam, 
  with other experimental parameters given in the text. Regions excluded by 
  existing data from the BaBar invisible search~\cite{Lees:2017lec}, DM scattering 
  at LSND~\cite{deNiverville:2011it}, E137~\cite{Bjorken:1988as,Batell:2014mga}, and
MiniBooNE~\cite{Aguilar-Arevalo:2017mqx}, as well as electron beam dumps E137~\cite{Bjorken:1988as} and 
Orsay~\cite{Davier:1989wz} are shown in gray.
The projections for an upgraded version of the SeaQuest experiment (dotted purple)~\cite{Berlin:2018tvf} and 
the Belle II invisible search ($20\,$ fb$^{-1}$, dotted/solid blue)~\cite{HeartyCV,Battaglieri:2017aum} are also shown.
  We have fixed $\alpha_D = 10^{-2}$, $m_\Ap / m_\pi =
  3$, $m_V/m_\pi = 1.8$, and $m_\pi/f_\pi = 3$ in computing experimental limits. 
Contours of the dark matter 
  self-interaction cross section per mass, $\sigma_\text{scatter} / m_\pi$, are shown as vertical gray dotted lines. 
  The dot-dashed gray contours denote regions excluded by measurements of the cosmic microwave background.
  The black solid (dashed) line shows the parameters for which hidden sector pions saturate the observed DM 
  abundance for $m_V/m_\pi = 1.8$ ($1.6$).
  \label{fig:simp_reach}
}
\end{figure}

\subsection{Strongly-Interacting Models\label{sec:simps}}
Until recently most light DM scenarios have focused on weak couplings in the hidden 
sector as described in the previous sections. Another generic possibility is that 
the dark sector is described by a confining gauge theory similar to our QCD~\cite{Strassler:2006im,Hochberg:2014kqa}. 
The low-energy spectrum then contains dark mesons, the lightest of which 
can make up the DM. The presence of heavier composite states, e.g. analogues of the SM 
vector mesons, and strong self-interactions can alter the 
cosmological production of DM~\cite{Berlin:2018tvf}. This leads to qualitatively different experimental targets compared
to those in the minimal models. Despite the large variety of possible scenarios featuring different 
gauge interactions and matter content, 
both visible and invisible signals appear to be generic in strongly interacting sectors. 
As a concrete example, we will focus on the model recently studied in Ref.~\cite{Berlin:2018tvf} with 
a $SU(3)$ confining hidden sector with $3$ light quark flavors, and a dark photon mediator. Therefore 
production of dark sector states occurs through the $A^\prime$ which then 
promptly decays either into dark pions and/or vector mesons. The 
dark pions and some of the vector mesons are either stable or long-lived 
and give rise to a missing momentum signal. The other set of 
vector mesons can decay to SM particles by mixing with the 
dark photon. Displaced decays of these states can give rise to a visible signal at an LDMX-style experiment.

The interactions of the dark photon with hidden sector 
pions $\pi$ and vector mesons $V$ relevant for 
fixed-target production are summarized by
\beq
\mathscr{L}\supset 
\frac{\epsilon}{2} F^\prime_{\mu\nu} F^{\mu\nu}
-\frac{g_D}{g_V} F^\prime_{\mu\nu} \tr Q V^{\mu\nu}
+ 2 i g_V \tr \left(V^\mu [\partial_\mu \pi,\pi]\right)
- \frac{3 g_D g_V}{8\pi^2 f_\pi} \epsilon^{\mu\nu\alpha\beta} F_{\mu\nu}^\prime\tr \left(Q V_{\alpha\beta} \pi\right).
\label{eq:simp_model}
\eeq
In the above Lagrangian, $\pi$ and $V$ are both $3\times3$ matrix fields, 
$f_\pi$ is the hidden sector pion decay constant, $g_D = \sqrt{4\pi\alpha_D}$  is the 
$U(1)_D$ gauge coupling, $g_V = m_V/(\sqrt{2} f_\pi)$ is 
the vector meson coupling, and $Q = \diag{(1,-1,-1)}$ is the 
matrix of $U(1)_D$ charges of the hidden sector quarks. 
The first term in Eq.~\ref{eq:simp_model} is the usual kinetic mixing between the SM photon and the $\Ap$, 
enabling production of hidden sector states through, e.g., electron bremsstrahlung. The second term 
is a kinetic mixing between the $\Ap$ and the vector mesons, allowing some $V$'s 
to decay directly to SM particles through an intermediate $\Ap$. The third term is the canonical interaction between 
vector mesons and pions; this interaction in combination with $\Ap-V$ kinetic mixing 
allows $\Ap$ to decay to $\pi\pi$ final states. The last 
term is an effective vertex that encodes anomalous decays $\Ap\rightarrow V \pi$.
Other interactions and processes are described in Refs.~\cite{Hochberg:2014kqa,Hochberg:2015vrg,Berlin:2018tvf}.
The Lagrangian in Eq.~\ref{eq:simp_model} makes it clear that a generic 
SIMP model with $\mAp > m_V > m_\pi$ will have a missing momentum  
signal from $\Ap\rightarrow \pi\pi$, and possibly from $\Ap\rightarrow V \pi$ if $V$ decays outside 
of the detector. The projected reach of an LDMX-style 
experiment to these invisible decay modes, along with existing constraints and representative 
projections for an upgraded version of the proton beam-dump SeaQuest and 
Belle II, is shown in Fig.~\ref{fig:simp_reach} 
for $\alpha_D = 10^{-2}$, $m_{\Ap} / m_\pi = 3$, $m_V/m_\pi = 1.8$, and $m_\pi/f_\pi = 3$.
 This figure also shows contours in $\mAp-\epsilon$ 
space where the hidden sector pions saturate the observed 
DM abundance for $m_V/m_\pi = 1.8$ ($1.6$) as the solid (dashed) black 
lines. In these models, the $\pi$ 
relic abundance is determined by semi-annihilations $\pi\pi\leftrightarrow V \pi$; since $m_V/m_\pi >1$ 
the forward reaction is Boltzmann suppressed. The final abundance of $\pi$ is therefore very sensitive to the ratio $m_V/m_\pi$,  
while the reach of collider and fixed-target experiments is not. In Fig.~\ref{fig:simp_reach} we assumed that 
hidden vector mesons that do not mix with the $\Ap$ decay invisibly to $\pi\pi$, 
while the analogues of the $\rho$ and $\phi$ decay into SM 
particles as described by the interactions in Eq.~\ref{eq:simp_model} (see Ref.~\cite{Berlin:2018tvf} for 
more details and generalizations of this scenario).
  
Finally, we note that SIMP models also generate visible signals (e.g. $\Ap\rightarrow \pi (V\rightarrow \ell^+\ell^-)$) 
at an LDMX-style experiment. This channel will be described in more detail in Sec.~\ref{sec:visible_signals}.
The projected reach of LDMX in this channel is also shown in Fig.~\ref{fig:simp_reach}.

\subsection{Freeze-In}
\label{sec:freezein}

In this section, we briefly discuss freeze-in production of DM~\cite{Dodelson:1993je, Hall:2009bx}. In this case, one typically assumes a negligible initial DM abundance arising from the epochs of inflation or reheating and that the dark sector never fully thermalizes with the SM. As a result, DM never carries the characteristically large comoving entropy density associated with thermalized radiation, and its abundance is instead slowly built up over time through feeble interactions with the SM bath. These cosmological scenarios generally invoke extremely small portal interactions that are hopelessly beyond the reach of detection in terrestrial experiments. However, in the limiting case of an ultra-light mediator, such as a dark photon of mass $m_\Ap \ll \text{keV}$, the large enhancement of DM-electron scattering at low-momentum transfer provides a detectable and cosmologically motivated target for future direct detection experiments~\cite{Battaglieri:2017aum}. 

Alternative variations can instead motivate large production rates at low-energy accelerators for low reheat temperatures and mediators much heavier than $10 \text{ MeV}$. We will illustrate this with a Dirac fermion, $\chi$, with unit charge under $U(1)_D$. We follow the semi-analytic procedure to solve the relevant Boltzmann equation outlined in, e.g., Ref.~\cite{Hall:2009bx}, to estimate the freeze-in production of $\chi$ through the process $e^+ e^- \to A^{\prime *} \to \chi \bar{\chi}$. If the dark photon mass is much larger than the reheat temperature of the universe, $m_{\Ap} \gg T_\text{RH}$, DM production is dominated at the earliest times (largest temperatures). We find that the final $\chi$ abundance is approximately 
\be
\label{eq:freezein}
\Omega_\chi h^2 \simeq 1.3 \times 10^{28} \times g_*^{-1/2} (T_\text{RH}) ~ g_{*S}^{-1} (T_\text{RH}) ~ \frac{ \alpha_\text{em} ~ \epsilon^2 \, \alpha_D \, m_\chi \, T_\text{RH}^3}{m_{\Ap}^4}
~,
\ee
where $g_*$ and $g_{* S}$ are the energy density and entropy density effective relativistic degrees of freedom. This is valid for $T_\text{RH} \lesssim 100 \text{ MeV}$, in which case similar contributions from muons are expected to be subdominant. Effects from the pre-thermal phase immediately following inflation are also not expected to significantly modify the estimate of Eq.~(\ref{eq:freezein}) for the dark photon model under consideration~\cite{Garcia:2018wtq}.
 
\begin{figure}[t!]
\includegraphics[width=9cm]{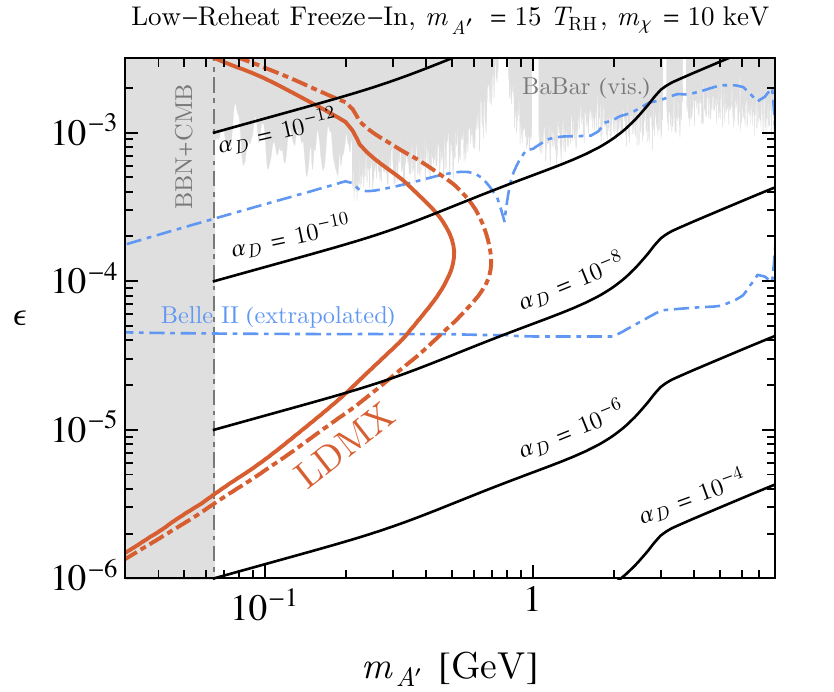}
\caption{
  LDMX sensitivity to the freeze-in scenario with a heavy dark photon and low-reheat temperature. The projected 
  reach of LDMX is shown as the solid red (dashed-dotted red) line for a tungsten (aluminum) target and a 8 (16) GeV beam.
  The correct relic abundance is obtained along the black contours for different choices of $\alpha_D$.
  The gray shaded regions are excluded by the BaBar resonance search~\cite{Lees:2014xha} and by 
  cosmological constraints on low reheating temperatures~\cite{deSalas:2015glj}.
  We also show the projected sensitivity of the Belle II monophoton search (blue dot-dashed) 
  as computed by rescaling the 20 fb$^{-1}$ background study up to 50 ab$^{-1}$ assuming statistics 
  limitation only~\cite{Battaglieri:2017aum,HeartyCV}. 
}
\label{fig:freezein}
\end{figure}

We explore a slice of parameter space in the $\epsilon-m_{\Ap}$ plane in Fig.~\ref{fig:freezein}. Along the black contours, the abundance of $\chi$ matches the observed DM energy density for various choices of $\alpha_D$. We have fixed $m_{\Ap} = 15 \, T_\text{RH}$ and $m_\chi = 1 \text{ keV}$ throughout. $m_{\Ap} \gg T_\text{RH}$ guarantees that on-shell $\Ap$ production via inverse-decays ($e^+ e^- \to \Ap$) followed by $\Ap \to \chi \chi$ is subdominant to the direct annihilation, $e^+ e^- \to A^{\prime *} \to \chi \chi$. Furthermore, DM masses significantly lighter than $\mathcal{O}(\text{keV})$ are constrained from considerations of warm DM~\cite{viel:2013apy}, although the exact strength of this bound warrants a dedicated study~\cite{Tongyan:2018}. We saturate this approximate lower bound, fixing $m_\chi = 10 \text{ keV}$ in Fig.~\ref{fig:freezein}, which from Eq.~(\ref{eq:freezein}) implies that larger portal couplings are necessary to acquire an adequate relic abundance. For $\epsilon \gtrsim 10^{-4}$ and $\alpha_D \lesssim 10^{-9}$, dark photon decays into SM leptons become non-negligible. In this case, searches for resonant pairs of leptons at BaBar restrict $\epsilon \lesssim 10^{-3}$~\cite{Lees:2014xha}. For a fixed freeze-in abundance of DM, Eq.~(\ref{eq:freezein}) implies that smaller values of $\epsilon$ correspond to larger $\alpha_D$. In this case, the $\Ap$ decays dominantly invisibly and efficient searches include those looking for missing momentum or energy at LDMX and Belle II. Also in Fig.~\ref{fig:freezein}, the reheat temperature is restricted to be larger than $T_\text{RH} \gtrsim 4.3 \text{ MeV}$ from considerations of nucleosynthesis and the CMB~\cite{deSalas:2015glj}. For a fixed $m_{\Ap} / T_\text{RH}$ ratio, this implies that $m_{\Ap} \gtrsim 65 \text{ MeV}$.

\section{Millicharged Particles and Invisible Decays of New Particles}
\label{sec:beyond_dm}
\subsection{Invisibly Decaying Dark Photons}
\label{sec:invisible-dark-photon}

\begin{figure}[t]
\includegraphics[width=3.2in,angle=0]{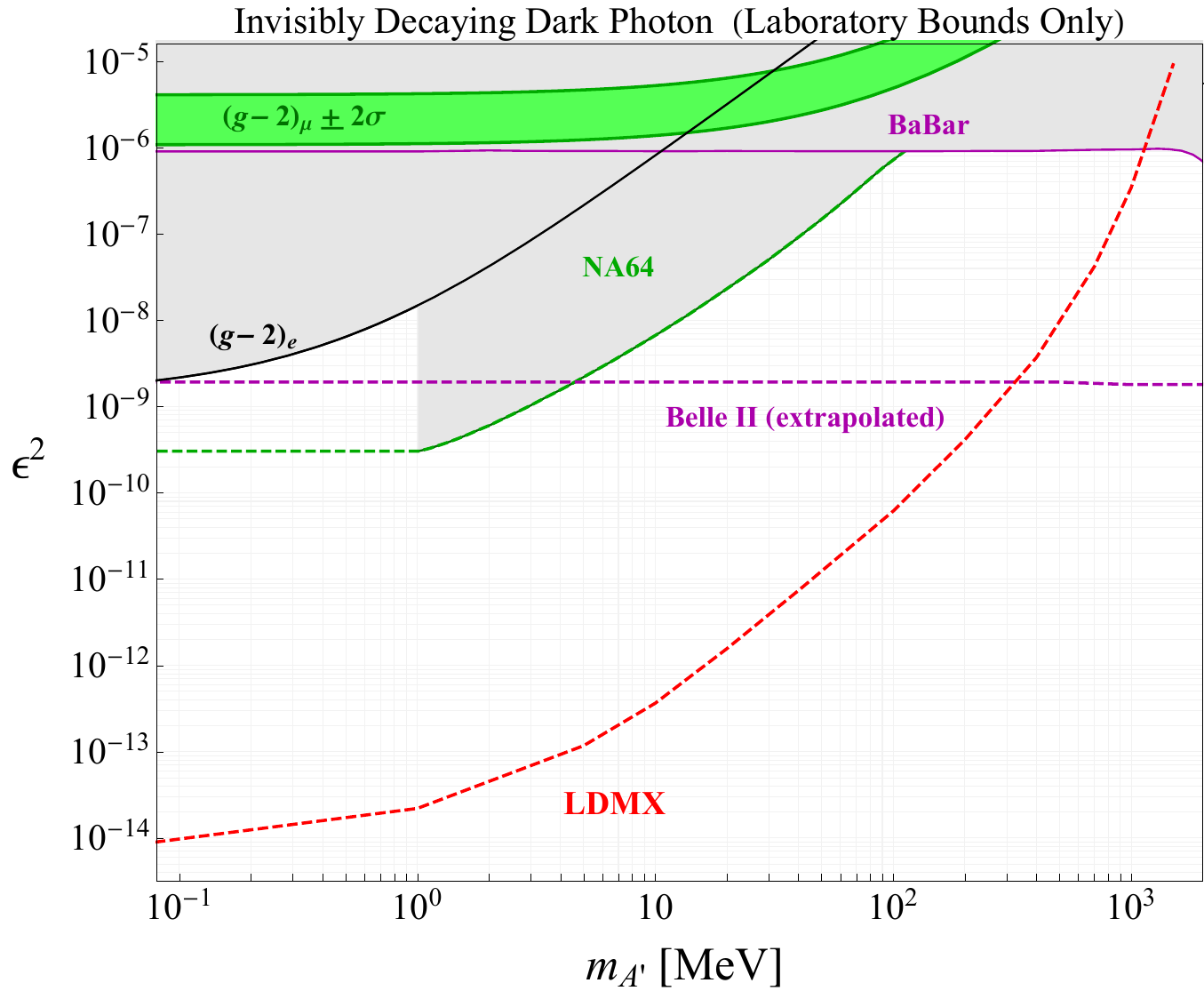}
\caption{LDMX Sensitivity to invisibly decaying dark photons. 
The dashed purple curve is the projected sensitivity $\gamma$+missing energy search at Belle II presented in 
 \cite{Battaglieri:2017aum} and computed by rescaling the 20 fb$^{-1}$ background study up to 50 ab$^{-1}$ \cite{HeartyCV}.
 The red dashed curve is the Phase 2 LDMX projection for a $10\%$ radiation length Tungsten target presented in  \cite{Battaglieri:2017aum} , which was 
 scaled up to $10^{16}$ EOT relative to a background study  
 with $4 \times 10^{14}$ EOT. The green dashed curve is the NA64 sensitivity projection
 assuming $10^{12}$ EOT  \cite{Battaglieri:2017aum}
}
\label{fig:invisible-dark-photon-epsilon-ma}
\end{figure}

The discussion in Sec.~\ref{sec:thermal-dm-section} considers an invisibly decaying dark photon mediator coupled
to various thermal DM candidates with the current interactions 
\beq
\mathscr{L} \supset  A^\prime_\mu \left(   \epsilon e J_{\rm EM}^\mu   +   g_D J_D^\mu \right),  
\eeq
where $J_{\rm EM}$ is the SM electromagnetic current and  $J_{D}^\mu$ is a dark sector current with coupling $g_D$ that
 allows $A^\prime$ to decay invisibly with a large branching fraction.
In this section, we consider the same $A^\prime$ particle, but interpret this signal agnostically with respect to the final
state decay products, which need not have any connection to DM 
as long as they are (meta)stable on the relevant experimental length scales.
In Fig.~\ref{fig:invisible-dark-photon-epsilon-ma}, we show the parameter space for this
scenario in the $\epsilon-m_{A^\prime}$ plane. Also shown are LDMX and NA64 projections taken from 
Ref.~\cite{Battaglieri:2017aum}.

\begin{figure}[htbp]
\includegraphics[width=3.7in,angle=0]{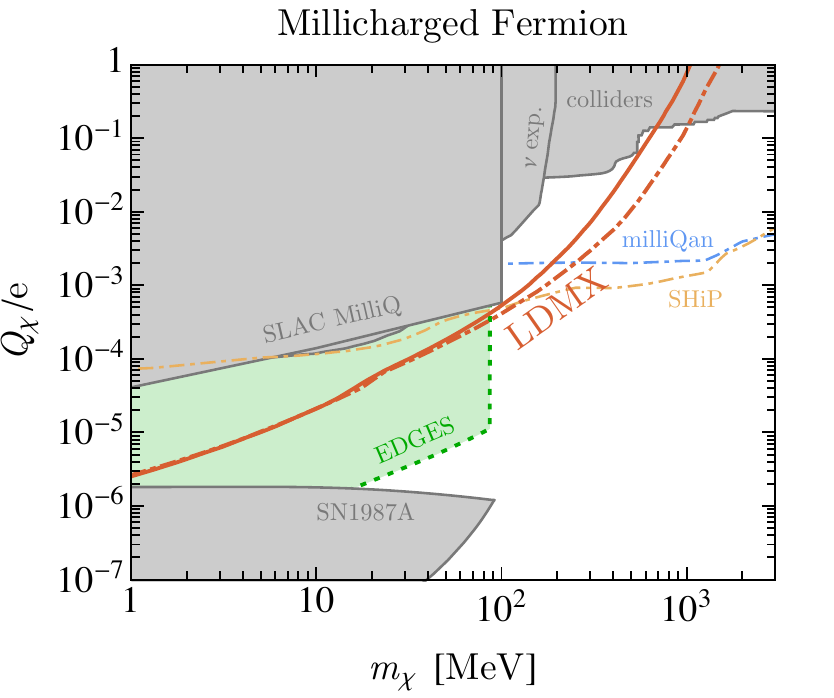}
\caption{LDMX sensitivity to Dirac fermion millicharged particles in the 
  $Q_\chi/e - m_\chi$ plane. The LDMX reach is shown as the solid (dot-dashed) red line for the configuration with a 8 (16) GeV electron beam on a tungsten (aluminum) target and $10^{16}$ EOT. Regions excluded by the SLAC MilliQ~\cite{Prinz:1998ua}, 
neutrino experiments~\cite{Magill:2018tbb}, supernova cooling~\cite{Chang:2018rso} and colliders are shown in gray.
Projected sensitivities of milliQan~\cite{Haas:2014dda} and SHiP~\cite{Magill:2018tbb} 
are shown as the blue and yellow dashed-dotted lines, respectively.
We expect that for $Q_\chi \sim e$ millicharged particles will deposit 
energy in the LDMX detector through ionization, thereby reducing the sensitivity of the missing 
momentum technique at large masses.
\label{fig:millicharge_sensitivity}
}
\end{figure}

\subsection{Millicharges}
Millicharged particles arise as the $\mAp\rightarrow 0$ limit of a 
dark photon coupled to $U(1)_D$ charges (i.e. the model described in Sec.~\ref{sec:invisible-dark-photon})~\cite{Holdom:1985ag}, or as fundamental particles 
with a small electromagnetic (EM) charge. 
In both cases, the effective Lagrangian for 
a millicharge $\chi$ is simply
\beq
\mathscr{L}\supset Q_\chi A_\mu \bar\chi \gamma^\mu \chi,
\eeq
where $Q_\chi \ll e$ is the EM charge of $\chi$ and we take $\chi$ to be a Dirac fermion. 
If $\chi$ is not associated with a $U(1)_D$ symmetry, then the discovery of a fundamental millicharged particle would refute the charge quantization 
principle~\cite{Foot:1990uf,Foot:1992ui} and inform us on related 
issues like the existence of monopoles and Grand Unification~\cite{Preskill:1984gd}.
Recently, relic millicharged particles have been proposed~\cite{Barkana:2018lgd}
as a possible explanation of the EDGES 21 cm signal~\cite{Bowman:2018yin}. Given the 
importance of millicharges in understanding of charge quantization and potential implications of the EDGES result, it is useful to 
search for these particles in the laboratory.
Pairs of $\chi$ particles can be produced in fixed-target experiments through an off-shell Bremsstrahlung 
photon. Once produced, the probability of millicharges to interact with the detector is 
suppressed by $(Q_\chi/e)^2 \ll 1$, so they are likely to escape the detector without depositing 
any energy. This means that such particles can be searched for in the missing momentum channel 
at an LDMX-like experiment. In Fig.~\ref{fig:millicharge_sensitivity}, we show 
the LDMX sensitivity to millicharged particles in the $Q_\chi/e - m_\chi$  plane for 
the setup with a 8 or 16 GeV electron beam, $10^{16}$ EOT, and tungsten (solid red line) and aluminum (dot-dashed red 
line) targets.
Existing constraints from the SLAC MilliQ and collider experiments~\cite{Prinz:1998ua}, neutrino experiments 
(LSND and MiniBooNE)~\cite{Magill:2018tbb}, and
supernova cooling~\cite{Chang:2018rso} are shown in gray. The region of parameter space that 
can explain the EDGES signal is highlighted in green~\cite{Berlin:2018sjs,Barkana:2018qrx}.
We note that LDMX can improve on the SLAC MilliQ and neutrino experiment results, and it can 
probe a significant portion of the EDGES-motivated parameter space. 
While we extend the LDMX curves to large masses and charges, 
we expect that for $Q_\chi \sim e$ millicharged particles will deposit 
energy in the detector through ionization. At this point $\chi$ 
behaves as a minimum-ionizing particle and so the missing momentum technique becomes inappropriate.
We also show the sensitivity of the proposed milliQan experiment at the LHC~\cite{Haas:2014dda} and the reach of 
the proposed SHiP experiment~\cite{Magill:2018tbb} as the dot-dashed blue and yellow lines, respectively.
Finally, we note that cosmological relic millicharges may be constrained 
from kinetic heating of galactic gas~\cite{Bhoonah:2018wmw}.
While this astrophysical bound is potentially extremely powerful, it 
is subject to uncertainties relating to cloud chemical composition and the resulting 
standard cooling rates, DM distribution, and the direction and 
magnitude of galactic magnetic fields.

\begin{figure}[t!]
\includegraphics[width=3.7in,angle=0]{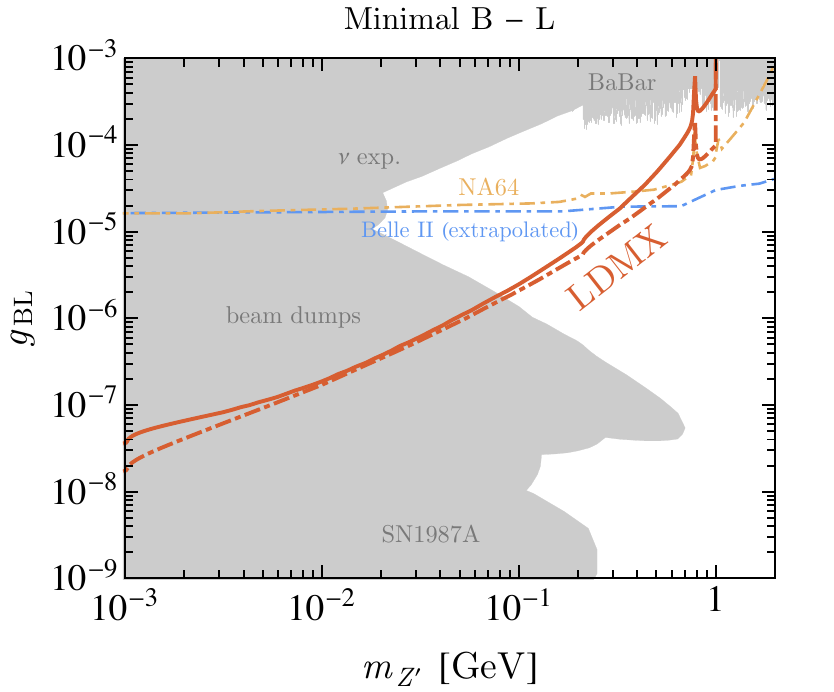}
\caption{LDMX sensitivity to $B-L$ gauge boson via its decay to neutrinos is 
  shown by the solid (dot-dashed) red line 
  in the $ g_{BL} - m_X$ plane for a 8 (16) GeV electron beam, $10^{16}$ EOT and a tungsten (aluminum) target. 
  Regions excluded by beam dumps~\cite{Bauer:2018onh,Ilten:2018crw}, neutrino 
scattering experiments~\cite{Deniz:2009mu, Bilmis:2015lja}, BaBar~\cite{Lees:2014xha} and SN1987A~\cite{Harnik:2012ni} are shaded in gray.
Projections for Belle II and NA64 are also shown as blue and yellow dashed-dotted lines, respectively.  
}
\label{fig:BL_sensitivity}
\end{figure}

\subsection{$B-L$ Gauge Bosons Decaying to Neutrinos}
\label{ssec:bl_neutrinos}
Unlike the minimal dark photon scenario in which the $\Ap$ is the lightest 
new state, $B-L$ gauge bosons ($\zp$) introduced in Sec.~\ref{sec:other-mediators} 
couple to neutrinos. This means that even this scenario with no 
additional states can be discovered in the missing momentum search due to the 
irreducible decay $\zp \rightarrow \nu \nu$. The LDMX sensitivity to this 
decay mode is shown in Fig.~\ref{fig:BL_sensitivity} for a 8 or 16 GeV 
electron beam, $10^{16}$ EOT, and tungsten and aluminum targets (solid red and dot-dashed red lines, 
respectively).
The existing constraints from beam dumps~\cite{Bauer:2018onh,Ilten:2018crw}, neutrino 
scattering experiments~\cite{Deniz:2009mu, Bilmis:2015lja}, and BaBar~\cite{Lees:2014xha} are shown in gray.
Projections for Belle II and NA64 are also shown as dot-dashed lines.

\subsection{Muonic Forces and $(g-2)_\mu$}
The longstanding $4 \sigma$ anomaly in the measured value of $(g-2)_\mu = 2a_\mu$  \be
 \label{eq:delta-amu}
\Delta a_\mu \equiv a_\mu^{{\rm exp}}  - a_\mu^{\rm theory} = (28.8 \pm 8.0) \times 10^{-10}.
\ee
is among the largest discrepancies in particle physics \cite{Patrignani:2016xqp}. 
 This result has motivated great interest in light (sub-GeV) weakly coupled particles. However, most models that explain this anomaly, e.g., dark photons with predominantly visible (see Sec.~\ref{sec:minimal-dark-photon}) 
 or invisible decay modes (see Sec.~\ref{sec:invisible-dark-photon}), have been excluded by laboratory measurements because
 they also predict sizable couplings to first generation particles -- see Ref.~\cite{alexander:2016aln} for a review.

The phenomenology of the $(g-2)_\mu$ anomaly
only requires interactions of the new states with muons.
Our benchmark phenomenological models therefore involve muon-philic 
particles with invisible and visible decay modes. 
Both possibilities can be tested with the missing momentum approach.
We define $S$ and $V$ to be scalar and vector particles with bilinear couplings to muons 
\be
 g_S S \overline{\mu} \mu \qquad {\rm (scalar)} ~~,~~~
 g_V V_\alpha \overline{\mu} \gamma^\alpha \mu \qquad {\rm (vector)},
 \label{eq:muon_specific_forces}
\ee
where $g_{S,V}$ are dimensionless couplings.
Both $S$ and $V$ can individually reconcile the $a_\mu = \frac{1}{2}(g-2)_\mu$ anomaly 
through their corrections to the $\mu-\gamma$ interaction vertex at loop-level~\cite{Pospelov:2008zw}
\be
\Delta a^{S}_{\mu}  \simeq  6.0 \times 10^{-10} \, \left( \frac{g_S}{10^{-4}} \right)^2  ~~,~~
\Delta a^{V}_{\mu}   \simeq  1.6 \times 10^{-9} \, \left( \frac{g_V}{10^{-4}} \right)^2 ,
\label{eq:g_minus_two_favored_couplings}
\ee
where we have taken $m_{S,V} \ll m_\mu$ for this estimate.
Note that axial-vector or pseudoscalar interactions shift $a_\mu$ in the opposite direction, so these particles would only increase the tension between theory and experiment in Eq.~(\ref{eq:delta-amu}).

\begin{figure}[t]
\includegraphics[width=3.in,angle=0]{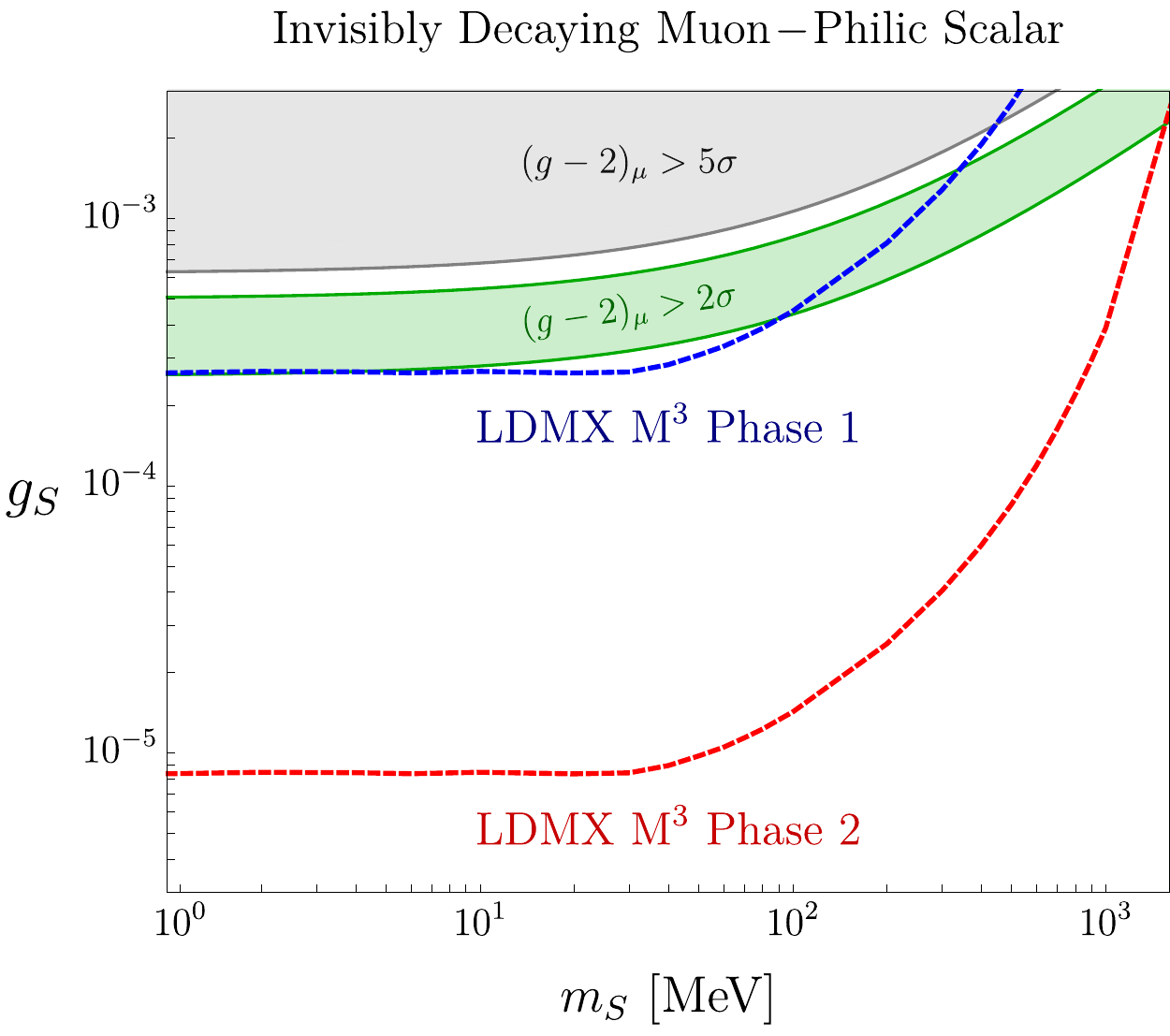}
\includegraphics[width=3.in,angle=0]{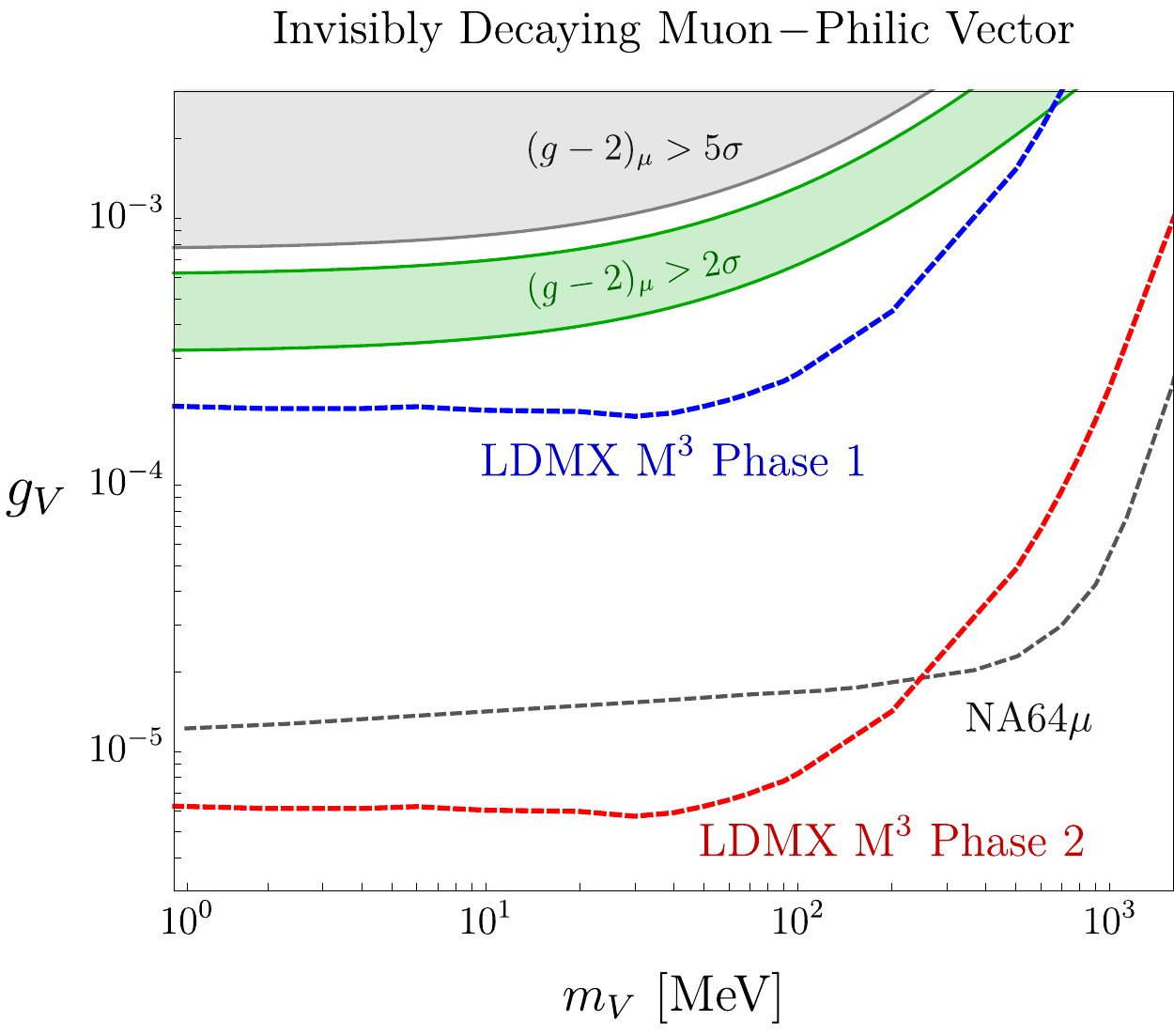}
\caption{ Parameter space for moun-philic invisibly decaying scalar ($S$) and vector ($V$) particles that decay invisibly. Both
plots are taken from Ref.~\cite{Kahn:2018cqs}. 
The green band represents the region where such particles resolve the $(g-2)_\mu$ anomaly.
The gray shaded region results in an unacceptably large correction to $(g-2)_\mu$. 
}
\label{fig:muon-focrce-phenomenological}
\end{figure}

In Fig.~\ref{fig:muon-focrce-phenomenological} we show the parameter space for which invisibly-decaying $S$ and
$V$ can explain the $a_\mu$ discrepancy to within $2\sigma$ (green band) and the region which 
overshoots the favored parameter space by more than $5\sigma$ (gray shaded region). Also shown
are the projections for a muon missing momentum ($M^3$) search using the LDMX setup
for Phase 1 ($10^{10}$ MOT, blue dashed) and Phase 2 ($10^{13}$ MOT, red dashed)~\cite{Kahn:2018cqs}. 

\begin{figure}[htbp]
  \centering
  \includegraphics[width=8cm]{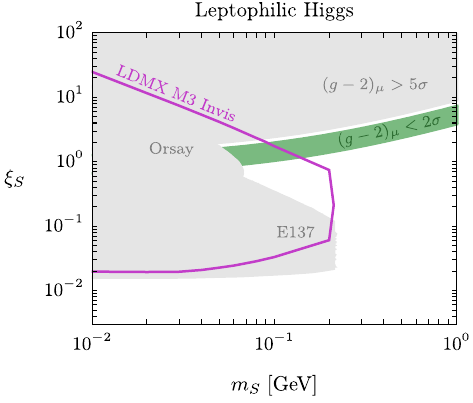}
  \caption{Muon beam LDMX sensitivity to a leptophilic Higgs~\cite{Batell:2016ove} in the missing momentum channel. The projection 
(purple line) is shown in the plane the scalar mass $m_S$ and the coupling $\xi_S$ (relative to the SM Yukawa $m_\ell/v$). 
The region of parameter space favored by the $(g-2)_\mu$ anomaly is shown in green. The gray regions are excluded 
by unacceptably large shifts to $(g-2)_\mu$, and by null results from the Orsay~\cite{Davier:1989wz} and E137~\cite{Bjorken:1988as} 
electron beam dumps.
  \label{fig:leptophilic_higgs}
}
\end{figure}

The scalar simplified model in Eq.~(\ref{eq:muon_specific_forces}) can be realized in UV-complete 
scenarios~\cite{Batell:2016ove,Batell:2017kty}. One such UV completion is a leptophilic scalar with mass-proportional interactions with leptons~\cite{Batell:2016ove}
\beq
\mathscr{L}\supset \xi_S \sum_{\ell=e,\mu,\tau} \left(\frac{m_\ell}{v}\right)S \bar \ell \ell, 
\eeq
where $\xi_S$ is the coupling strength relative to the SM Yukawas $m_\ell/v$ and $v=246$ GeV is the SM Higgs vacuum expectation value.
This set of interactions follows naturally from a limit of the two-Higgs-doublet model with an additional 
singlet field~\cite{Batell:2016ove}.
In the notation of Eq.~(\ref{eq:muon_specific_forces}), $S$ couples to muons with the strength 
$g_S = \xi_S (m_\mu/v)$ and so the leptophilic scalar can resolve the $(g-2)_\mu$ 
anomaly for $\xi_S \sim 1$ (see Eq.~(\ref{eq:g_minus_two_favored_couplings})). 
If $S$ is the lightest new state, it must decay back to the SM after production. 
For $m_S < 2m_\mu$, the only available channel is $S\rightarrow \bar e e$ which has a rate 
suppressed by $m_e/v \approx 2\times 10^{-6}$. The corresponding boosted decay length 
is long compared to the size of LDMX, suggesting that many scalars will decay outside of the 
detector leading to a missing momentum signal. The signal rate is estimated 
by requiring that $S$ decays beyond the HCAL which extends to $\sim 315\; {\rm cm}$ after the target.
In Fig.~\ref{fig:leptophilic_higgs} we show the sensitivity of the muon beam version of LDMX to 
these decays with a 15 GeV beam and $10^{13}$ MOT. 
Given the small, but non-zero coupling to electrons, high-intensity electron beam experiments, such as Orsay~\cite{Davier:1989wz} and E137~\cite{Bjorken:1988as}, 
are also sensitive to displaced $S$ decays; the corresponding exclusions are shown in gray. 
We find that the nominal electron beam LDMX luminosity is not high enough to cover 
new parameter space. However, the $(m_\mu/m_e)$-enhanced coupling to muons implies 
that the muon beam version of LDMX would be the ideal method of testing this 
parameter space, including parts of the $(g-2)_\mu$-favored region (shown as a green band in Fig.~\ref{fig:leptophilic_higgs}).
The decays of $S$ become prompt for $m_S > 2m_\mu$ when the $S\rightarrow \mu^+ \mu^-$ channel is kinematically available; this sets an upper 
limit in $m_S$ on the sensitivity of the displaced decay channel, which is visible both in the LDMX and E137 regions in Fig.~\ref{fig:leptophilic_higgs}.

\section{Visibly Decaying Dark Photons, Axions, and Strongly Interacting Dark Sectors}
\label{sec:visible_signals}
If DM is heavier than the mediator, its relic abundance 
is determined by dark sector interactions alone~\cite{Pospelov:2007mp}. This means that 
there is no sharp target for the coupling with SM particles. 
However, this class of \emph{secluded} DM models 
gives rise to a different set of signals that can also 
be searched for at missing momentum experiments. 
If a mediator is produced, only decays to SM particles are 
kinematically allowed.
Weakly-coupled mediators tend to be long-lived and can 
travel macroscopic distances before decaying, leading to, e.g., 
displaced electromagnetic showers. An LDMX-like detector can then 
be used as a fully-instrumented, short baseline beam dump that can search 
for displaced, visible energy depositions similar 
to the proposal of Refs.~\cite{Gninenko:2013rka,Andreas:2013lya}.
This technique 
tests mediator physics independent of the nature of DM, 
so it is interesting in its own right as a probe of 
generic dark sectors. Below, we consider two 
minimal mediator models (the dark photon and 
axion-like particles), showing that the visible 
search channel can probe new territory.

Visible signals also arise in non-minimal dark sectors that 
contain long-lived particles decaying to the SM. 
This occurs, for example, in models of strongly 
interacting dark sectors where a subset of hidden sector vector 
mesons are long-lived and must decay to the SM due to kinematic 
constraints. As already discussed 
in Sec.~\ref{sec:simps}, these scenarios also provide missing momentum signals. 
Simultaneous measurements of the invisible and visible channels 
therefore can shed light on the nature of these dark sectors.
Visible decays can also be observed in models of inelastic DM (iDM) when a heavier dark state de-excites to a lighter state 
through a three-body decay. These final states were considered in Sec.~\ref{sec:pseudo-dirac-large-delta} 
for DM splittings 
for which the decay length of the excited state exceeded the size of the detector. Note that similarly to some 
decays in the SIMP theories described above, iDM visible decays 
are not resonant, having a stable DM particle in the final state 
that escapes the detector.

For the purpose of demonstrating the power 
of the displaced-shower technique, we will focus on an 
LDMX-like experiment with a $0.1$ radiation length (r.l.) Tungsten target separated by 
$\sim 15\; {\rm cm}$ from a $40$ r.l. electromagnetic Tungsten-based sampling calorimeter (with actual thickness of $\sim 28\; {\rm cm}$) and 
a large hadronic calorimeter (the space between the target and ECAL is 
occupied by tracking layers needed for recoil electron $p_T$ measurement -- see Fig.~\ref{fig:ldmx_setup}). We will take the total length of the 
ECAL/HCAL assembly to be 3 meters. 
These are not final parameters of the detector design, but rather 
realistic and representative values for this class of experiments.
Unless otherwise noted, we 
will show experimental reach for the proposed Phase II LDMX run with 
8 GeV and 16 GeV electron beams (representative of the 
accelerator capabilities of DASEL/JLAB and of CERN, respectively) and $10^{16}$ EOT. 
We also note that the search for visible displaced decays does not 
necessitate single-electron tracking needed for the missing momentum 
program; thus it is conceivable to increase the effective luminosity of the 
experiment by, e.g., using a thicker target and a higher-current beam.
We will take the signal 
region for the visible decays to be between $\zmin \sim 43\;{\rm cm}$ and 
$\zmax\sim 315\; {\rm cm}$, corresponding to energy depositions 
in the HCAL for the setup described above. The shield length $\zmin$ is chosen to maximize sensitivity 
to short lifetimes, while mitigating the late-photon conversion background (see below).
As we show below, the most interesting sensitivity of this experiment lies in the 
short-lifetime region, so the precise choice of $\zmax$ is not important.
We also require the recoil electron to have $E_e^{\mathrm{rec}} \geq 0.3\Ebeam$ after the 
target.

The short baseline of the 
experiment means that there are penetrating backgrounds that must be rejected, many of
which are closely related to the backgrounds limiting the missing momentum search channel~\cite{izaguirre:2014bca}. 
A photo-nuclear reaction can fake missing energy if all particles in the final state are undetectable. The same reaction can fake a displaced-decay signal if a single particle with  nearly the full beam energy, instead of being missed completely, propagates forward in the detector and then interacts, faking an EM shower close to the beam energy.  
Such fake showers could be produced by high-EM-fraction showers of a hard neutron, or by displaced decays of neutral kaons.  
The missing-energy background appears to be dominated by single-neutron final states of photo-nuclear reactions. As these are also the reactions that send their energy most forward, and in which the true energy of the neutron is closest to the incident photon energy (as is expected of a displaced-decay signal), we expect them to be even more dominant in a displaced search.  As discussed in Ref.~\cite{izaguirre:2014bca}, the dominant such reactions are $\gamma p \rightarrow \slashed{\pi}^+ + n$ and $\gamma n \rightarrow \slashed{\pi}^0 + n$, where the pion is backward-going and therefore not detected in the ECAL.
The resulting forward-going neutron can penetrate deep inside the ECAL and/or HCAL before initiating a hadronic shower.
As measured in Ref.~\cite{Anderson:1969bq}, this back-scattering reaction has a rate of about $2 \mu\rm{b}\; ({\rm GeV}/E_\gamma)^3$ per nucleon, or a single-nucleon yield of $\sim 2\cdot 10^{-8}$ ($3 \cdot 10^{-9}$) per incident photon at 8 (16) GeV.  Thus, we expect about 6 million (800\,000) such events per $10^{16}$ electrons on target. It should be noted that these results neglect nuclear screening and the possibility of subsequent interactions, both of which will decrease the true number of single-neutron events. In addition, there is some probability of the back-scattered pion (or nucleons that it kicks off) to interact with the detector leading to a vetoable signal at the photo-nuclear interaction site --- current LDMX studies suggest that this may provide a factor of $\sim 10$ further reduction. 
While an $O(1)$ fraction of the remaining single-neutron events would shower in the range $43\;{\rm cm} < z <  315\;{\rm cm}$, this background can be further mitigated by two handles.
First, the longitudinal and transverse shapes of hadronic and electromagnetic showers are quite different. While shower shape can provide up to $\sim 10^{-5}$ rejection of hadrons in  this energy range (see e.g. \cite{BOEZIO2006111}), this performance cannot be expected from the LDMX HCAL, which has far less segmentation. Nonetheless, it is plausible that 2--3 orders of magnitude rejection of hadronic showers may be possible, especially given that the known energy of the outgoing particle (whether an ALP or a neutron) can be used as a constraint to reject high-EM-fraction showers originating from reactions like $n+n\rightarrow n \,n\, \pi^0$ with an energetic $\pi^0$.  
An additional handle that is quite powerful (adding another $\sim 2$ orders of magnitude rejection) in LDMX is the $p_T$ distribution of the recoiling electron -- photo-nuclear processes are dominated by low $p_T$ events, while the production of a 10--100 MeV late-decaying particle will lead to a significant spread in $p_T$ (as in the case of the invisible signal -- see~\cite{izaguirre:2014bca}).

We briefly mention other background processes which seem less likely to be limiting: multi-hadron final states are far more numerous, but also rejectable (for example, for an $n\pi^+\pi^-$ final state to fake the signal, the pions must both be missed entirely by the detector, and the neutron must have an upward fluctuation in shower energy, in addition to the EM-like shower and high recoil $p_T$ discussed above).  
Displaced decays of neutral kaons and $\Lambda^0$ baryons are always accompanied by another $s=1$ state. For example $\phi$ meson decays 
produce $K_L K_S$ pairs. If the associated state is short lived ($K_S$ in this case), its decay at the site 
of the photo-nuclear reaction leaves an energy deposition that can be vetoed. Even processes that produce two $K_L$'s offer the prospect of detecting the decay or interaction of the second $K_L$, and have a lower rate than the single-neutron process described above.  

Given these considerations, we will assume that the hadronic backgrounds described above can be efficiently mitigated (though not necessarily fully rejected), through hard energy deposition in the ECAL at the photo-nuclear interaction point, through shower shape rejection in the ECAL and HCAL, and through $p_T$ selections in the tracker. The remaining background is then late secondary photon conversion. We estimate 
that this process will produce $\sim 9$ (880) signal-like events 
for $10^{16}$ ($10^{18}$) EOT. For $10^{16}$ EOT, single electron tracking is still feasible, so recoil electron $p_T$ may be used 
to reject photonuclear and late-conversion backgrounds.
In a high-luminosity phase ($10^{18}$ EOT), this discriminator is not available.
In the long-lifetime region (corresponding to small couplings) 
signal events will be linearly distributed in the 
longitudinal direction, while the late conversion background is 
exponentially falling. Thus, in this region of parameter space 
the background can be reduced by \emph{increasing} $\zmin$.
However, we emphasize that a careful background study is needed 
to determine the potential reach in the visible channel. The 
various sensitivity projections below show 
contours of $14$ and $930$ events, corresponding to $95$\% C.L. reach given the late conversion background in the nominal and 
high-luminosity configurations, respectively.
We emphasize that even if there is background contamination at the level of $\sim 10^3$ events, LDMX can 
still probe significant new regions of parameter space. 
We also note that the visible channel offers additional handles that can be used for background 
rejection, such as full beam energy reconstruction (for the models with no missing energy in the final state) and exponential variation of 
the signal rate with $\zmin$ (in the short life-time regime). 

\subsection{Minimal Dark Photon}
\label{sec:minimal-dark-photon}
A dark photon produced via the kinetic mixing portal
\beq
\mathscr{L} \supset -\frac{\epsilon}{2} F_{\mu\nu}^\prime F^{\mu\nu}, 
\eeq
must decay back to SM particles if it is the lightest state in the dark sector (as, for example, in 
models of secluded DM~\cite{Pospelov:2007mp}). 
Depending on its lifetime, this decay 
can occur inside or outside of the detector, leading to both 
visible and invisible signals at an LDMX-like experiment.

\begin{figure}[t!]
  \centering
  \includegraphics[width=0.49\textwidth]{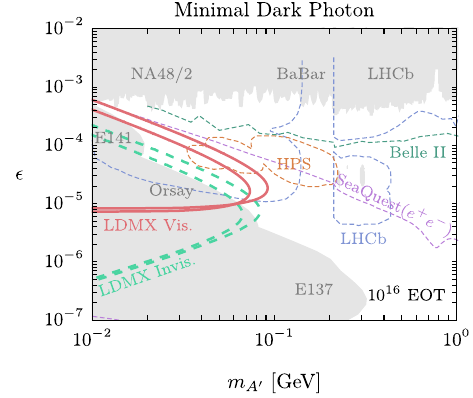}
  \includegraphics[width=0.49\textwidth]{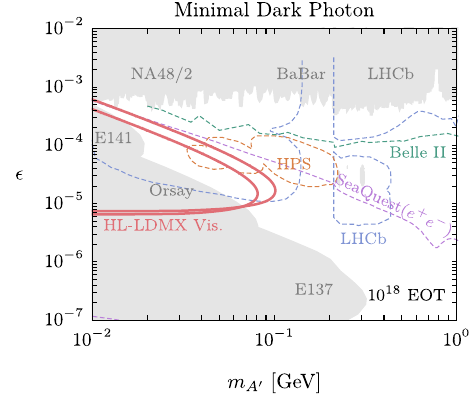}
  \caption{Sensitivity of an LDMX-style experiment to visibly-decaying dark photons for $10^{16}$ (left panel) and $10^{18}$ (right panel) EOT. 
  The solid red lines show the 95\% C.L. reach of a 
  search for late decays inside of the detector (assuming late $\gamma$ conversion background), while the green-dashed lines correspond to the missing momentum channel where 
the dark photon decays outside of the detector. 
  In both cases, the two sets of lines correspond to $8$ and $16$ GeV beams, with $\Ebeam=16$ GeV having 
  slighter better reach in mass.
  The high-luminosity configuration ($10^{18}$ EOT) must forgo single electron tracking, so the missing momentum 
search (and the use of $p_T$ as a background discriminant in the visible channel) is not possible.
  Existing constraints from E141, Orsay and E137 beam-dump 
  experiments~\cite{Andreas:2012mt}, NA48/2~\cite{Batley:2015lha}, LHCb~\cite{Aaij:2017rft} and BaBar~\cite{Lees:2014xha}
  are shown in gray. Projected sensitivities of HPS (orange)~\cite{Battaglieri:2017aum}, an upgraded version of SeaQuest~\cite{Berlin:2018pwi} (purple), Belle II (green, $50\;\mathrm{ab}^{-1}$ integrated luminosity)~\cite{Battaglieri:2017aum} and LHCb (blue)~\cite{Ilten:2015hya,Ilten:2016tkc} are shown as 
  thin dashed lines (see text for details).
  \label{fig:ldmx_reach_minimal_dp}
}
\end{figure}

Dark photon production at electron beam fixed-target experiments 
can be estimated using Weiszacker-Williams approximation~\cite{Bjorken:2009mm}, 
while the full kinematics and geometric acceptances can be implemented in a Madgraph simulation as in Ref.~\cite{Andreas:2012mt}.
Since a fully instrumented beam-dump experiment like LDMX has $\mathcal{O}(1)$ 
acceptance, these methods are in good agreement with each other. 
However, both estimates neglect the creation of $A^\prime$ from 
secondary positrons produced in the target; this contribution enhances the 
yield in certain regions of parameter space~\cite{Marsicano:2018krp}.
We neglect this contribution in estimating the experimental yields,
since the sensitivity gain is expected to lie in the long-lifetime regime, 
whereas the short baseline beam dump scenario studied here is most 
powerful in the short-lifetime region. The total event yield at an LDMX-style 
experiment can be approximated as 
\beq
N_{\rm sig} \approx N_{A^\prime} \times \left(e^{-\zmin/\gamma c\tau_{\Ap}} - e^{-\zmax/\gamma c\tau_{\Ap}}\right),
\eeq
where $N_{\Ap}$ is the total $\Ap$ yield and the second factor is the probability 
of $\Ap$ to decay between $\zmin$ and $\zmax$, $\gamma$ is the 
typical $\Ap$ boost and $c \tau_{\Ap}$ is its proper decay length.
The total dark photon yield for $10^{16}$ EOT is then approximately
\beq
N_{\Ap} \approx 7\times \left(\frac{\epsilon}{10^{-5}}\right)^2 \left(\frac{100\;{\rm MeV}}{\mAp}\right)^2,
\eeq
while the $\Ap$ decay length can be estimated to be
\beq
\gamma c\tau_{A^\prime} \approx 65\;\mathrm{cm}\times\left( \frac{E_{A^\prime}}{8\; {\rm GeV}  }\right) \left(\frac{10^{-5}}{\epsilon}\right)^{2} \left(\frac{100\;{\rm MeV}}{m_{A^\prime}}\right)^2,
\label{eq:dark_photon_lifetime}
\eeq
where we normalized the $A^\prime$ energy at production to the nominal 
LDMX Phase II beam energy (recall that for $m_{A^\prime} > m_e$, the dark 
photon carries away most of the beam energy~\cite{Bjorken:2009mm}). 
This lifetime is in the interesting range for an LDMX-style experiment for both visible 
and missing-momentum signals. We show the projected sensitivity 
of Phase II of LDMX to this scenario in Fig.~\ref{fig:ldmx_reach_minimal_dp} for $8$ and $16$ GeV beams along with 
existing constraints from beam dump experiments~\cite{Andreas:2012mt}, 
NA48/2~\cite{Batley:2015lha}, LHCb~\cite{Aaij:2017rft} and BaBar~\cite{Lees:2014xha}.
There are many on-going and proposed searches for the minimal $\Ap$ scenario targeting different 
regions of parameter space. We show the sensitivity of the following representative subset in Fig.~\ref{fig:ldmx_reach_minimal_dp}: 
the displaced vertex search at HPS~\cite{Battaglieri:2017aum}, displaced decays at an upgraded version of SeaQuest~\cite{Berlin:2018pwi}, dilepton resonance search at Belle II, and LHCb $D^*$ and inclusive searches~\cite{Ilten:2015hya,Ilten:2016tkc}. 
The Belle II reach is 
  estimated from the BaBar result~\cite{Lees:2014xha} by a simple rescaling, assuming $50\;\mathrm{ab}^{-1}$ integrated luminosity and 
  a better invariant mass resolution as described in Refs.~\cite{HeartyCV,Battaglieri:2017aum}.
A more complete list of planned and upcoming experiments can be found in Refs.~\cite{alexander:2016aln,Battaglieri:2017aum}.

\begin{figure}[t!]
  \centering
  \includegraphics[width=0.45\textwidth]{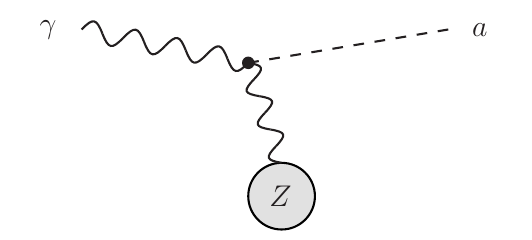}
  \includegraphics[width=0.45\textwidth]{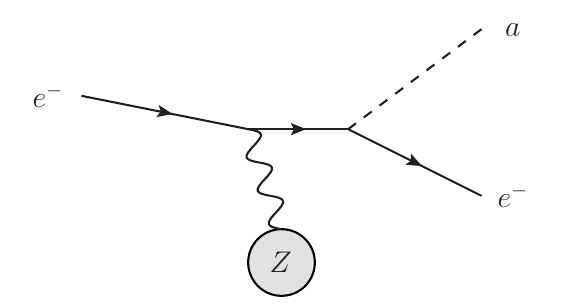}
\caption{
  Axion-like particle production at an electron fixed-target experiment in the photon (left) and electron-coupling (right) 
 dominated regimes. In the left panel, a secondary bremsstrahlung photon undergoes 
 Primakoff conversion in the electric field of a nucleus. In the right panel, an axion is emitted as bremsstrahlung 
 radiation in an electron-nucleus collision.
  \label{fig:alp_production_fixed_target}
}
\end{figure}

\subsection{Axion-like Particles}
New pseudo-scalar bosons interacting 
with pairs of SM gauge bosons are commonly called axion-like particles (ALPs).
ALPs arise as pseudo-Nambu-Goldstone bosons (pNGBs) of spontaneously broken 
global symmetries, and as zero-modes of antisymmetric tensor fields in string theory~\cite{Jaeckel:2010ni}.
We follow the notation of Ref.~\cite{Dobrich:2015jyk} and 
parametrize the low-scale interactions of an ALP $a$ with photons and electrons as  
\beq
\mathscr{L} \supset \frac{1}{4\Lambda_\gamma} a F_{\mu\nu} \widetilde{F}^{\mu\nu}
+ \frac{\partial_\mu a }{\Lambda_e} \bar e \gamma_\mu \gamma_5 e.
\label{eq:alp_model}
\eeq
In realistic models both types of couplings are present (as well as interactions 
with other fermions), with the electron coupling $\Lambda_e \sim f_a$ and the photon 
interaction generated at one loop, such that 
\beq
\frac{1}{\Lambda_\gamma} \sim \frac{\alpha}{4\pi f_a}, 
\eeq
i.e., the fundamental scale $f_a$ is \emph{smaller} than $\Lambda_\gamma$ by a loop factor. 
We will consider the photon and electron couplings as independent and 
investigate the limiting cases where only one of the two interactions dominates.
The main production mechanism in fixed-target experiments (with a sufficiently thick target) 
is through secondary photons in the photon-coupling dominated case and via 
direct bremsstrahlung in the electron case. The two processes are shown in Fig.~\ref{fig:alp_production_fixed_target}.
The lab-frame decay length is given by
\beq
  \gamma c\tau_a = \begin{cases} 
    32\;\mathrm{cm}\times\left( \frac{E_a}{8\;    {\rm GeV}   }\right) \left(\frac{\Lambda_\gamma}{10^4\;   {\rm GeV}  }\right)^2 \left(\frac{100\;{\rm MeV}}{m_a}\right)^4 & \gamma\text{-dominated} \\
    15\;\mathrm{cm}\times\left( \frac{E_a}{8\;   {\rm GeV}  }\right) \left(\frac{\Lambda_e}{10^2\;  {\rm GeV} }\right)^2 \left(\frac{100\;{\rm MeV}}{m_a}\right)^2 & e\text{-dominated}.
  \end{cases}
\label{eq:alp_lifetime}
\eeq
Eq.~\ref{eq:alp_lifetime} demonstrates
that in both limits, the ALP lifetime is in the experimentally interesting range for an 
LDMX-style experiment. The total yield of ALPs can also be estimated as
\beq
N_a \approx \begin{cases}
  90 \times \left(\frac{10^4\;   {\rm GeV} }{\Lambda_\gamma}\right)^2& \gamma\text{-dominated} \\
    8\times \left(\frac{100\;  {\rm GeV} }{\Lambda_e}\right)^2 \left( \frac{100\;{\rm MeV}}{m_a}\right)^2 & e\text{-dominated}.
\end{cases}
\eeq

\begin{figure}[htbp]
  \centering
  \includegraphics[width=0.49\textwidth]{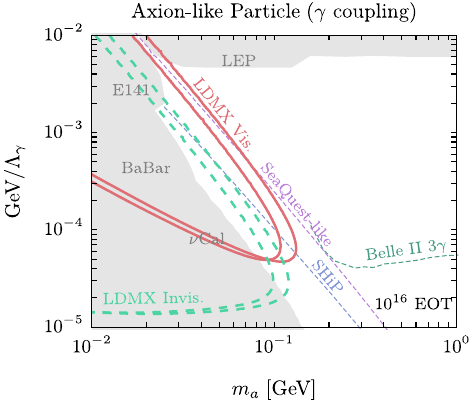}
  \includegraphics[width=0.49\textwidth]{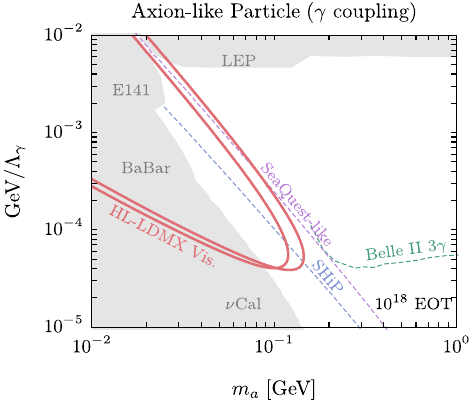}\\
  \vspace{0.4cm}
  \includegraphics[width=0.49\textwidth]{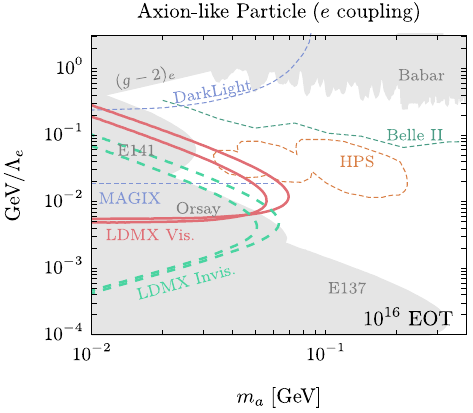}
  \includegraphics[width=0.49\textwidth]{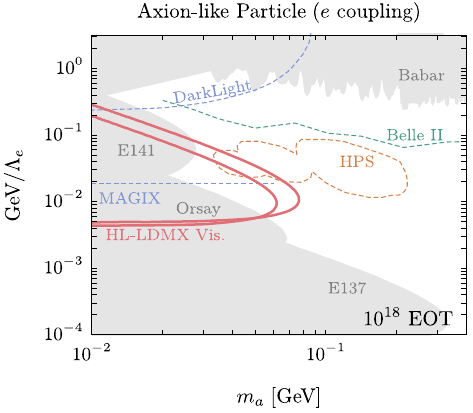}
\caption{Sensitivity of an LDMX-style experiment to axion-like particles (ALPs) dominantly coupled to photons (top row) or electrons (bottom row) via late-decay and invisible channels. 
  The solid red lines show the 95\% C.L. reach of a 
  search for late decays inside of the detector (assuming late $\gamma$ conversion background), while the green-dashed lines correspond to the missing momentum channel where 
the ALP decays outside of the detector.
  In both cases, the two sets of lines correspond to $8$ and $16$ GeV beams, with $\Ebeam=16\;  {\rm GeV} $ having 
  slighter better reach in mass; the left (right) column assumes $10^{16}$ ($10^{18}$) EOT.
  The high-luminosity configuration ($10^{18}$ EOT) must forgo single electron tracking, so the missing momentum 
search (and the use of $p_T$ as a background discriminant in the visible channel) is not possible.
  In the top row, recasts of constraints from beam dump experiments E141, E137, $\nu$Cal, 
  and the BaBar monophoton search from Ref.~\cite{Dolan:2017osp}, and LEP~\cite{Jaeckel:2015jla} are shown as gray regions. 
  Projections for SHiP~\cite{Dobrich:2015jyk}, a SeaQuest-like experiment with sensitivity to $\gamma\gamma$ final states~\cite{Berlin:2018pwi}, Belle II 3 photon search ($50\;\mathrm{ab}^{-1}$ integrated luminosity)~\cite{Dolan:2017osp} 
  are shown as thin dashed lines. In the bottom row, existing constraints from E141, Orsay, BaBar~\cite{Lees:2014xha} and electron $g-2$ 
  are shaded in gray, while the estimated sensitivities of DarkLight~\cite{balewski:2014pxa}, HPS~\cite{Battaglieri:2017aum}, MAGIX~\cite{Denig:2016dqo,Battaglieri:2017aum} 
  and Belle II are indicated as thin dashed lines.
  \label{fig:ldmx_reach_alp}
}
\end{figure}

We show the sensitivity of an LDMX-style experiment to photon- and electron-coupled 
ALPs in Fig.~\ref{fig:ldmx_reach_alp} along with existing constraints and future projections.
For the photon-coupled ALPs (top row) we show recasts of constraints from beam dump experiments E141, E137, $\nu$Cal, 
  and the BaBar monophoton search from Ref.~\cite{Dolan:2017osp}, and LEP~\cite{Jaeckel:2015jla} as gray-shaded regions. 
  Projections for SHiP~\cite{Dobrich:2015jyk}, a SeaQuest-like experiment sensitive to $\gamma\gamma$ final states~\cite{Berlin:2018pwi} and the proposed Belle II 3 photon search ($50\;\mathrm{ab}^{-1}$ integrated luminosity)~\cite{Dolan:2017osp} 
  are shown as thin dashed lines. For the electron-coupled ALPs (bottom row) existing constraints come 
  from E141, Orsay, BaBar~\cite{Lees:2014xha} and electron $g-2$; we also show the estimated sensitivity of DarkLight~\cite{balewski:2014pxa}, 
  HPS~\cite{Battaglieri:2017aum}, MAGIX~\cite{Denig:2016dqo,Battaglieri:2017aum} and Belle II indicated as thin dashed lines.

Recently Ref.~\cite{Alves:2017avw} pointed out that a QCD axion with 
a mass of $\lesssim 30\;{\rm MeV}$ and $\Lambda_e \sim 0.1-1\;   {\rm GeV}  $ \emph{might} still be viable due to model-dependence 
of existing constraints and hadronic uncertainties. This QCD axion window lies precisely in the 
region of parameter space that will be probed by planned experiments shown in Fig.~\ref{fig:ldmx_reach_alp}. 
However, due to the low scale associated with Pecci-Quinn symmetry breaking, UV completions of 
this class of axion models can significantly alter certain bounds, such as the $(g-2)_e$ exclusion 
in Fig.~\ref{fig:ldmx_reach_alp} - see Ref.~\cite{Alves:2017avw}.
It is also important to note that certain completions of the simplified model in Eq.~\ref{eq:alp_model} 
can lead to exotic Higgs decays to $Za$ and $aa$, and $Z$ boson decays to $\gamma a$~\cite{Bauer:2017ris}. 
For large enough couplings, these processes can provide complementary coverage to the existing beam-dump constraints.
In particular, if the ALP-photon coupling arises in the electroweak-preserving phase from interactions with hypercharge and $W$ bosons, 
one expects comparable ALP couplings to $\gamma\gamma$ and $Z\gamma$, unless the latter is tuned to be small. 
In this case, part of the $\gamma$-coupled ALP parameter space with $m_a \lesssim 100\;{\rm MeV}$ and $\Lambda_\gamma \lesssim 1\;{\rm TeV}$
has been probed by LHC searches for $Z\rightarrow \gamma(a\rightarrow \gamma\gamma)$~\cite{Bauer:2017ris}.

\subsection{Strongly-Interacting Models}
\label{sec:visible_simps}
Strongly interacting dark sectors introduced in Sec.~\ref{sec:simps} 
can be searched for both in missing momentum and late visible decay 
channels. The missing momentum signal was already discussed in Sec.~\ref{sec:simps}. 
The visible signal arises from the decay chain 
$\Ap\rightarrow V \pi$, followed by $V\rightarrow \ell^+ \ell^-$.
The interactions leading to this process are specified in Eq.~\ref{eq:simp_model}.
In particular, the visible decay of the vector meson $V$ is made possible by the 
$V-\Ap$ kinetic mixing (an analogue of the $\rho-\gamma$ mixing in the SM), which is 
suppressed by $g_D/g_V$, where $g_D < 1$ is a small perturbative $U(1)_D$ gauge coupling, while 
$g_V = m_V/(\sqrt{2} f_\pi)$ is a coupling strength in the strongly-interacting sector 
and can easily be of order a $\sim$ few (for example in the SM, $g_\rho \approx 6$).
This means that the $V$ lifetime is enhanced compared to a minimal dark photon of a similar mass 
(as long as $m_V < 2m_\pi$). Thus, for moderate values of the SM-$\Ap$ kinetic mixing $\epsilon \gtrsim 10^{-4}$ 
a subset of hidden sector vector mesons will be long-lived on the length-scale of an LDMX-like experiment.
Because their decay length is not uniquely determined by $\epsilon$, these states can be abundantly produced 
while having a long lifetime (in contrast to minimal models where $\epsilon$ controls both the production 
and decay rates).
The projected reach of a search for the displaced decay $V\rightarrow \ell^+ \ell^-$ for a representative choice of parameters 
is shown in Fig.~\ref{fig:simp_reach} as 
purple solid and dashed lines, corresponding to 8 and 16 GeV beam energies. The shaded regions are excluded 
by existing data, while dotted lines show the sensitivity of future collider and 
beam-dump experiments as described in Sec.~\ref{sec:simps}.

\section{Conclusions}
\label{sec:conclusions}

In this paper, we investigated a broad range of sub-GeV dark sector scenarios and evaluated the sensitivity of small-scale accelerator experiments and applicable direct detection efforts. Our focus was on the keV-GeV mass range, and our primary goal was to understand the range of new physics sensitivity provided by the inclusive missing momentum measurement proposed by the accelerator-based Light Dark Matter eXperiment (LDMX). Our analysis revealed that LDMX is sensitive to a very broad range of important but unexplored thermal and non-thermal dark matter, very weakly coupled millicharges, and invisible decays of vector and scalar mediator particles. We also analyzed sensitivity to the production and {\it visible} decays of new particles. We found that LDMX can explore many decades of new coupling and mass territory for well-motivated mediator particles like axions with either photon or electron couplings, visibly decaying dark photons and other gauge bosons, and Higgs-like or simplified model inspired scalars. 

LDMX both complements and dramatically expands on the sensitivity provided by other near term accelerator opportunities like Belle II and ongoing direct detection opportunities. When combined, important new territory in nearly all of the scenarios of keV-GeV dark matter and light dark sector mediator particles discussed in~\cite{Battaglieri:2017aum} can be explored, with exciting discovery prospects. Among these, many of the simplest and most compelling, such as direct SM freeze-out of light dark matter, can be explored with a satisfying degree of breadth and sensitivity. 

There are several directions for future work. First, LDMX and other accelerator experiments hold the promise to measure many aspects of dark sector physics in the case of a discovery. For example, kinematic measurements in LDMX could be used to measure the mass and coupling of the mediator particle responsible for dark matter scattering. In addition, measurements of the production of dark matter through off-shell mediator exchange could independently be used to measure the dark matter mass and the coupling of the mediator to dark matter. Similar measurements in Belle II and beam dump experiments could be used to independently measure a subset of such parameters, and direct detection experiments could be used to corroborate the overall mass scale and expected scattering rate, as well as verify the cosmological stability of the newly discovered particles. It would be interesting to understand, even roughly, how well such measurements could be performed, and to understand the discriminating power of the overall program of experiments discussed here and in the US Cosmic Visions Report~\cite{Battaglieri:2017aum}. 

\begin{acknowledgments}
We would like to thank members of the LDMX Collaboration for many important experimental insights that directly aided in the theoretical studies presented in this paper. We especially thank Torsten Akesson, Owen Colegrove, Giulia Collura, Valentina Dutta, Bertrand Echenard, Joshua Hiltbrand, David Hitlin, Joseph Incandela, John Jaros, Robert Johnson, Jeremiah Mans, Takashi Maruyama, Jeremy McCormick, Omar Moreno, Timothy Nelson, Gavin Niendorf, Reese Petersen, Ruth Pottgen, Nhan Tran, and Andrew Whitbeck.  AB, NB, PS and NT are supported by the U.S. Department of Energy under Contract No. DE-AC02-76SF00515. 
Fermilab is operated by Fermi Research Alliance, LLC, under Contract No. DE-AC02-07CH11359 with the US Department of Energy.
Part of this work was completed at the Kavli Institute for Theoretical Physics, which is supported in part by the National Science Foundation under Grant No. NSF PHY-1748958. We also thank David Morrissey, David McKeen, Maxim Pospelov, Tom Rizzo, Gustavo Marques Tavares and Yu-Dai Tsai for valuable discussions. NB thanks TRIUMF for hospitality during the completion of this work.
\end{acknowledgments}

\bibliographystyle{JHEP}
\bibliography{bibliography}

\end{document}